\documentclass[aip,pof,graphicx]{revtex4-1}

\usepackage{natbib}
\usepackage{epsfig}
\usepackage{subeqn}
\usepackage{subfigure}
\usepackage{graphicx}
\usepackage{bm}
\usepackage{color}

\begin{document}

\title {Coupled Ostrovsky equations for internal waves in a shear flow}

\author{A. Alias}
\affiliation{ School of Informatics and Applied Mathematics, Universiti Malaysia Terengganu,  21030 Terengganu, Malaysia}
\affiliation{Department of Mathematical Sciences, Loughborough University, Loughborough LE11 3TU, UK}

\author{R. H. J. Grimshaw}
\affiliation{Department of Mathematical Sciences, Loughborough University, Loughborough LE11 3TU, UK}

\author{K.R. Khusnutdinova}
\thanks{Corresponding author. Electronic mail: K.Khusnutdinova@lboro.ac.uk.}
\affiliation{Department of Mathematical Sciences, Loughborough University, Loughborough LE11 3TU, UK}

\date{\today}

\begin{abstract}
 In the context of  fluid flows, the coupled Ostrovsky equations arise when two distinct linear long wave modes have nearly coincident phase speeds in the presence of background rotation. In this paper, nonlinear waves in a stratified fluid in the presence of shear flow are investigated both analytically, using techniques from asymptotic perturbation theory, and through numerical simulations. The dispersion relation of the system, based 
 on a three-layer model of  a stratified shear flow, reveals various dynamical behaviours, 
  including the existence of unsteady and steady envelope wave packets.
\end{abstract}


\maketitle 


\bigskip

\section{Introduction}
It is widely known that the Korteweg-de Vries (KdV) equation, with various extensions, is a canonical model for the description of 
 the nonlinear  internal waves  that are commonly observed in the oceans, 
see the reviews \citet{Grimshaw01, Helfrich06, GOSS98} and references therein. 
The KdV equation is  developed for weakly nonlinear long waves, and importantly in the 
context of this paper, is derived on the assumption that the dynamics is dominated by a {\bf single} linear long wave mode.
When background rotation is included, the KdV equation is replaced by  the Ostrovsky equation, see \citet{Ostrovsky78, Leonov81, Helfrich07, Grimshaw85, Grimshaw13a}, given by, in a reference frame moving with the linear long wave phase speed, 
\begin{equation}\label{O}
\displaystyle \{A_t  +  \nu AA_x + \lambda A_{xxx}\}_x  =  \gamma A   ,
\end{equation}
where $\gamma$ is the rotation coefficient, and $\nu$ and $\lambda$ are the nonlinearity and dispersion coefficients, respectively.
Here, $A(x,t)$ is the amplitude of the linear long wave mode $\phi (z)$  corresponding
to the linear long wave phase speed $c$, which is determined from the modal equations
\begin{eqnarray}
& &\displaystyle (\rho_0 W^2 \phi_{z} )_z + \rho_0 N^2 \phi = 0 \,, \label{modal0} \\
& &\displaystyle \phi = 0  \quad \text{at} \quad z =-h\,, \quad \hbox{and} \quad
W^2 \phi_{z }= g\phi \quad \text{at} \quad z =0 \,. \label{modal0bc}
\end{eqnarray}
Here $\rho_{0} (z)$ is the stable background density stratification, 
$\rho_0 N^2 = -g\rho_{0z}$, $W=c-u_{0}$ where $u_{0}(z)$ is the background shear flow, and it is assumed that
there are no critical levels, that is $W\ne 0$ for any $z$ in the flow domain. 
The coefficients  are given by 
\begin{eqnarray}
 I\nu =  3\, \int_{-h}^0\rho_0 W^2 \phi_z^3\, dz  \,, \quad
 I\lambda =  \int_{-h}^0\rho_0 W^2 \phi^2\, dz \,,  \quad 
 I\gamma =    f^2 \int^{0}_{-h} \, \rho_{0} \Phi \phi_{z}  \, dz \,, \label{coeff} 
 \end{eqnarray}
 where
 \begin{eqnarray}
 I =  2\, \int_{-h}^0 \rho_0 W \phi_z^2 \, dz \,, \quad 
\rho_0 W\Phi  = \rho_0 W\phi_{z}  -(\rho_{0}u_{0})_z \phi  \,, \label{Phi} 
\end{eqnarray} 
and $f$ is the Coriolis parameter. Note that when there is no shear flow, that is $u_0 (z) \equiv 0$, then $\Phi \equiv \phi_z $ and 
$\gamma = f^2/2c $;  in this case $\lambda \gamma >0$. 

\begin{figure}[htbp]
\begin{center}
\includegraphics[height=1.8in,width=2.5in]{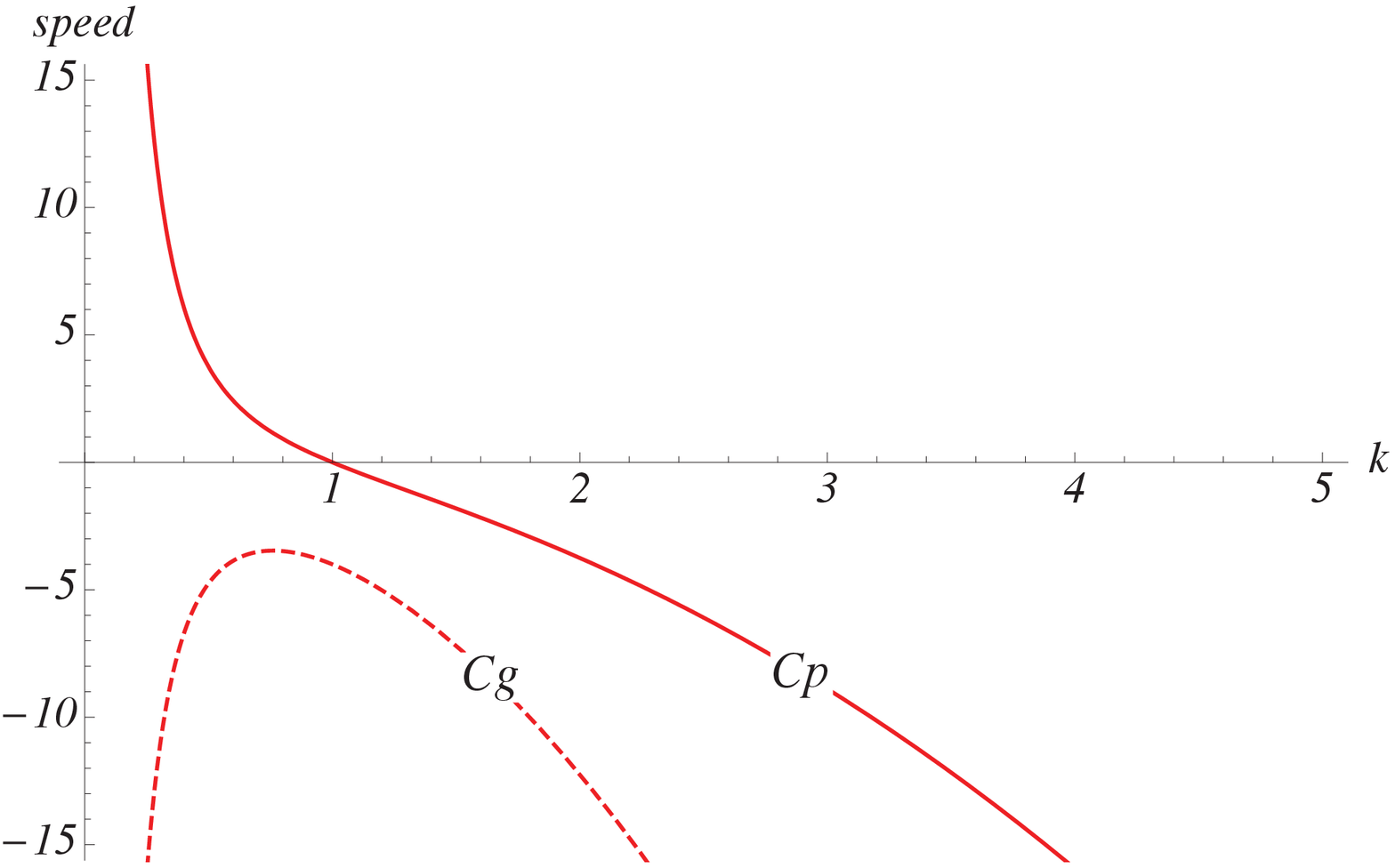}
\includegraphics[height=2in,width=3in]{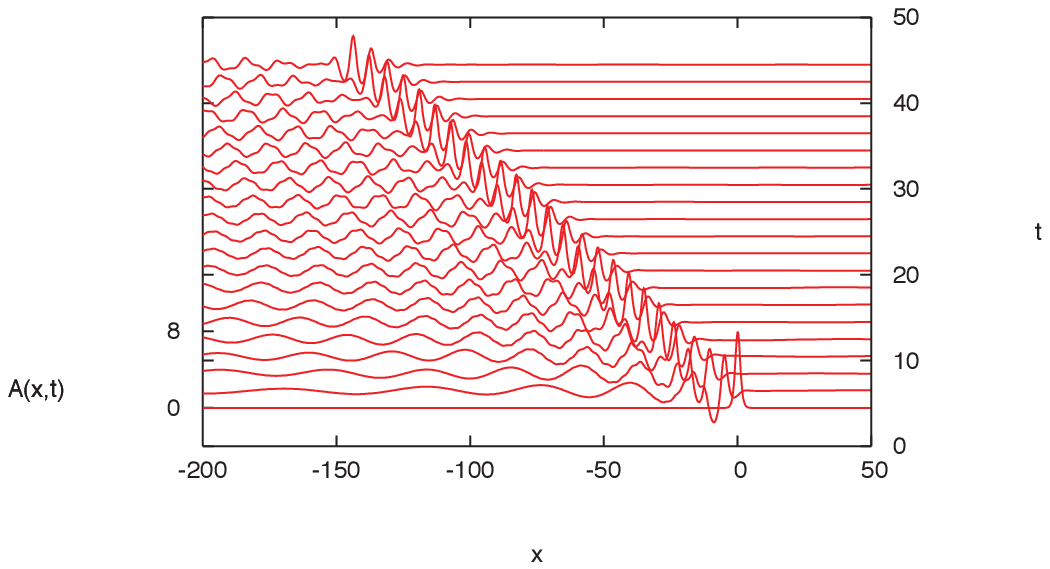}
 \caption{(Left): Dispersion relation for the  Ostrovsky equation (\ref{O}) when $\lambda \gamma >0$ with $\lambda=\gamma=1$. (Right): Numerical solution of the Ostrovsky equation for an initial condition given by a KdV solitary wave with amplitude $8$ at $x=0$. }
\label{fig:DR_OE_unsteady}
\end{center}
\end{figure}

\begin{figure}[htbp]
\begin{center}
\includegraphics[height=1.8in,width=2.5in]{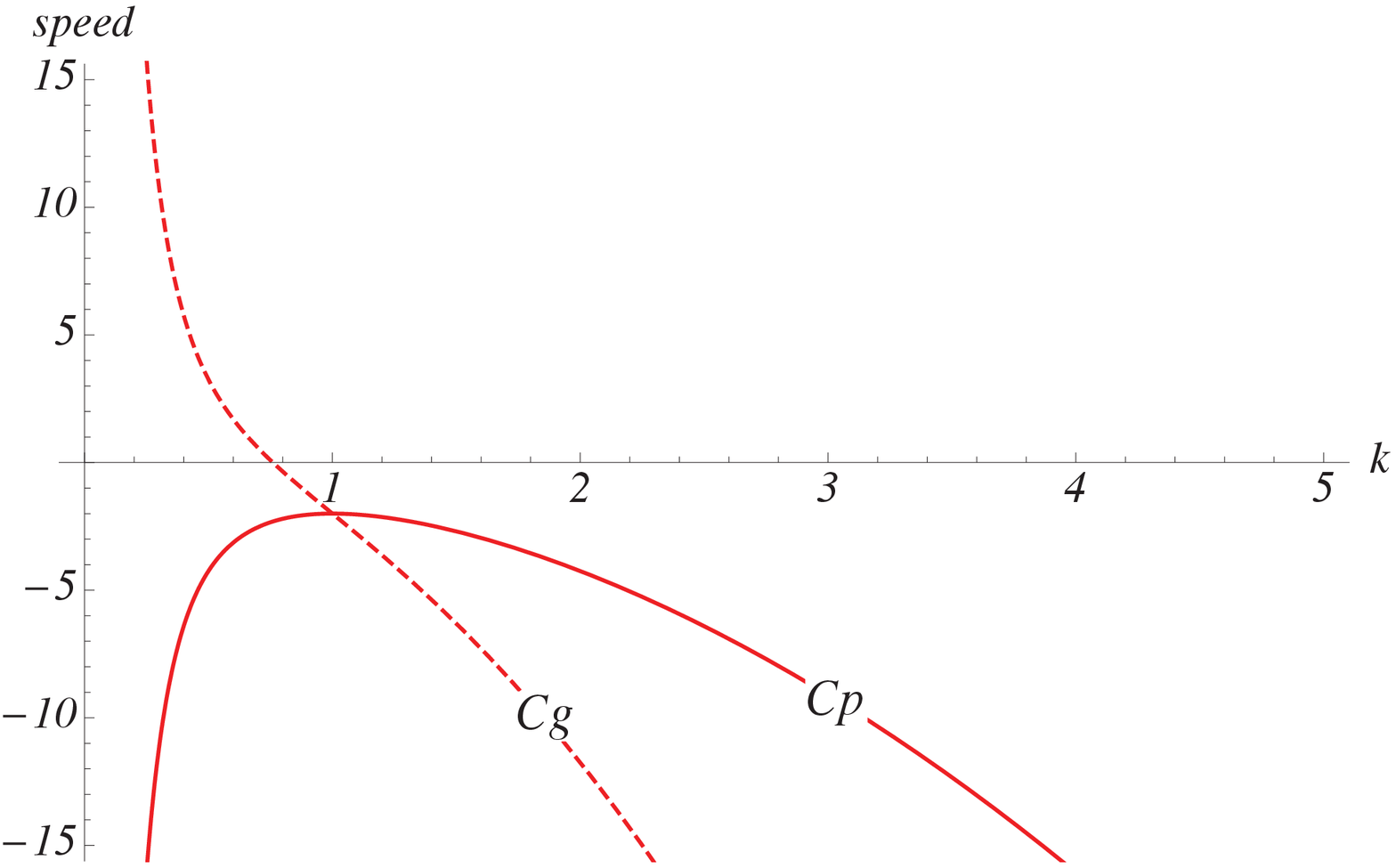}
\includegraphics[height=2in,width=3in]{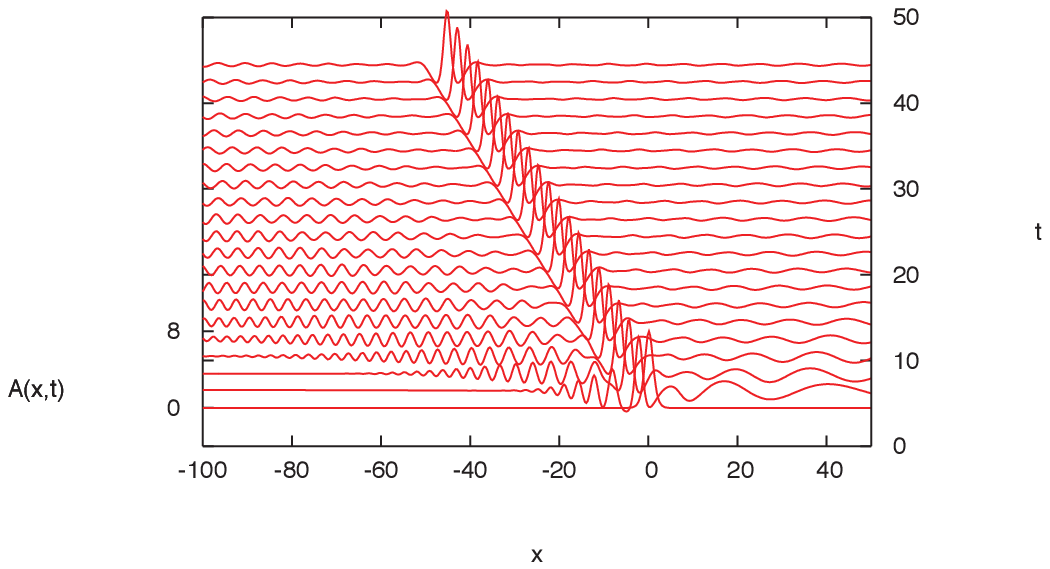}
 \caption{(Left): Dispersion relation for the  Ostrovsky equation (\ref{O}) when $\lambda \gamma <0$ with $\lambda=-\gamma=1$.  (Right): Numerical solution of the Ostrovsky equation for an initial condition given by a KdV solitary wave with amplitude $8$ at $x=0$. }
\label{fig:DR_OE_steady}
\end{center}
\end{figure}

The effect of the Earth's rotation for the time evolution of an internal wave becomes important 
when the wave propagates for several inertial periods. For oceanic  internal  waves, in the absence of a shear flow,  
$\lambda \gamma > 0$,  
and then it is known that there are no steady solitary wave solutions of equation (\ref{O}),
see \citet{Grimshaw12} and the references therein. Recently, it was established that
 the long-time effect of rotation in this case is the destruction of the 
initial  internal solitary wave by the radiation of  small-amplitude inertia-gravity waves, 
and the emergence of a  propagating unsteady nonlinear wave packet, associated with the extremum of the group speed, 
see \citet{Grimshaw12,GHO98,Helfrich07,Grimshaw08}. The same phenomenon was observed independently by 
\citet{YK01} in the context of waves in solids. Indeed, the discrete model  in \citet{YK01} can be related to a two-directional generalisation of the Ostrovsky equation derived in \citet{G96}. A typical linear dispersion curve and numerical simulation is shown in Figure \ref{fig:DR_OE_unsteady}. On the other hand, when  $\lambda \gamma < 0$ 
the Ostrovsky equation (\ref{O}) can support steady envelope wave 
packets, associated with an extremum of the phase speed, see \citet{GS} and \citet{OS}. 
Here a typical case is shown in Figure 
\ref{fig:DR_OE_steady}.  We note that  \citet{OS} derived
this case for magneto-acoustic waves in a rotating plasma.  
Although this case is not relevant to the ocean in the absence of current shear,
as a by-product of the analysis presented here, 
we will show that sufficiently strong shear near a pycnocline may lead to situations where $\lambda \gamma < 0$.

It is known that for internal waves it is possible for the phase speeds of different modes to be nearly coincident, 
and then there will be a resonant transfer of energy between the waves, see \citet{E61}. In this case,  the KdV equation is replaced by two coupled KdV equations, describing a strong interaction between internal solitary waves of different modes, see \citet{GG84,G13}.  Various families of solitary waves are supported by coupled KdV equations depending on the structure of 
 the linear dispersion relation: pure solitary waves, generalised solitary waves and envelope solitary waves, see the review  \citet{G13}.
 In \citet{Alias2013} we extended the derivation of the coupled KdV equations  to take account of background rotation, and also a background
shear flow. We found that then the single Ostrovsky equation (\ref{O}) is replaced  by two coupled Ostrovsky equations, 
each equation having both  linear and nonlinear coupling terms, given by
\begin{eqnarray}
&& I_1 (A_{1\tau}  + \mu_1 A_1 A_{1s} + \lambda_1 A_{1sss}  - \gamma_1 B_1)  \nonumber \\
&& \qquad \qquad \qquad  + \nu_{1} [A_{1}A_{2}]_s  + \nu_2 A_2 A_{2s} + \lambda_{12} A_{2sss} -  \gamma_{12} B_2 =0  \,,  \label{OE1} \\
&& I_2 (A_{2\tau}  + \mu_2 A_2 A_{2s} + \lambda_2 A_{2sss} + \Delta A_{2s} - \gamma_2 B_2 )  \nonumber \\
&& \qquad \qquad \qquad + \nu_{2} [A_{1}A_{2}]_s  + \nu_1 A_1 A_{1s} + \lambda_{21} A_{1sss} -\gamma_{21} B_1 =0 \,,  \label{OE2}
\end{eqnarray}
where ${B_1}_s = A_1, {B_2}_s = A_2$.  The derivation of (\ref{OE1},\ref{OE2}) from the fully nonlinear Euler equations  is briefly described in subsection \ref{derivation}, and more fully in \citet{Alias2013}. 
Coupled Ostrovsky equations  also arise in the context of  waves in layered elastic waveguides, see \citet{KSZ09,KM11}. 
Thus, this model belongs to the class of canonical mathematical models for nonlinear waves, 
inviting a detailed study of the dynamics of its solutions. 

In our previous paper \citet{Alias2013} we examined in detail the case when there is no background shear flow, and then
the coefficients $\gamma, \nu $ vanish and $\beta = \mu$, leading to a simplification of the underlying linear dispersion relation.
In this paper, we restore the background shear flow, and find that the range of dynamical behaviours is then greatly extended.
The rest of the paper is organised as follows. In section \ref{derivation} we briefly  overview the derivation of a pair of coupled Ostrovsky equations  from the complete set of equations of motion  for an inviscid, incompressible, density stratified fluid with boundary conditions appropriate to an oceanic situation, using the asymptotic multiple-scales expansions.  The effect of background shear is examined using a three-layer model in section \ref{3layerflow}.
 In section \ref{dispersion} we analyse various cases for the linear dispersion relation. 
 In section \ref{numerics},  based on the analysis of the linear dispersion relation, 
 we present some numerical simulations using a pseudo-spectral method. Some conclusions are drawn in section \ref{discussion}.

Our results show that a background shear flow allows for  configurations when initial KdV solitary-like waves in the coupled system  are destroyed, 
and replaced by a variety of nonlinear envelope wave packets. Two principal types are found;
first there are unsteady envelope wave packets, which constitute a two-component counterpart of the 
outcome for the single Ostrovsky equation (\ref{O}) with $\lambda \gamma > 0$ and are 
associated with an extremum for the group velocity; second, there are steady wave packets, 
which are not found for  the single Ostrovsky equation with $\lambda \gamma > 0$, are associated with an 
extremum in the phase velocity,  and constitute a two-component counterpart of the outcome for the single 
Ostrovsky equation (\ref{O}) when $\lambda \gamma < 0$.  
Overall, the dynamics of solutions of the coupled equations is much more complicated. 
However, the main features of the complex dynamics observed in numerical simulations can be classified and explained in terms of the  behaviour of the relevant dispersion curves.

 \section{Coupled Ostrovsky equations}

\subsection{Derivation}\label{derivation}

We consider the  two-dimensional flow of an inviscid, incompressible fluid on an $f$-plane. 
In the basic state the fluid has a density stratification $\rho_{0}(z)$, a corresponding pressure $p_{0}(z)$ such that $p_{0 z}=-g \rho_{0}$ and a horizontal shear flow $u_0(z)$ in the $x$-direction. When $u_{0} \ne 0$, this basic state is maintained by a body force.
Then the equations of motion relative to this basic state are given by
\begin{eqnarray}
 \rho_0(u_{t }+u_0u_x+wu_{0z})+p_x &=& 
 -(\rho_0+\rho)(uu_x+ wu_z - fv)-\rho(u_{t }+u_0u_x+wu_{0z})
  \,, \label{eq1}\\
 \rho_0(v_{t }+u_0v_x + fu) + \rho f u_{0} &=& -(\rho_0+\rho)(uv_x+wv_z )-\rho(v_{t }+u_0v_x) -\rho f u \, , \label{eq2} \\
p_z+g\rho &=& -(\rho_0+\rho)(w_{t }+(u_0+u)w_x+ww_z) \,, \label{eq3} \\ 
 g(\rho_{t }+u_0\rho_x)-\rho_0N^2w &=& -g(u\rho_x+w\rho_z) \,, \label{eq4}  \\
 u_x+w_z &=& 0 \,. \label{eq5}
\end{eqnarray}
Here, the terms $(u_0+u,v,w)$ are the velocity components in the $(x,y,z)$ directions, $\rho_0+\rho$ is the density, $p_0+p$ is the pressure, $t$ is time, $N(z)$ is the buoyancy frequency, defined by $\rho_0N^2=-g\rho_{0z}$ and $f$ is the Coriolis frequency. 
The  free surface and rigid bottom boundary conditions to the above problem are given by
\begin{eqnarray} 
 p_0+p=0 \qquad \text{at}   \qquad z=\eta \,,  \label{bC7} \\
 \eta_{t } + (u_0+u)\eta_x=w \qquad \text{at}   \qquad z=\eta \,, \label{bd3} \\
 w=0 \qquad \text{at}  \qquad z=-h \,. \label{bC6} 
\end{eqnarray}
The constant $h$ denotes the undisturbed depth of the fluid, and the function $\eta$ denotes the displacement  of the free surface from its undisturbed position $z=0$. A new variable $\zeta$ denotes the vertical particle displacement,  which is related to the vertical speed, $w$. It is defined by the equation
\begin{eqnarray}
 \zeta_{t }+(u_0+u)\zeta_x+w\zeta_z=w\label{VPD},
 \end{eqnarray}
 and satisfies the boundary condition
 \begin{eqnarray}
 \zeta = \eta \qquad \text{at} \qquad z = \eta. \label{bd4}
 \end{eqnarray}
 
The system of coupled Ostrovsky equations is derived using the Eulerian formulation, following a similar strategy to the  derivation of coupled KdV equations using the Lagrangian formulation in \citet{GG84, G13}; 
the full derivation can be found in \citet{Alias2013}.
At the leading linear long wave order, and in the absence of any rotation, 
the solution for $\zeta $ is given by an  expression of the form $A(x-ct)\phi(z)$ where the modal function is given by
(\ref{modal0}, \ref{modal0bc}).  In general there is an infinite set of solutions for [$\phi(z), c$].
Here we consider  the case when there are  two modes with 
nearly coincident speeds $c_1 =c$ and $c_2 = c + \epsilon^2 \Delta $, $\epsilon \ll 1$, where 
$\Delta $ is the detuning parameter.  Importantly, we assume that the  modal functions $\phi_1 (z), \phi_2 (z)$
are {\it distinct}, and each satisfy the system (\ref{modal0}, \ref{modal0bc}), that is
\begin{eqnarray}
& & (\rho_0 W_{i}^2 \phi_{iz} )_z + \rho_0 N^2 \phi_i = 0 \,, \quad i =1, 2 \label{modal} \\
& & \phi_i = 0  \quad \text{at} \quad z =-h\,, \quad \hbox{and} \quad
W_{i}^2 \phi_{iz }= g\phi_i  \quad \text{at} \quad z =0 \,. \label{modalbc}
\end{eqnarray}
Here $W_{i} = c_i  - u_{0}(z)$ where $c_i$ is the long wave speed corresponding
 to the mode $\phi_i (z), i = 1,2.$
In the sequel, $W_{i} = W =c -u_{0} (z)$ with an error of order $\epsilon^2 $.

Next we introduce the scaled variables
\begin{eqnarray}
\tau=\epsilon \alpha t  \,, \quad   s=\epsilon (x-c t) \,, \quad f = \alpha \tilde{f}
\end{eqnarray}
where $\alpha=\epsilon^2$ and seek a solution in the form of asymptotic multiple - scales expansions
 \begin{eqnarray}
 (\zeta, u, \rho,  p) &=&  \alpha  (\zeta_1 , u_1 , \rho_1 , p_1 )+ 
\alpha^2  (\zeta_2 , u_2 , \rho_2 , p_2 ) + \cdots \,, \label{exp1} \\
 \qquad (w, v) &=& \alpha \epsilon  (w_1 , v_1) + 
\alpha^2 \epsilon  (w_2 , v_2) + \cdots \,. \label{exp2} 
\end{eqnarray}
Substituting these expansions into the system (\ref{eq1}) - (\ref{eq5}), 
and assuming that two waves $A_1$ and $A_2$ are present at the  leading order, we obtain
\begin{eqnarray}
\zeta_1 &=& A_1 (s, \tau )\phi_1 (z) + A_{2} (s, \tau ) \phi_2 (z) \,,  \label{zeta1}\\
u_1 &=& A_{1} \{W \phi_1 \}_z  + A_{2} \{W \phi_2 \}_z  \,,  \\
w_1 &=& - A_{1s}W \phi_1   - A_{2s}W \phi_2  \,, \\
p_1 &=& \rho_0 A_1 W^2 \phi_{1z} +  \rho_0 A_2 W^2 \phi_{2z} \,, \label{p1}\\
g\rho_1 &=& \rho_0 N^2 \zeta_1 \,, \\
v_1 &=&  \tilde{f}(B_1\Phi_1 + B_2 \Phi_2 ) \,, \quad \rho_0 W\Phi_{1,2}  = \rho_0 W\phi_{1z,2z}  -(\rho_{0}u_{0})_z \phi _{1,2} \,, 
\quad B_{1s,2s}=A_{1,2} \,. \label{v1} 
\end{eqnarray}
Importantly, the exact solution of the linearised equations should contain the exact expressions  $W_1$ and $W_2$ 
in the terms related to the first and second waves, respectively, rather than just $W$. 
This difference between the exact and leading order solutions necessitates the introduction of 
correction terms at the next order, in order to recover the distinct modal equations for the functions $\phi_1$ and $\phi_2$.

Collecting terms of the second order for each equation, and calculating the correction terms originating from the leading order, 
the following equations are obtained,
\begin{eqnarray}
 \rho_0(-Wu_{2s}+u_{0z}w_2)+p_{2s} &=& -\rho_0(u_{1\tau}+u_1u_{1s}+w_1u_{1z}) + \rho_1(Wu_{1s}-u_{0z}w_1)
 + \rho_0 \tilde{f} v_1, \, \, \label{EQ11}\\
 \rho_0(\tilde{f} u_2 -Wv_{2 s}) + \rho_{2}\tilde{f} u_{0}  &=& -\rho_0 (v_{1\tau}  + u_{1}v_{1s} + w_{1}v_{1z}) + \rho_1Wv_{1s} - \rho_1 \tilde{f} u_1   \,,  \label{EQ22}\\
p_{2z}+g\rho_2 &=& \rho_0 Ww_{1s} + 2 \Delta A_2 \lbrace \rho_0W\phi_{2z}\rbrace_z  \,, \label{EQ33}\\
 - gW\rho_{2s}-\rho_0N^2w_2&=&-g(\rho_{1\tau}+u_1\rho_{1s}+w_1\rho_{1z})  \,, \label{EQ44}\\
 u_{2s}+w_{2z}&=& 0 \,,  \label{EQ55} \\
 W \zeta_{2s} + w_2 &=& \zeta_{1\tau} + u_1 \zeta_{1s} + w_1 \zeta_{1z}\,. \label{EQ66}
 \end{eqnarray}
 Similarly, the boundary conditions (\ref{bC6}) - (\ref{bd3}), (\ref{bd4}) yield 
 \begin{eqnarray}
&& w_2 = 0 \quad \mbox{at} \quad z = -h\,, \label{bC62}\\
 &&p_2 -\rho_0 g \eta_2 + p_{1z} \eta_1-\frac{1}{2}\rho_{0z} g \eta_1^2
 - 2  \Delta \rho_0 W\phi_{2z}   A_2 = 0  \quad \mbox{at} \quad z = 0 \,, \label{Pz0} \\
&& w_2 + w_{1z} \eta_1 - \eta_{1\tau} + W \eta_{2s} - u_{0z} \eta_1 \eta_{1s} - u_1 \eta_{1s} = 0 \quad \mbox{at} \quad z = 0\,, \label{bd32}\\
 &&\zeta_2 + \zeta_{1z} \eta_1 - \eta_2 = 0 \quad \mbox{at} \quad z = 0 \,. \label{bd42}
 \end{eqnarray}
 Eliminating all variables in favour of $\zeta_2 $ yields 
 \begin{eqnarray}
&&  \lbrace \rho_0 W^2 \zeta_{2sz} \rbrace_z + \rho_0 N^2 \zeta_{2 s}=M_2 \quad \text{at} \quad -h<z<0  \,, \label{MM} \\
&&  \zeta_2 =0 \quad \text{at} \quad z=-h \,, \quad  
\rho_0 W^2\zeta_{2sz}-\rho_0 g \zeta_{2s}=N_2 \quad \text{at}\quad z=0 \,, \label{NN}
 \end{eqnarray}
 where $M_2 , N_2 $ are  known expressions containing terms in $A_{i}$ and their derivatives.
 The full expressions can be found in \citet{Alias2013}.

 Two compatibility conditions need to be imposed
 on the system (\ref{MM}, \ref{NN}), given by
 \begin{eqnarray}\label{comp}
 \int^{0 }_{-h} \, M_2 \phi_{1,2} \,dz - [N_2 \phi_{1,2}]_{z=0 }  =0 \,,
 \end{eqnarray}
where $\phi_{1,2}$ are evaluated at the leading order. 
These compatibility conditions lead to the coupled Ostrovsky equations
\begin{eqnarray}
&& I_1 (A_{1\tau}  + \mu_1 A_1 A_{1s} + \lambda_1 A_{1sss}  - \gamma_1 B_1)  \nonumber \\
&& \qquad \qquad \qquad  + \nu_{1} [A_{1}A_{2}]_s  + \nu_2 A_2 A_{2s} + \lambda_{12} A_{2sss} -  \gamma_{12} B_2 =0  \,,  \label{O1} \\
&& I_2 (A_{2\tau}  + \mu_2 A_2 A_{2s} + \lambda_2 A_{2sss} + \Delta A_{2s} - \gamma_2 B_2 )  \nonumber \\
&& \qquad \qquad \qquad + \nu_{2} [A_{1}A_{2}]_s  + \nu_1 A_1 A_{1s} + \lambda_{21} A_{1sss} -\gamma_{21} B_1 =0 \,,  \label{O2}
\end{eqnarray}
where ${B_1}_s = A_1, {B_2}_s = A_2$, and the coefficients are given by
\begin{eqnarray}
I_i\mu_i &=&  3\int^{0}_{-h}\, \rho_0 W^2 {\phi_i}^{3}_z \, dz \,, \label{mu} \quad \quad 
I_i \lambda_i = \int^{0}_{-h}\, \rho_0 W^2 {\phi_i}^2 \, dz  \,, \label{lambda} \\
I_i &=& 2\int^{0}_{-h} \, \rho_0 W {\phi_i}^{2}_z \, dz \,, \label{I} \quad \quad \quad
 \lambda_{12} = \lambda_{21} =  \int_{-h}^{0}\, \rho_0 W^2 \phi_{1}\phi_{2} \, dz \,, \\
 \nu_{1} &=&  3\int_{-h}^{0}\, \rho_0 W^2 \phi^{2}_{1z}\phi_{2z} \, dz \,, \quad  \quad 
 \nu_{2} =  3\int_{-h}^{0}\, \rho_0 W^2 \phi^{2}_{2z}\phi_{1z} \, dz \,, \\
\label{gamma}
I_{i} \gamma_{i} &=&   \tilde{f}^2 \int^{0}_{-h} \, \rho_{0} \Phi_{i} \phi_{iz}  \, dz \,, \quad  \quad  \quad
\label{gamma1}
\gamma_{ij} = \tilde{f}^2  \int^{0}_{-h} \, \rho_{0} \Phi_{i} \phi_{jz} \, dz  \,. 
\end{eqnarray}
Here $i, j = 1, 2$.  \\
We scale the dependent and independent variables as  
\begin{equation}\label{scale}
A_1 = \frac{u}{ \mu_1 }\,, \quad A_2 =\frac{ v}{ \mu_2  } \,,
\quad s = \lambda^{1/2}_1 X  \,, \quad \tau = \lambda^{1/2}_1 T    \,, 
\end{equation}
assuming that $\lambda_2 > 0, \lambda_1 \ne 0, \mu_{1,2} \ne 0$ without loss of generality.  
Then  equations (\ref{O1}, \ref{O2})  take the form
\begin{eqnarray}\
&&(u_{T} + u u_X + u_{XXX} + n  (uv)_X + m v v_X + \alpha v_{XXX})_X = \beta u + \gamma v, \label{CO1}\\
&&(v_{T} + v v_X + \delta  v_{XXX} + \Delta   v_X + p  (uv)_X + q  u u_X + \lambda u_{XXX})_X =  \mu v + \nu u, \label{CO2}
\end{eqnarray}
where
\begin{eqnarray}
&& n = \frac{\nu_1}{I_1 \mu_2}, \quad m = \frac{\mu_1 \nu_2}{I_1 \mu_2^2}, \quad 
\alpha = \frac{\lambda_{12} \mu_1}{\lambda_1 I_1 \mu_2}, 
\quad \beta = \gamma_1 \lambda_1 , \quad \gamma = \frac{\gamma_{12} \mu_1 \lambda_1}{I_1 \mu_2 } , \nonumber \\
&& \delta = \frac{\lambda_2}{\lambda_1}, \quad p = \frac{\nu_2}{I_2 \mu_1}, \quad 
q = \frac{\mu_2 \nu_1}{I_2 \mu_1^2}, \quad \lambda = \frac{\lambda_{21} \mu_2}{\lambda_1 I_2 \mu_1}, \quad \mu = \gamma_2 \lambda_1 , 
\quad \nu = \frac{\gamma_{21} \mu_2 \lambda_1}{I_2 \mu_1}. \label{coeffnew}
\end{eqnarray}
Here,  
\begin{equation}\label{coeffrel}
\frac{q}{n} =\frac{p}{m}=\frac{\lambda }{\alpha }= \frac{\gamma_{12} \nu }{\gamma_{21} \gamma} =
\frac{I_1 \mu^{2}_2 }{I_2 \mu^{2}_1 }   \,, \quad 
\frac{\alpha \lambda }{\delta } = \frac{\lambda^{2}_{12}}{\lambda_{1} \lambda_{2} I_1 I_2 }  < 1 \,. 
\end{equation}

Here, the scaled variables $u$ and $v$, and the coefficient $p$ should not be confused with the velocity components and the pressure. Note that with this scaling (\ref{scale}), the scaled variables $X, T$  have dimensions of $C^{-1/2}, C^{-3/2}$ respectively, where $C$ is a velocity scale, i.e $m \, s^{-1}$. The dependent variables $u$ and $v$ have the dimension of $C$.  The coefficients  $n, m, \alpha, \delta , p, q, \lambda $ are dimensionless, while $\beta , \gamma, \mu, \nu $ have dimensions of $C^{2}$, and $\Delta$ has the dimension of $C$. The wavenumber $k$ and $c_p, c_g$ have the dimensions of $C^{1/2}$ and $C$ respectively. In the sequel we omit writing these dimensions for the scaled variables, but write the unscaled physical parameters in dimensional form.

\subsection{Three-layer flow with shear}\label{3layerflow}

\begin{figure}[htbp]
\begin{center}
\includegraphics[height=2in,width=3.5in]{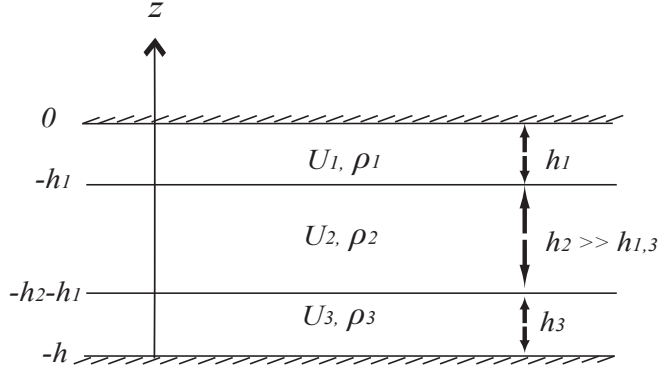}
 \caption{A schematic representation of the three-layer model with shear flow.}
\label{fig:0}
\end{center}
\end{figure}

As an illustrative example with sufficient parameters to explore several cases of interest, we consider a three-layer  background flow, 
$-h< z < 0$, with interfaces at $z=-h_2 - h_1, z=-h_1 $, and 
$h=h_1 + h_2 +h_3 $, shown in Figure \ref{fig:0}. Here, $\rho_0$ and $u_0$ are piecewise-constant density and 
velocity fields, respectively, and they are represented using the Heaviside step-function as follows,
$$\rho_0 (z) = \rho_3 +  (\rho_2  - \rho_3 )H(z + h_2 + h_1 ) + (\rho_1  -\rho_2 )H(z+h_1 ) \,, $$
$$ u_{0} (z) = U_3 +  (U_2  - U_3 )H(z + h_2 +h_1 ) + (U_1  -U_2 )H(z+h_1 )   \,.$$
This three-layer flow  model is not meant to be realistic in the strict sense but is used here  as a guide
for appropriate values of the parameters. The model is chosen to yield explicit formulae, but
can be regarded as a simplification of a background flow with smooth density and shear profiles across the interfaces. In 
the long wave limit we consider we expect this  piecewise model to yield coefficients close to those
which would come from such a smooth model. It is also pertinent to note that this background flow is 
subject to Kelvin-Helmholtz instability, but these arise as short waves, which may occur in reality, 
but are excluded in the long wave system we study here,  due to the large  separation of scales.
With rigid boundaries at $z=-h, 0$,  the modal functions are given by
\begin{eqnarray}\label{mode}
& \phi = A_3 \frac{h + z}{h_3} \,, \quad -h < z <  -h_2 -h_1   \,, \\
& \phi = A_1 \frac{h_1 +h_2  + z}{h_2} - A_3 \frac{h_1 + z}{h_2} \,,\quad -h_2 - h_1 < z < -h_1 \,, \\
& \phi =  -A_1 \frac{z}{h_1} \,, \quad -h_1 < z <  0 \,.
\end{eqnarray}

The modal functions are normalized so that $\phi =A_{1,3}$ at $z=-h_{1}, -h_{1}-h_{2}$.
At each interface there is the jump relation,
$$ [\rho_0 (c-u_{0})^2 \phi_z ] = g[\rho_0 ]\phi \,. $$
This yields the system
\begin{eqnarray}
\{\frac{\rho_{1} (c-U_{1})^2}{h_1 } + \frac{\rho_{2} (c-U_{2})^2}{h_2 }  - g(\rho_2 -\rho_1)\}A_1 - \frac{\rho_{2} (c-U_{2})^2}{h_2 }A_3 &=& 0 \,, \label{A1}\\
 - \frac{\rho_{2} (c-U_{2})^2}{h_2 }A_1 + \{\frac{\rho_{3} (c-U_{3})^2}{h_3 } + \frac{\rho_{2} (c-U_{2})^2}{h_2 }  - g(\rho_3 -\rho_2)\}A_3  &=& 0 \,, \label{A_3} 
\end{eqnarray}
which can be written as
\begin{eqnarray}
&& D_1 A_1 - EA_3 =0 \,, \quad - EA_1 + D_3 A_3 = 0 \,, \label{sys} \\
&& D_1 = \frac{\rho_{1} (c-U_{1})^2}{h_1 } + \frac{\rho_{2} (c-U_{2})^2}{h_2 }  - g(\rho_2 -\rho_1) \,, \label{C6} \\
&& D_3 = \frac{\rho_{3} (c-U_{3})^2}{h_3 } + \frac{\rho_{2} (c-U_{2})^2}{h_2 }  - g(\rho_3 -\rho_2) \,, \label{C7 } \\
&& E=\frac{\rho_{2} (c-U_{2})^2}{h_2 } \,. \label{E}
\end{eqnarray}
Without loss of generality we put $U_2 =0$ henceforth.

The dispersion relation, determining the speed $c$ is then given by
\begin{equation}\label{dispfull} 
D_1 D_3 = E^2 \,. 
\end{equation}
A resonance with two distinct modes requires that $D_1 = D_3 =E =0$ simultaneously.  There are two cases,
either $c=0$ or $h_2 \gg h_{1,3}$. The first case contains implicit critical layers, and hence is not considered here.
The second case is,
\begin{equation}\label{resonanceb}
 h_2 \gg h_{1,3}  \,, \quad  c = U_1 \pm \{\frac{gh_1 (\rho_2 - \rho_1 )}{\rho_1 }\}^{1/2} 
= U_3 \pm \{\frac{gh_3 (\rho_3 - \rho_2)}{\rho_3 }\}^{1/2} \,. 
 \end{equation}
 For given densities $\rho_{1,2,3}$ and layer depths $h_{1,3}$, these determine the allowed shear $U_1 - U_3  $.
There are four cases, 
but in the sequel we consider only the right-propagating waves, choosing $c >0$, which then imposes a constraint
on the allowed choices for $U_1 - U_3 $.   

The modal functions and their derivatives are given by
 \begin{eqnarray}\label{mode1}
 \phi_1 =  0\,, \quad  \phi_{1z} = 0 \,, \quad  & -h < z <  -h_2 -h_1   \,, \\
 \phi_1 =  \frac{h_1 +h_2  + z}{h_2} \,,\quad  \phi_{1z} = \frac{1}{h_2 } \,, \quad & -h_2 - h_1 < z < -h_1 \,, \\
\phi_1 =  - \frac{z}{h_1} \,, \quad \phi_{1z} = -\frac{1}{h_1 } \,, \quad   & -h_1 < z <  0 \,;
\end{eqnarray}
\begin{eqnarray}\label{mode2}
 \phi_2 =  \frac{h + z}{h_3} \,, \quad \phi_{2z} = \frac{1}{h_3 } \,, \quad  & -h < z <  -h_2 -h_1   \,, \\
 \phi_2 = - \frac{h_1 + z}{h_2} \,,\quad \phi_{2z} = -\frac{1}{h_2 }\,, \quad &  -h_2 - h_1 < z < -h_1 \,, \\
\phi_2  = 0 \,, \quad  \phi_{2z} = 0 \,, \quad  &-h_1 < z <  0 \,.
\end{eqnarray}
Note that here the subscripts $1,2$ on the modal functions should not be confused with the subscripts $1,2,3$ for each layer. 
Now all coefficients in the coupled Ostrovsky equations can be calculated, taking into account that $h_2 \gg h_{1,3}$,
where appropriate:
\begin{eqnarray}
I_1\mu_1&=&  -\frac{3 \rho_1(c-U_1 )^2 }{h_{1}^{2}}  \,, \quad  I_2\mu_2 =  \frac{3 \rho_3(c-U_3 )^2 }{h_{3}^{2}} \, , \label{mu} \\
I_1\lambda_1 &=& I_2 \lambda_2 = \frac{c^2 \rho_2 h_2 }{3} \,, \label{lambda} \\
I_1 &=& \frac{2\rho_1 (c-U_1 )}{h_1 } \,, \quad I_2 =  \frac{2\rho_3 (c-U_3 )}{h_3 } \,, \label{I} \\
 \lambda_{12} &=& \lambda_{21} =  \frac{c^2 \rho_2\ h_2 }{6}  \,, \\
 \nu_{1} &=&  \nu_2  = 0 \,.
 \end{eqnarray}
For the coefficients $\gamma_{1,2,12}$ we must evaluate $\Phi_{1,2}$:
\begin{equation}\label{Phi}
\Phi_{1,2}  = \phi_{1z,2z}  -\frac{(\rho_{0}u_{0})_z }{\rho_0 W} \phi _{1,2} \,,
\end{equation}
\begin{eqnarray}
I_{i} \gamma_{i} &=&   \tilde{f}^2 \int^{0}_{-h} \, \rho_{0} \Phi_{i} \phi_{iz}  \, dz \,, \label{gamma} \\
\gamma_{ij} &=& \tilde{f}^2  \int^{0}_{-h} \, \rho_{0} \Phi_{i} \phi_{jz} \, dz  \,. \label{gamma1}
\end{eqnarray}
Here $\rho_0, W =c -u_0 $ are piecewise constant, so the second term in (\ref{Phi})
behaves like a $\delta $-function.  Specifically, write
$$ \Phi_{1,2}  = \phi_{1z,2z} + \{[\log{|W|}]_z  - \frac{u_0 }{W}[\log{\rho_{0}}]_z \} \phi _{1,2}\,, $$
where the last term can be ignored in the Boussinesq approximation,  but is kept here, and we treat $\log |W|$ and $\log \rho_0$ as piecewise - constant functions.
The derivatives of $[\cdots ]$  are  $\delta $-functions, leading to the product of  a $\delta $-function with a 
discontinuous function in (\ref{gamma}, \ref{gamma1}). 
In order to evaluate these expressions we first note that $\phi_{1z}, \phi_{2z}$ are zero except in the upper and bottom layer
respectively, where they are constants,  and  also $\Phi_{1}=0 $ in the bottom layer, and $\Phi_2 = 0 $ in the top layer. Hence
\begin{eqnarray}
I_{1} \gamma_{1} &=&  - \frac{\rho_1 \tilde{f}^2 }{h_1}\int^{0}_{-h} \, \Phi_{1}H(z+h_1)  \, dz  = 
 \frac{\rho_1 \tilde{f}^2 }{h_1}\{1 + \frac{1}{2}\log{[\frac{ |W_2 |}{|W_1 |}]}  -  \frac{U_1 }{2W_1 }\log{[\frac{ \rho_2}{\rho_1 }]} \} \,,
\label{gamma2} \\  
I_{2} \gamma_{2}  &=&    \frac{\rho_3 \tilde{f}^2}{h_3 } \int^{0}_{-h } \,   {\Phi_{2}}H(-z -h_1 -h_2 ) dz =
 \frac{\rho_3\tilde{f}^2 }{h_3}\{1 + \frac{1}{2}\log{[\frac{ |W_2 |}{|W_3 |}]}  - 
 \frac{U_3 }{2W_3 }\log{[\frac{ \rho_2 }{\rho_3 }]} \} \,, \label{gamma3} \\
\gamma_{12} &=& \gamma_{21}  = 0 \,.  \label{gamma4}
\end{eqnarray}
Here we have used the expression that when a $\delta $-function  multiplies a discontinuous function $f(x)$,
$$ \int \, f(x) \delta (x) dx = \frac{1}{2}(f(0+)+f(0-)) \,. $$

Next we  let $g_{1} = g(\rho_2-\rho_1 )/\rho_1$, $ g_{3} = g(\rho_3-\rho_2)/\rho_3 $ and use
the Boussinesq approximation that otherwise $  \rho_1 \approx \rho_2 \approx \rho_3$. 
 We then obtain that,
\begin{eqnarray}
n&=&m=p=q=\gamma=\nu=0\, , \\
\delta&=&\frac{h_3  (c-U_1)}{h_1 (c-U_3)} = -2\alpha \,,  \quad \lambda=-\frac{1}{2}\,, \\ 
\beta&=&{\frac{c^2 h_1 h_2 \tilde{f}^2 [1+\frac{1}{2}\log{|c/(c-U_1) |}]}{12 (c-U_1)^2} }\,, \\
\mu&=&{\frac{c^2 h_1 h_2\tilde{f}^2 [1+\frac{1}{2}\log{|c/(c-U_3) |}]}{12 (c-U_1)(c-U_3)} }\,, \\
\hbox{so that} \quad \mu &=& \beta F \,, \quad F = 
\{\frac{c-U_1 }{c-U_3 }\}\frac{1+\frac{1}{2}\log{|c/(c-U_3) |}}{1+\frac{1}{2}\log{|c/(c-U_1) |}} \,. 
\end{eqnarray}
Then there are four possibilities according to the value of $c$,
 \begin{eqnarray}\label{rescon}
& \hbox{Case}  \,1: \quad c = U_1 + \sqrt{g_1 h_1} = U_3+  \sqrt{g_3 h_3} \,, \\
& \hbox{Case} \,  2:  \quad c = U_1 - \sqrt{g_1 h_1} =U_3- \sqrt{g_3 h_3} \,, \\
& \hbox{Case}  \, 3:  \quad c = U_1 + \sqrt{g_1 h_1} =U_3-  \sqrt{g_3 h_3} \,, \\
& \hbox{Case}  \, 4:  \quad c = U_1 - \sqrt{g_1 h_1}    =  U_3+  \sqrt{g_3 h_3} \,.
\end{eqnarray}
Bearing in mind that a piecewise-constant shear flow is a simplified model of a continuous shear flow, then
 in order to avoid an implicit  critical layer, we choose 
$ c > \hbox{max} [U_1 , 0, U_3 ] \,, $
where we recall that we have set $U_2 = 0$. This condition then implies that only Case 1 is allowed.

In full detail, for Case 1,
\begin{eqnarray}
\delta&=&\sqrt{\frac{g_1 h_3}{g_3 h_1}} =-2\alpha \,, \quad \lambda = - \frac{1}{2} \,, \label{C1a}\\
\beta&=&\frac{h_2  \tilde{f}^2 (\sqrt{g_1 h_1}+U_1)^2 [1+\frac{1}{2}\log{|(\sqrt{g_1 h_1}+U_1)}/\sqrt{g_1 h_1}|]}{12 g_1} \,, \label{C1b} \\
\mu &=& \beta F \,, \quad F = \sqrt{\frac{g_1 h_1}{g_3 h_3}} 
\frac{[1+\frac{1}{2}\log{|(\sqrt{g_3 h_3}+U_3)/\sqrt{g_3 h_3}|}]}{[1+\frac{1}{2}\log{|(\sqrt{g_1 h_1}+U_1)/\sqrt{g_1 h_1}|}]} \,. \label{C1c} 
\end{eqnarray}
Note that $\beta > 0$ unless $U_1 $ is such that:
$$ | 1+ \frac{U_1}{\sqrt{g_1 h_1 }} |< e^{-2}  \,, \quad 
-1 <  \frac{U_1}{\sqrt{g_1 h_1 }} < e^{-2} -1= -0.865  \,, $$
when $\beta < 0$.  Similarly $\mu > 0$ unless $U_3 $ is such that:
$$  |1+ \frac{U_3}{\sqrt{g_3 h_3 }}| < e^{-2}  \,, \quad  
-1 <  \frac{U_3}{\sqrt{g_3 h_3 }} < e^{-2} -1 = -0.865  \,,  $$
when $\mu < 0$. Here we have used the condition for the exclusion of an implicit critical layer.
Note also  that  $U_1 , U_3 $ are constrained by the resonance condition (\ref{rescon}).
Nevertheless, all four possibilities can be realised, that is; Case A: $\beta > 0, \mu >0$, Case B: $\beta > 0, \mu < 0$, 
Case C: $\beta < 0, \mu >0$, Case D: $\beta <  0, \mu < 0$.

Specifically, we  choose $\tilde{f}=5 \times 10^{-3} \, s^{-1}$ and choose $g_{1,3}$ of the order $10^{-1} \leftrightarrow 10^{-3}  \, m \,s^{-2}$. The upper layer and lower depths $h_{1,3}$ are chosen to be of order $50 \leftrightarrow 1000 \, m$.
Next,  we choose $U_1$ and use the resonance condition (\ref{rescon}) to determine the value of $U_3$,
since $U_1$ and $U_3$ are not independent.  
Finally, $h_2$ is a free parameter, so $\beta$ can be chosen arbitrarily, but then $\mu =\beta F$ is determined.
Typically we choose $\beta $ so that $h_2 \gg h_{1,3} $ but of order $4 \leftrightarrow 6 \, km$.
For instance, choose $U_1 = 1 \, m s^{-1} $,  $h_1 = 50 \, m$, $g_1 = 0.1\, m s^{-2}$,  and then $\beta >0$; 
in this case, also $\mu >0$ when $\sqrt{g_3 h_3 } < 23.97$, and  $\mu < 0 $ when 
$\sqrt{g_3 h_3 } >  23.97$, on using the resonance condition (\ref{rescon}) to determine $U_3 = 3.236 - \sqrt{g_3 h_3 }$.  
Alternatively, choose  $U_1 = - 1.8 \, m s^{-1}$, $h_1 = 500 \, m$, $
 g_1 =0.01\, m s^{-2}$, and again $\beta > 0$,
but now $U_3 = 0.436 - \sqrt{g_3 h_3 }$, so that $\mu < 0$ when $\sqrt{g_3 h_3 } >  3.22$, a more realistic value. 
Next, choose $U_1 $ so that $-1 < U_1 /\sqrt{g_1 h_1 }< -0.865 $, for instance $U_1 = - 1.8 \, m s^{-1}$,  $h_1 = 800 \, m$, $g_1 =0.005\, m s^{-2} $
and then $\beta < 0 $; in this case $U_3 = 0.2 - \sqrt{g_3 h_3 }$, so that  $\mu  > 0$ when 
$\sqrt{g_3 h_3 } <1.48$, and $\mu < 0$ when 
$\sqrt{g_3 h_3 } > 1.48$. Alternatively, we can choose $U_1=-1.4 \, m s^{-1}$, $h_1=1000\, m$, $g_1=0.0025 \, m s^{-2}$ 
and then again $\beta<0$; but now $U_3 = 0.181 - \sqrt{g_3 h_3 }$, so that  $\mu  > 0$ when 
$\sqrt{g_3 h_3 } <1.34$, and $\mu < 0$ when $\sqrt{g_3 h_3 } > 1.34$.
 Although these velocities are quite large, note that they scale with $\sqrt{g_1 h_1 }$ and $\sqrt{g_3 h_3}$ and
would be somewhat smaller and more realistic if $g_1, g_3 $ were reduced by a factor of $10^{-1}$ to $10^{-3}$.

Finally in this section, we would like to point out that the type of the current model considered here
can also lead to the anomalous version of the single Ostrovsky equation when $\lambda \gamma < 0$.
In particular, we  show  that the two-layer reduction of this three-layer model obtained 
by taking the  $h_2 \gg h_{1,3}$, that is  a single shallow layer with the density $\rho_1$ and current $U_1$ 
overlying a deep layer with the density $\rho_2$ and zero  current can lead to this 
anomalous situation. Indeed, for this special case, the dispersion relation determining the speed $c$ is again given by
(\ref{dispfull}), where we now let $h_2 \gg h_{1,3}$, so that $E=0$, and then, for a single mode, either $D_{1}=0$ or $D_{3}=0$. 
Here, we  choose $D_{1}=0$ and with now  $D_{3} \neq 0$, it follows from (\ref{sys}) that $A_3 = 0$,  
$A_1$ is arbitrary and we set $A_1 =1$. 
Then  the modal function $\phi$ obtained from (\ref{mode}) is given by:
\begin{eqnarray}\label{modeone}
 \phi =  0\,, \quad  \phi_{z} = 0 \,, \quad   -h < z <  -h_2 -h_1   \,, \\
 \phi =  \frac{h_1 +h_2  + z}{h_2}  \,,\quad  \phi_{z} =\frac{1}{h_2 }  \,, \quad  -h_2 - h_1 < z < -h_1 \,, \\
 \phi =  - \frac{z}{h_1} \,, \quad \phi_{z} = -\frac{1}{h_1 } \,, \quad    -h_1 < z <  0 \,,
\end{eqnarray}
Without loss of generality, we put $U_2=0$ henceforth. Then using the limit 
$h_2 \gg h_1 $, the speed $c$ is given by:
\begin{equation}\label{speedone}
c_{1,2} = U_1 \pm (g^{\prime }h_1 )^{1/2}  
\quad \hbox{where} \quad g^{\prime } = \frac{g(\rho_2 - \rho_1 )}{\rho_1 } \,.
 \end{equation}
Note that the third layer is not involved at all. Indeed, this analysis goes through in a similar manner
when there is only one interface (the upper interface) and $h_2 $ is finite, but we will not show the details here.
To avoid an implicit critical layer, we must choose $c > \hbox{max}[U_1 , 0] $,
or $c < \hbox{min}[U_1, 0 ] $.
The first case, denoted as the positive mode propagating to the right, 
holds provided that $U_1 + (g^{\prime }h_1 )^{1/2} > 0$ , and the latter,
denoted as the negative mode propagating to the left, holds  provided that $U_1 - (g^{\prime }h_1 )^{1/2} <0$.

Now all coefficients in the Ostrovsky equation (\ref{O})  can be calculated, taking into account that $h_2 \gg h_1$,
\begin{eqnarray}
I\nu&=& 
 - \frac{3 \rho_1 g^{\prime} }{h_1 }  \,, \label{nu} \\
I\lambda &=&
\frac{c^2 \rho_2 h_2 }{3} \,,\label{lambda}  \\
I \gamma &=&  
\frac{\rho_1 \tilde{f}^2 }{h_1}(1 + \frac{1}{2}\log{\frac{ |W_2 |}{|W_1 |}}  -  \frac{U_1 }{2W_1 }\log{\frac{ \rho_2}{\rho_1 }}) \,, \\
 I &=&
  \frac{2 \rho_1 W_1}{h_1} \,. \label{I} 
\label{gammaone} 
 \end{eqnarray}
Note that $I> 0$, so that $\nu < 0, \lambda >0$,  for the  mode to the right, and $I < 0$, so that $\nu > 0, \lambda < 0$, for the  mode to the left.
As expected $\nu \lambda < 0$ for both modes, which describe waves of depression. 
In the Boussinesq approximation when $\rho_1 \approx \rho_2 $, we obtain
\begin{equation}
I \gamma = 
 \frac{\rho_1 \tilde{f}^2 }{h_1}\{1 + \frac{1}{2}\log{[\frac{ |c |}{|c-U_1|}]}  \} \,.
\label{gammatwo} \\  
\end{equation}
Thus, $I\gamma > 0$ unless $U_1 $ is such that
\begin{equation}\label{gammatwo}
\frac{|c|}{\sqrt{gh_1 }}  = |1 \pm \frac{U_1 }{\sqrt{gh_1 }}| < e^{-2} \,,
\end{equation}
\begin{eqnarray}\label{gammathree}
\hbox{that is}\,,  \quad &  -1 <  \frac{U_1}{\sqrt{g_1 h_1 }} < e^{-2} -1= -0.865  \,, \\
\hbox{or}\,, \quad & 1 > \frac{U_1}{\sqrt{g_1 h_1 }} > 1 -  e^{-2}  = 0.865  \,, 
\end{eqnarray}
 for the mode to the right and left respectively.   Here we have also used the condition for the  
 exclusion of an implicit critical layer.
 Note that the two modes are essentially the same, so it is enough to consider the mode to the right.
   Then unless  (\ref{gammathree}) holds, $\lambda \gamma > 0$ and we have the typical 
   Ostrovsky equation with only  unsteady wave packet solutions.
   But if instead (\ref{gammathree}) holds then $\lambda \gamma < 0$ and we have the anomalous Ostrovsky equation
   for which there is a steady envelope wave packet solution. Let us also note that in the case of a two-layer fluid 
   with finite depths $h_1$ and $h_2$ as mentioned above,  the condition (\ref{gammatwo}) holds but 
   $e^{-2}$ is replaced with $ e^{-2 \kappa}$, where $\kappa =  h / (h_2 - h_1)$, yielding similar results.

\begin{figure}[htbp]
\begin{center}
\includegraphics[height=2in,width=3in]{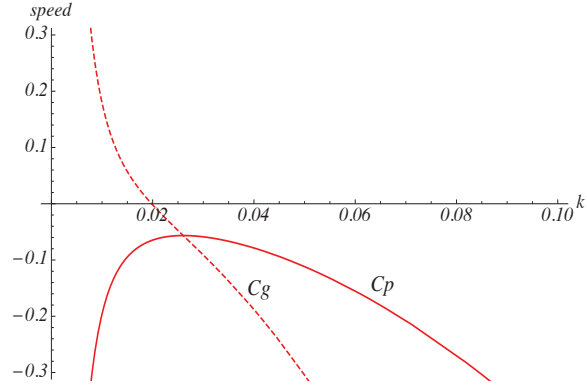}
 \caption{Dispersion relation for the single Ostrovsky equation when $\lambda \gamma <0$. }
\label{fig:DR_OE}
\end{center}
\end{figure}

A typical dispersion curve is shown in Figure \ref{fig:DR_OE}, where 
 $\nu = -4.7\, \hbox{x}\,10^{-3} \, , \lambda =41.64\, \, \hbox{and}\, \gamma=-1.9\, \hbox{x}\,10^{-5}\,,$  
 when setting  $h_1 = 0.1 \,km,  h_2 \approx 3.0\, km$, $U_1=-0.3 \,m \, s^{-1}$, $\rho_1=1\, kg\, m^{-3} \,  \hbox{and}\,\rho_2=1.0001\, kg\, m^{-3}$.  
 There exists a spectral gap for the phase speed, which has a maximum value $c=-0.057$ at $k=0.026$. The group velocity is positive as $k \to 0$, but negative  as $k \to \infty $, and at the point of maximum phase speed, the phase and group velocities are equal.  Hence a steady wave packet can exist.

\begin{figure}[htbp]
\begin{center}
\includegraphics[height=7cm,width=10cm]{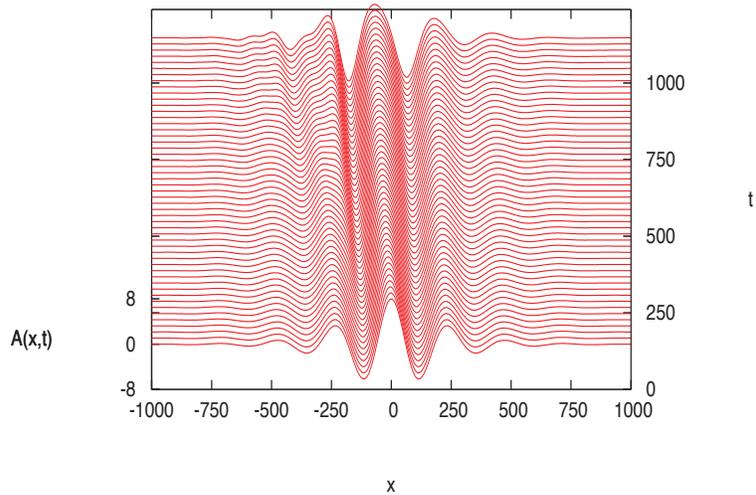}
 \caption{Numerical simulation for the Ostrovsky equation (\ref{O}) when $\lambda \gamma <0$ using the wave packet initial condition (\ref{AppendixIC}) with $k= 0.026\, ,A_0 =8, K_0=0.25 \ k$ and $ V_0=1$. }
\label{fig:OE_3d}
\end{center}
\end{figure}

\begin{figure}[h]
\begin{center}
\includegraphics[height=13cm,width=12cm]{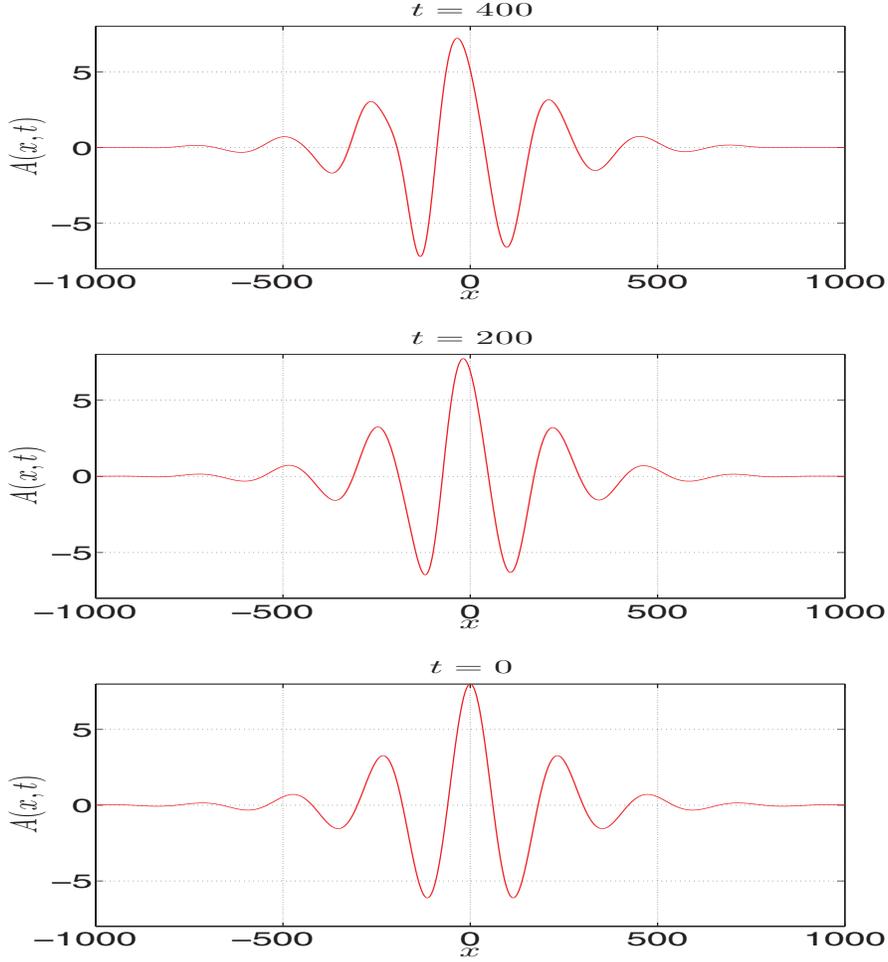}
 \caption{Same as Figure \ref{fig:OE_3d}, but a cross-section at  $\tau=0\, , 200\, , 400$.}
\label{fig:OE_2d}
\end{center}
\end{figure}

A typical numerical result is shown in Figures \ref{fig:OE_3d} and \ref{fig:OE_2d} using a wave packet initial condition:
\begin{eqnarray}\label{AppendixIC}
A(x,0)&=&  V_0 A_0 \, \hbox{sech}(K_0 X) \, \hbox{cos}(k X)\,, 
\end{eqnarray}
where $V_0=1\, , A_0=8\,, K_0=0.25\, k$ and $k=0.026$. The solution is dominated by a steady wave packet, as expected, with the speed $-0.069$, which is in good agreement with the theoretical value.

\section{Linear dispersion relation} \label{dispersion}

The structure of the linear dispersion relation determines the possible solution types.
It is obtained by seeking solutions of the linearised equations 
 in the form
\begin{equation}\label{sinus}
 u = u_0 e^{i k(X-  c_pT)} + c.c., \quad v = v_0 e^{i k (X - c_pT)} + c.c. \,, 
 \end{equation}
where $k$ is the  scaled  wavenumber, $c_p (k)$ is the phase speed and $c.c. $ denotes the complex conjugate. 
This leads to 
\begin{eqnarray}
 (c_p - C_1 (k) )u_0  + (\alpha k^2  - \frac{\gamma }{k^2 }) v_0  &=&  0 \,, \label{disp1} \\
 (\lambda k^2 - \frac{\nu }{k^2 } ) u_0 + (c_p - C_2 (k) )v_0  &=& 0 \,, \label{disp2} \\
 \text{where} \quad C_1 (k) = -k^2 + \frac{\beta }{k^2 } \,, \quad
 C_2 (k) &= & \Delta - \delta k^2 + \frac{\mu }{k^2 } \,.  \label{disp0}
\end{eqnarray}
The determinant of this $2 \times 2$ system yields the dispersion relation
\begin{equation}\label{disp}
(c_p - C_1 (k))(c_p-C_2 (k) ) = D(k) = (\alpha k^2  - \frac{\gamma }{k^2 }) (\lambda k^2 - \frac{\nu }{k^2 } ) \,.
\end{equation}
Solving this dispersion relation we obtain the two branches of the dispersion relation, 
\begin{equation}\label{dispsol}
c_p = c_{p1,p2} = \frac{C_1 + C_2}{2} \pm \frac{1}{2}\{4D + (C_1 - C_2)^2 \}^{1/2} \,. 
\end{equation}
Here $C_{1,2} (k)$ are the linear phase speeds of the uncoupled Ostrovsky equations, 
obtained formally by setting the coupling term $D(k) = 0$. If $D(k) > 0$ for all $k$,  
then both branches are real-valued for all wavenumbers $k$, 
and the linearised system is spectrally stable. 
Here  $\gamma =  \nu =0$ and $\alpha \lambda > 0$
so that $D (k)= \alpha \lambda k^4 > 0$ for all $k$.

Consider now Case 1, where $c >0, I_1 > 0, I_2 > 0$, and so 
 $\lambda_{1,2}> 0 $,  so that $\delta > 0$,   and $0 < \alpha \lambda =\delta /4$. 
 Also we recall that $\Delta < 0$
without loss of generality. The main effect of the background shear is that
now $\beta \ne \mu$, and indeed each can be either positive or negative. Then (\ref{dispsol}) takes the form
\begin{equation}\label{dispsol1}
c_p=c_{p1,p2} =   \frac{\beta+\mu}{ 2 k^2 } + \frac{\Delta}{2} - \frac{(1+\delta )k^2 }{2}
\pm \frac{1}{2} \sqrt{[\frac{\beta-\mu}{k^2 } - \Delta -(1-\delta) k^2 ]^2 + 4 \alpha \lambda k^4 }.  
\end{equation}
The group velocities are given by $c_{g} = d(k c_p)/dk $,
\begin{eqnarray}\label{gv}
c_g=c_{g1, g2} &=& -\frac{\beta+\mu }{2 k^2} + \frac{\Delta}{2} - \frac{3(1+\delta )k^2 }{2} \nonumber \\ && \pm
\frac{(\Delta + (1-\delta) k^2-\frac{\beta-\mu}{k^2}) (\frac{3}{2}(1-\delta)k^2+\frac{1}{2}(\Delta+\frac{\beta-\mu}{k^2}))+ 6\alpha \lambda k^4 }
{\sqrt{(\Delta + (1-\delta) k^2-\frac{\beta-\mu}{k^2} )^2 + 4 \alpha \lambda k^4 }} \,.
\end{eqnarray}

Next it is useful to examine the limits $k \to 0, \infty $. Thus
\begin{eqnarray}
&& c_{p1,p2}  \to \frac{ F_{1,2}}{k^2 } \,, 2F_{1,2} = \beta + \mu \pm |\beta - \mu | \qquad \qquad  \qquad  \quad\qquad \hbox{as} \quad k \to 0 \,, \label{lim0} \\
&& c_{p1,p2} \to E_{1,2} k^2 \,,  2E_{1,2} = -(1+\delta ) \pm \{(1-\delta )^2 + 4\alpha \lambda \}^{1/2} 
\quad \hbox{as} \quad k \to \infty \,. \label{lim1}
\end{eqnarray}
\begin{eqnarray}
&& c_{g1,g2}  \to -\frac{F_{1,2}}{k^2 }  \, \, \, \quad \hbox{as} \quad k \to 0 \,, \label{lim0} \\
&& c_{g1,g2} \to 3E_{1,2} k^2  \quad \hbox{as} \quad k \to \infty \,. \label{lim1}
\end{eqnarray}
Note that since $0 < \alpha \lambda < \delta$, $E_2 < E_1 < 0$. 
One can see that there are  four possibilities of qualitatively different behaviour of the dispersion relation, depending on the signs of the coefficients $\beta$ and $\mu$, as Case A: $\beta > 0, \mu >0$, Case B: $\beta > 0, \mu < 0$, Case C: $\beta < 0, \mu >0$, Case D: $\beta <  0, \mu < 0$.

\begin{figure}[h]
\begin{center}
\includegraphics[height=2in,width=3.8in]{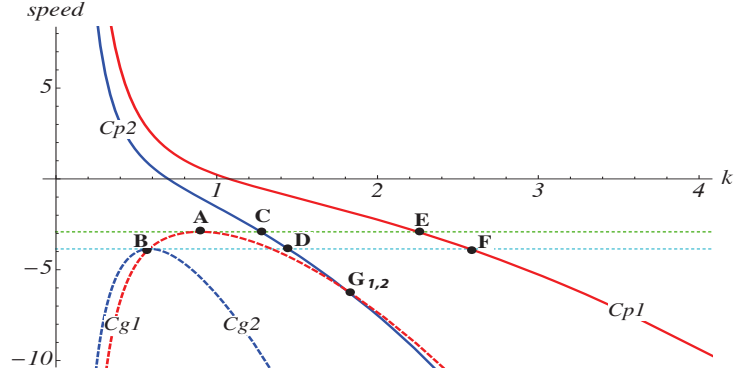}
 \caption{Typical dispersion curve for Case A with $\delta=1.414 \,, \alpha=-0.707 \,, \lambda=-0.5 \, , \Delta=-0.5 \,  , \beta=1\, $ and $\mu=0.604\, $.}
\label{fig:1}
\end{center}
\end{figure}

\medskip
\noindent
{\bf Case A}:  $\beta > 0 , \mu > 0$.  Then $F_1 = \hbox{max}[\beta , \mu] >  F_2 = \hbox{min}[\beta , \mu ] > 0$.
There is no spectral gap in either mode, and this case is similar to the situation without any background shear, 
discussed in our previous paper, \citet{Alias2013}. But there is now a significant difference
since here $\beta \neq \mu $ due to the effect of the background shear flow. 
 A typical dispersion curve is  shown in Figure \ref{fig:1},  where 
 $\beta =1 \,, \mu =0.604 \,, \Delta=-0.5 \,, \delta = 1.414\,, \alpha = -0.707\,, \lambda = -0.5$  
 when setting  $h_1 = 50 \, m\,,  h_2 \approx 3.9 \,km\,, h_3=100\, m\,$, $
 g_1 = g_3= 0.1\, m s^{-2}$\,, $U_1=1 \,m \, s^{-1}\, , U_3=0.074\, m\, s^{-1}\,,\rho_1=0.99\, \rho_2$ and $\rho_3=1.01\, \rho_2$.
Here, and in the subsequent plots of dispersion curves, the letters $A, B, \cdots $ indicate the turning points and 
possible resonant points, identified for comparison with our numerical results.
For both modes the group velocities are negative for all $k$, and each has a  single turning point at $k=k_{m1, m2}$
respectively. In general it is possible that there are $0, 2, 4, \cdots $ turning points for $c_p $ where $dc_{p}/dk = 0 $
and  $c_p = c_g $. Each such turning point can generate a generalised envelope solitary wave, see \citet{GI03} for instance.
Further it is also possible that there are $1, 3, 5, \cdots $ turning points for $c_{g} $ where $dc_{g}/dk =0$, and each such
turning point is expected to generate an unsteady wave packet analogous to those found by \citet{Grimshaw08} 
for the single Ostrovsky equation. Figure 2 shows the simplest case when there are $0,1$ turning points respectively. 
But since there are four independent parameters $\beta , \mu , \Delta , \alpha \lambda $ 
(note that $\delta = -2\alpha , \lambda = -0.5$, see (\ref{C1a})) in the expressions
(\ref{dispsol1}, \ref{gv}) for $c_p, c_g $ respectively, we cannot rule out the possibility that other ``non-typical'' cases
may occur. Even though the expressions (\ref{dispsol1}, \ref{gv}) are explicit, a full exploration of the
$4$-dimensional parameter space is beyond our present scope. Nevertheless an asymptotic expansion in the 
parameter $\alpha \lambda \ll 1$ described below confirms that only the typical case arises in this asymptotic regime.

\begin{table}[h] \begin{center}
\begin{tabular}{|c|c|c|c|}
\hline $Point $ & $Speed$ & $Wavenumber, k$  & $Ratio, u_0/v_0$  \\ 
\hline  $A$  & $-2.912$$\vert_{max \,Cg_1}$& $0.895$ & $ 3.692$$\vert_{Cp_1}$ \\
\hline  $B$  & $ -3.854 $$\vert_{max \, Cg_2} $& $0.584$ & $ -0.132$$\vert_{Cp_2}$\\
\hline   $C$ & $ -9.628 $$\vert_{Cg_2}$& $1.274$ & $-0.602$$\vert_ {Cp_2}$  \\
\hline   $D$ & $ -12.131$$\vert_{Cg_2} $& $1.446$ & $-0.659$$\vert_{Cp_2}$  \\
\hline   $E$ & $ -9.135 $$\vert_{Cg_1}$& $2.251$ & $1.829$$\vert_{Cp_1}$  \\
\hline   $F$ & $ -11.786 $$\vert_{Cg_1}$& $2.574$ & $1.788 $$\vert_{Cp_1}$  \\
\hline   $G_{1,2}$ & $ -6.118$$\vert_ {Cg_1}$& $1.806$ & $1.938$$\vert_{Cp_1}$  \\
	   	 & $ -18.501 $$\vert_{Cg_2}$&  & $-0.730$$\vert_{Cp_2}$  \\
\hline 
\end{tabular}\end{center}
\caption{Values of the group speed, wavenumber and ratio, calculated using the phase speed, at each point in Figure \ref{fig:1}.}
\label{Table1}
\end{table}

\begin{figure}[h]
\begin{center}
\includegraphics[height=2in,width=3.8in]{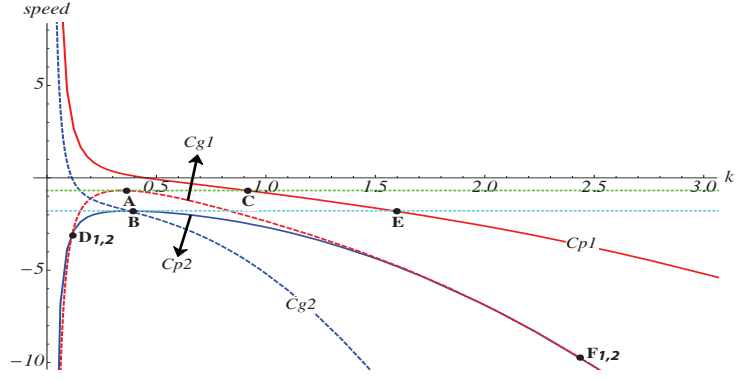}
 \caption{Typical dispersion curve for Case B with $\delta=1 \,, \alpha=-0.5 \,, \lambda=-0.5 \,, \Delta=-1.5 \, ,\beta=0.04\, $ and $\mu=-0.02 \, $.}
\label{fig:2}
\end{center}
\end{figure}

\medskip
\noindent
{\bf Case B}:  $\beta > 0 , \mu < 0$. Then $F_1 = \beta > 0, F_2 = \mu< 0$. 
 A typical dispersion curve is shown in Figure \ref{fig:2}, where 
 $\beta = 0.04 \, , \mu =-0.02\, , \Delta=-1.5 \,,\delta = 1\, , \alpha = -0.5\,, \lambda = -0.5$  
 when setting  $h_1 = 500 \, m\,,  h_2 \approx 5.5\, km\,, h_3=1000\, m$\,, 
 $g_1 =0.01\, ms^{-2}\,, g_3= 0.02\, m s^{-2}$\,, $U_1=-1.8 \,m \, s^{-1}$\,, $U_3=-4.036 \,m \, s^{-1}\,$, $\rho_1=0.999\, \rho_2$ and $\rho_3=1.002\, \rho_2$\,. 
There is no spectral gap in mode $1$, and the group velocity is negative for all $k$ with a turning point at $k=k_{m1}$.
But mode $2$ has a spectral gap, as the phase speed has a maximum value, $c_{s2}$ at $k=k_{s2}$.
For this mode the group velocity is positive as $k \to 0$ and  negative 
as $k \to \infty $.  At the value $c_{p2} = c_{s2}$, the phase and group velocities are equal, 
and then this mode $2$ can support a steady wave packet.
However, this wave packet lies in the spectrum of mode $1$, and hence may  decay by radiation into mode $1$;
strictly, it is a generalised solitary wave.
Here, in general it is possible that there are $0, 2, 4, \cdots $ turning points for $c_p $ for mode $1$,
and $1,3,5, \cdots $ for mode $2$.  Further it is also possible here that there are $1, 3, 5, \cdots $ turning points for $c_{g} $ 
in mode $1$, and $0, 2, 4, \cdots $ for mode $2$. However, the asymptotic expansion in the 
parameter $\alpha \lambda \ll 1$ described below confirms that only the typical case $0, 1, 1, 0$ 
of turning points arises in this asymptotic regime.

\begin{table}[h]  \begin{center}
\begin{tabular}{|c|c|c|c|} 
\hline $Point $ & $Speed$ & $Wavenumber, k$  & $Ratio, u_0/v_0$  \\ 
\hline  $A$  & $ -0.683$$\vert_{max \, Cg_1}$& $0.345$ & $ 33.696$$\vert_{Cp_1}$ \\
\hline  $B$ & $ -1.785 $$\vert_{max \,Cp_2=Cg_2}$& $0.372$ & $-0.036$$\vert_{Cp_2}$\\
\hline   $C$ & $ -2.0430 $$\vert_{Cg_1}$& $0.914$ & $4.012$$\vert_{Cp_1}$  \\
\hline  $D_{1,2}$  & $ -2.987 $$\vert_{Cg_1}$& $0.117$ & $871.768$$\vert_{Cp_1} $   \\
		&$ -0.068 $$\vert_{Cg_2}$&		&$-0.001$$\vert_{Cp_2} $\\	
\hline  $E$  & $-4.676$$\vert_{Cg_1}$& $1.583$ & $ 1.779$$\vert_{Cp_1}$   \\
\hline  $F_{1,2}$  & $ -9.722 $$\vert_{Cg_1}$& $2.433$ & $1.287 $$\vert_{Cp_1} $   \\
		&$-27.297 $$\vert_{Cg_2} $		&	&$-0.778$$\vert_{Cp_2} $\\
\hline 
\end{tabular}\end{center}
\caption{Values of the group speed, wavenumber and ratio, calculated using the phase speed, at each point for Figure \ref{fig:2}.}
\label{Table2}
\end{table}

\begin{figure}[h]
\begin{center}
\includegraphics[height=2in,width=3.8in]{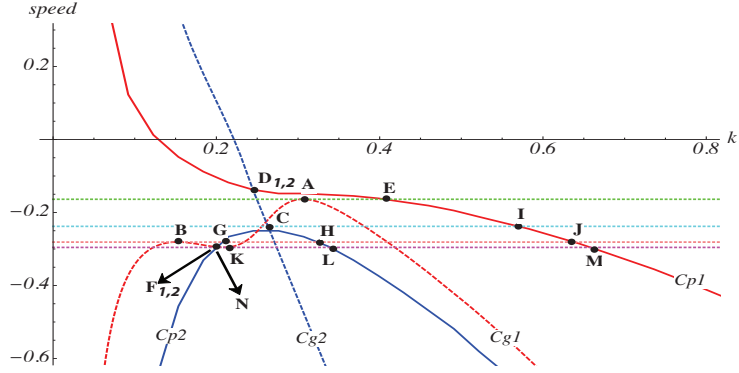}
 \caption{Typical dispersion curve for Case C with $\delta=1.414 \,, \alpha=-0.707 \,, \lambda=-0.5  \, , \Delta=-0.1\, ,\beta=-0.01\,$ and $\mu=0.002\, $.}
\label{fig:3}
\end{center}
\end{figure}

\medskip
\noindent
{\bf Case C}: $\beta < 0 , \mu > 0$ . Then $F_1 = \mu >0 , F_2 = \beta < 0$. 
A typical dispersion curve set is  shown in Figure \ref{fig:3}, 
where  $\beta = -0.01 \, , \mu =0.002 \, , \Delta=-0.1\, $, $\delta = 1.414\,, \alpha = -0.707\,, \lambda = -0.5$
when setting $h_1 = h_3=800\, m\,,  h_2 \approx 4.0\, km\,$, $g_1 =0.005\,m \,s^{-2}\,$, $g_3= 0.0025\, m s^{-2}\,$,  $U_1=-1.8 \,m \, s^{-1}\,$, $U_3=-1.214 \,m \, s^{-1}, $ $\rho_1=0.9995\, \rho_2$ and $\rho_3=1.00025\, \rho_2$.
At first glance, this is overall similar to case B because there is no spectral gap in mode $1$, and the group velocity is negative for all $k$;  but now the group velocity $c_{g1}$ 
has three turning points, a global maximum at $A$, a local minimum at $K$ and a local  
maximum at $B$. This is not the simplest case, where we would expect only one turning point,
but we display it here as potentially there could be energy focussing associated with each of these turning points, 
and the consequent emergence of three unsteady nonlinear wave packets.
As in case B, mode $2$ has a spectral gap, as the phase speed has a maximum at $C$;
the group velocity is positive as $k \to 0$ and  negative as $k \to \infty $.  At this point,
the phase and group velocities are equal, and so then this mode $2$ can support  a steady wave packet.
However, this wave packet lies in the spectrum of mode $1$, 
and hence may  decay by radiation into mode $1$.   

\begin{table}[h]\begin{center}
\begin{tabular}{|c|c|c|c|} 
\hline $Point $ & $Speed, $ 	& $Wavenumber, k$  & $Ratio, u_0/v_0$  \\ 
\hline  $A$  & $ -0.164$$\vert_{max\, Cg_1}$	& $0.306$ 	& $ 1.309$$\vert_{Cp_1}$ \\
\hline  $B$  & $ -0.281$$\vert_ {max\, Cg_1}$	& $0.152$ 	& $0.040$$\vert_ {Cp_1}$\\
\hline   $C$ & $ -0.238 $$\vert_{max \, Cp_2=Cg_2}$& $0.259$ 	& $-2.164$$\vert_ {Cp_2}$  \\
\hline  $D_{1,2}$  & $ -0.263$$\vert_ {Cg_1}$	& $0.245$ 	& $0.472$$\vert_{Cp_1} $   \\
		&$ -0.137 $$\vert_{Cg_2}$	&			&$-2.994$$\vert_{Cp_2} $\\	
\hline  $E$  & $-0.273$$\vert_{Cg_1}$	& $0.404$ 	& $ 1.898$$\vert_ {Cp_1}$   \\
\hline  $F_{1,2}$  & $ -0.294$$\vert_{Cg_1}$	& $0.199$ 	& $0.149 $$\vert_{Cp_1} $   \\
		&$0.1081$$\vert_ {Cg_2} $	&			&$-9.497$$\vert_{Cp_2} $\\
\hline  $G$  & $0.075$$\vert_{Cg_2}$	& $0.206$ 	& $ -7.874$$\vert_ {Cp_2}$   \\
\hline  $H$  & $-0.623$$\vert_{Cg_2}$	& $0.326$ 	& $ -0.932 $$\vert_{Cp_2}$   \\
\hline  $I$    & $-0.577$$\vert_{Cg_1}$	& $0.571$ 	& $1.944$$\vert_{Cp_1}$   \\
\hline  $J$  & $-0.722$$\vert_{Cg_1}$	& $0.638$	 	& $1.914 $$\vert_{Cp_1}$   \\
\hline  $K$  & $-0.296$$\vert_{min \, Cg_1}$	& $0.209$ 	& $ 0.191$$\vert_ {Cp_1}$   \\
\hline  $L$  & $-0.681$$\vert_{Cg_2}$	& $0.339$ 	& $-0.870$$\vert_{Cp_2}$   \\
\hline  $M$  & $-0.770$$\vert_{Cg_1}$	& $0.659$ 	& $1.904 $$\vert_{Cp_1}$   \\
\hline  $N$  & $0.111$$\vert_{Cg_2}$	& $0.199$ 	& $-9.651$$\vert_ {Cp_2}$   \\
\hline 
\end{tabular}\end{center}
\caption{Values of the group speed, wavenumber and ratio, calculated using the phase speed, at each point for Figure \ref{fig:3}.}
\label{table3}
\end{table}

\begin{figure}[h]
\begin{center}
\includegraphics[height=2in,width=3.8in]{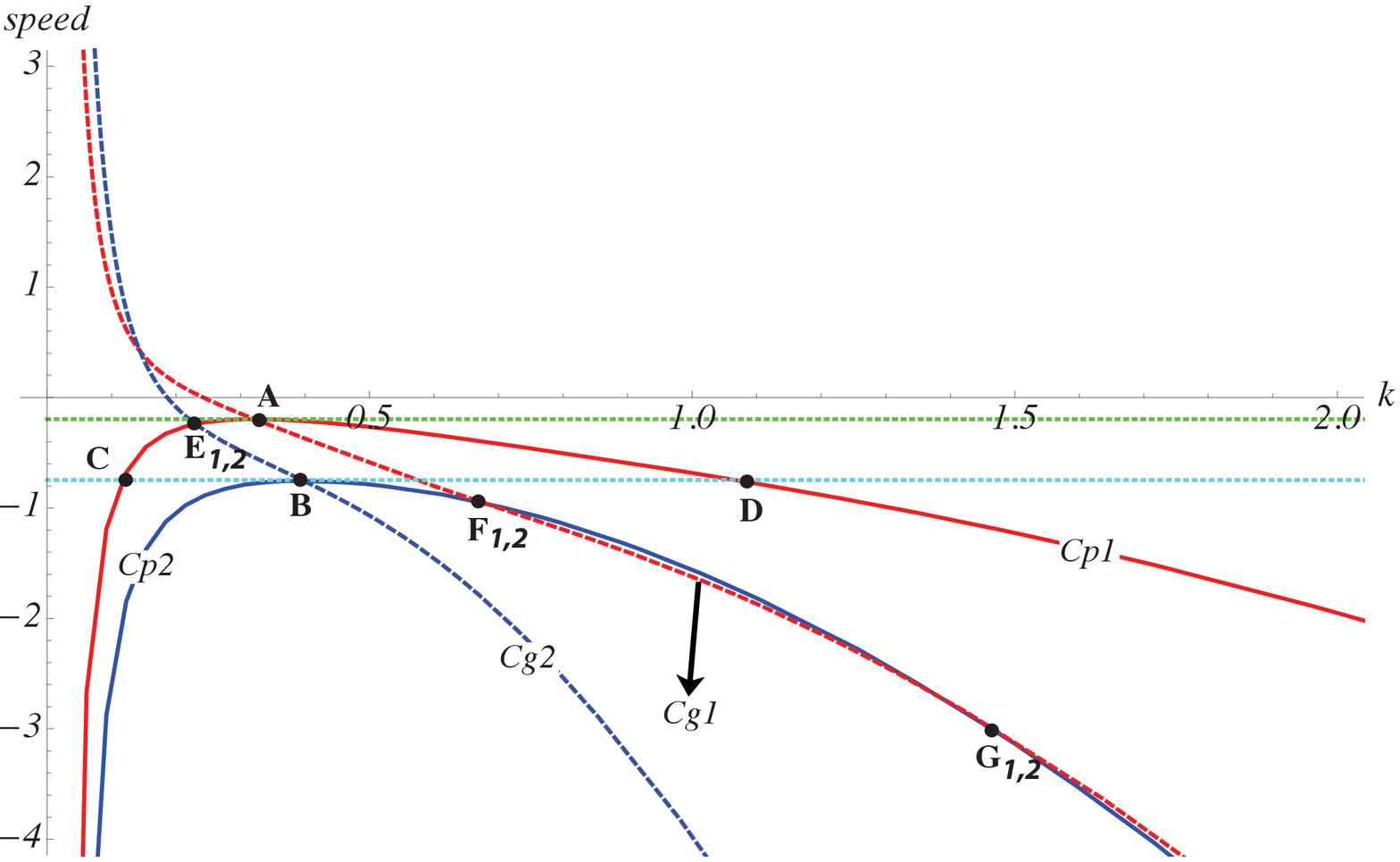}
 \caption{Typical dispersion curve for Case D with $\delta=0.707 \,, \alpha=-0.354 \,, \lambda=-0.5 \, ,\Delta=-0.5\, , \beta=-0.01\,$ and   $\mu=-0.02 \, $.}
\label{fig:4}
\end{center}
\end{figure}

\medskip
\noindent
{\bf Case D}:  $\beta < 0 , \mu < 0$. Then $F_2 = \hbox{min}[\beta , \mu ] <  F_1 =\hbox{max}[\beta , \mu ] < 0$. 
A typical dispersion curve for this case is  shown in Figure \ref{fig:4}, 
where $\beta =-0.01 \, , \mu =-0.02 \, ,  \Delta=-0.5\, $, 
$\delta = 0.707, \alpha = -0.354, \lambda = -0.5$ when setting
$h_1 = h_3=1000\, m,  h_2 \approx 4.4 \, km\,$, $ g_{1}=0.0025\, m \, s^{-2}\,,  g_{3}= 0.005\, m  \,s^{-2} \,, U_1=-1.4 \,m \, s^{-1}\,, U_3=-2.055 \,m \, s^{-1}\,,\rho_1=0.9998\, \rho_2$ and $\rho_3=1.0005\, \rho_2$. Now both modes have phase  speeds with maxima $c_{s1}, c_{s2}$ at $k =k_{s1}, k_{s2}$,
denoted by the points $A,B$ respectively. For both modes,
the group velocity is positive as $k \to 0$, but negative  as $k \to \infty $, and at the point of maximum phase speed,
the phase and group velocities for each mode  are equal.  Hence a steady wave packet can exist for each mode, 
but will be radiating for mode $2$ .

\begin{table}[h]\begin{center}
\begin{tabular}{|c|c|c|c|} 
\hline $Point $ & $Speed$ & $Wavenumber, k$  & $Ratio, u_0/v_0$  \\ 
\hline  $A$  & $ -0.197 $$\vert_{max\,Cp_1=Cg_1}$	& $0.322$ & $ 10.973$$\vert_{Cp_1}$ \\
\hline  $B$  & $ -0.747$$\vert_{max\,Cp_2=Cg2}$	& $0.395$ & $-0.105 $$\vert_{Cp_2}$\\
\hline  $C$ & $ 0.692$$\vert_{Cg_1}$	& $0.117$ & $180.200$$\vert_{Cp_1}$  \\
\hline  $D$  & $ -1.781$$\vert_{Cg_1}$	& $1.066$ & $ 1.011$$\vert_{Cp_1}$   \\
\hline  $E_{1,2}$  &$0.033$$\vert_{Cg_1}$ 	& $0.231$ & $25.339$$\vert_{Cp_1}$   \\
	       & $-0.241$$\vert_{Cg_2}$	&		    &$ -0.028$$\vert_{Cp_2}$	\\
\hline  $F_{1,2}$  & $ -0.931$$\vert_{Cg_1}$	& $0.664$ &$ 2.117$$\vert_{Cp_1}$   \\
	       &$-1.761$$\vert_{Cg_2}$	&		    &$ -0.334$$\vert_{Cp_2}$ \\
\hline  $G_{1,2}$ & $-2.894$$\vert_{Cg_1}$	& $1.438$ &  $ 0.794$$\vert_{Cp_1}$  \\
	       &$-8.182$$\vert_{Cg_2}$	&		    &$ -0.892$$\vert_{Cp_2}$ \\
\hline 
\end{tabular}\end{center}
\caption{Values of the group speed, wavenumber and ratio, calculated using the phase speed, at each point for Figure \ref{fig:4}.}
\label{Table4}
\end{table}

As indicated above we use an asymptotic expansion with $\epsilon = \alpha \lambda \ll 1$ 
to find all turning points explicitly.  From (\ref{disp}), since here $\gamma = \nu =0$, 
\begin{equation}\label{dispnew}
(c_p - C_1 (k))(c_p-C_2 (k) ) = \epsilon k^4 \,, \quad C_1(k)=-k^2+\frac{\beta}{k^2} \,, \quad  C_2(k)=\Delta-\delta k^2 +\frac{\mu}{k^2} \,, 
\end{equation}
where $\epsilon = \alpha \lambda \ll 1 $. Note that the effective expansion parameter is $\epsilon k^4 $ and so
this can only be valid when $k$ is also  sufficiently small, say $k < 1$. Expanding in powers of $\epsilon $ then yields
\begin{eqnarray}
&&c_{pa}=C_1 +  \frac{\epsilon k^4}{C_1-C_2 } -   \frac{\epsilon^2 k^8}{(C_1-C_2)^3} + \cdots \,, \nonumber \\
&&c_{pb} =C_2 -  \frac{\epsilon k^4}{C_1-C_2 } +   \frac{\epsilon^2 k^8}{(C_1-C_2)^3} + \cdots \,. \nonumber \\
\end{eqnarray}
The derivatives are given by
\begin{eqnarray}
&&  c_{pak} = -2k - \frac{2\beta }{k^3 }    + \epsilon \{\frac{4k^3 }{C_1 - C_2 } - \frac{k^4 (C_{1k} - C_{2k})}{(C_1 - C_2 )^2 } \}  + \cdots  \,, \label{tp1} \\
&&  c_{pbk} =  - 2\delta k  - \frac{2\mu }{k^3 } - \epsilon \{\frac{4k^3 }{C_1 - C_2 } - \frac{k^4 (C_{1k} - C_{2k})}{(C_1 - C_2 )^2 } \} + \cdots \,. \label{tp2}
\end{eqnarray}
The corresponding group velocities are found from $c_g = c_p + kc_{pk}$: 
\begin{eqnarray}
&& c_{ga} =  -3k^2-\frac{\beta}{k^2} +  \epsilon \{\frac{5k^4 }{C_1 - C_2 } - \frac{k^5 (C_{1k} - C_{2k})}{(C_1 - C_2 )^2 } \}  
+ \cdots  \,, \label{gv1} \\
&& c_{gb} = \Delta -3 \delta k^2-\frac{\mu}{k^2}  
- \epsilon \{\frac{5k^4 }{C_1 - C_2 } - \frac{k^5 (C_{1k} - C_{2k})}{(C_1 - C_2 )^2 }\} + \cdots \,, \label{gv2}\\
&& c_{gak}=-6k+\frac{2\beta}{k^3} + \epsilon \{\frac{20k^3 }{C_1 - C_2 } - \frac{10k^4 (C_{1k} - C_{2k})}{(C_1 - C_2 )^2 } 
- k^5  \{\frac{(C_{1k} - C_{2k})}{(C_1 - C_2 )^2 }\}_k  \}  + \cdots  \,, \label{gv3}  \\
&& c_{gbk}=-6 \delta k+\frac{2\mu}{k^3} - \epsilon \{\frac{20k^3 }{C_1 - C_2 } - \frac{10k^4 (C_{1k} - C_{2k})}{(C_1 - C_2 )^2 } 
- k^5  \{\frac{(C_{1k} - C_{2k})}{(C_1 - C_2 )^2 }\}_k  \} + \cdots \,. \label{gv4}
\end{eqnarray}
The turning points for $c_p $ can now be found by equating (\ref{tp1}, \ref{tp2}) to zero, and those for $c_{g}$ found by equating
(\ref{gv3}, \ref{gv4}) to zero.  Consistently with this asymptotic expansion, the solutions for $k$ are sought in the form
 $k=k_0+\epsilon k_1+\epsilon^2 k_2+\dots$ by collecting the $O(1)$ and $O(\epsilon)$ terms.  
 Then, we obtain  the following formal asymptotic solutions:

 \begin{eqnarray*}
 &&c_{pak} = 0: \quad k=k_0 + \epsilon \frac{k_0^9 (k_0^4 (-1 + \delta) - 2 k_0^2 \Delta + 
   3 (\beta - \mu))}{(k_0^4 - 3 \beta) (\beta + 
   k_0^4 (-1 + \delta) - k_0^2 \Delta - \mu)^2}+  \dots \quad \, , k_0= \sqrt[4]{-\beta} ; \\ \nonumber 
 &&c_{pbk} = 0: \quad k=k_0 - \epsilon \frac{k_0^9 (k_0^4 (-1 + \delta) - 2 k_0^2 \Delta + 
   3 (\beta - \mu))}{(k_0^4 \delta - 3 \mu) (\beta + 
   k_0^4 (-1 + \delta) - k_0^2 \Delta - \mu)^2}+  \dots \quad \, , k_0= \sqrt[4]{\frac{-\mu}{\delta}} ; \\ \nonumber 
 && c_{gak} = 0: \quad k=k_0+ \epsilon k_0^{9}
\lbrace \frac{ 3 k_0^8 (-1 + \delta)^2 - 
   9 k_0^6 (-1 + \delta) \Delta + 21 (\beta - \mu)^2+27k_0^2 \Delta (-\beta + \mu)}{3 (k_0^4 + \beta) (\beta + k_0^4 (-1 + \delta) - 
 k_0^2 \Delta - \mu)^3} \\ && \nonumber \qquad \qquad \qquad +\frac{ 
   2 k_0^4 (4 \beta (-1 + \delta) + 5 \Delta^2 + 
      4 \mu (1- \delta)}{3 (k_0^4 + \beta) (\beta + k_0^4 (-1 + \delta) - 
 k_0^2 \Delta - \mu)^3}\rbrace + \dots \quad \, , k_0=\sqrt[4]{\frac{\beta}{3}} ; \\ \nonumber 
  && c_{gbk} = 0: \quad k=k_0- \epsilon k_0^{9}
\lbrace \frac{ 3 k_0^8 (-1 + \delta)^2 - 
   9 k_0^6 (-1 + \delta) \Delta + 21 (\beta - \mu)^2+ 27k_0^2 \Delta (-\beta + \mu)}{3 (k_0^4 \delta + \mu) (\beta + k_0^4 (-1 + \delta) - k_0^2 \Delta - \mu)^3} \\ && \nonumber \qquad \qquad \qquad -\frac{2 k_0^4 (4 \beta (-1 + \delta) + 5 \Delta^2 + 
      4 \mu (1- \delta)}{3 (k_0^4\delta+ \mu) (\beta + k_0^4 (-1 + \delta) - 
 k_0^2 \Delta - \mu)^3}\rbrace + \dots \quad \, , k_0=\sqrt[4]{\frac{\mu}{3 \delta}} .
\end{eqnarray*}

The outcomes for each case are described below.

\medskip
\noindent
{\bf Case A}: $\beta > 0, \mu > 0 $.  Here we put $a=1,b=2$ and find that both $c_{p1k} <0 $ and $c_{p2k}<0$. 
Thus there are no turning points for $c_{p1}$ and $c_{p2}$ in this approximation. However, $c_{g1k}=0$ 
yields  just one turning point $k=k_0+\epsilon k_1\approx 0.868$  for the parameter values of Figure \ref{fig:1},
compared to the exact   value $0.895$. 
Also $c_{g2k}=0$ yields just one turning point $k=k_0+\epsilon k_1\approx 0.574$, compared to the exact   value $0.584$.

\medskip
\noindent
{\bf Case B}: $\beta > 0, \mu < 0 $. Here we again put  $a=1,b=2$ and find that $c_{p1k}<0$ and so
there is no turning point for $c_{p1}$. However, there is a single turning point for $c_{p2}$,
given by  $c_{p2k}=0$,  $k=k_0+\epsilon k_1\approx 0.372$, for the parameter values of Figure \ref{fig:2} 
compared to the exact value $0.372$. Next, there is a single turning point for $c_{g1}$ when $c_{g1k}=0$ 
gives $k=k_0+\epsilon k_1\approx 0.345$, compared to the exact  value  $0.345$. Since $c_{g2k}<0$, 
there are no turning points for  $c_{g2}$.

\medskip
\noindent
{\bf Case C}: $\beta < 0, \mu > 0$. Here we put $a=2,b=1$ and find that $c_{p1k}<0$ and so there is no turning
 point for $c_{p1}$. However, there is a single turning point for $c_{p2}$, given by  $c_{p2k}=0$,
 $k=k_0+\epsilon k_1\approx -0.408$ for the parameter values of Figure \ref{fig:3}, compared to the exact value of $0.259$. 
 However, we note here that $k_{0}=0.316$ and  the correction term $| \epsilon k_1 |$ is much too large, 
 indicating that the asymptotic expansion is not at all  useful in this case.  Next there is a single 
 turning point for $c_{g1}$ and $c_{g1k}=0$ gives $k=k_0+\epsilon k_1\approx 0.151$, 
 compared to the exact value of $0.152$ that is point $B$ in Figure \ref{fig:3}.
 However, we note here there also exists a minimum point $K$ in Figure \ref{fig:3} at $k=0.209$, 
 and a maximum point $A$ at $k=0.306$ which are not found by this asymptotic analysis. 
 Since $c_{g2k}<0$ there are no stationary points in $c_{g2}$.

\medskip
\noindent
{\bf Case D}: $\beta < 0, \mu < 0$.  Here we put  $a=1,b=2$. There are turning points for both 
$c_{p1}, c_{p2}$ and $c_{p1k} =0, c_{p2k}=0$ yield $k=k_0+\epsilon k_1\approx 0.322, 0.392$, respectively,
for the parameter values of Figure \ref{fig:4},  compared to the exact values of $0.322, 0.395$.
Here both $c_{g1k}<0$ and $c_{g2k}<0$ and hence there are  no turning points in both $c_{g1}$ and $c_{g2}$.

\section{Numerical simulations}\label{numerics}

 In this section we present some  results from numerical simulations of the scaled equations (\ref{CO1},\ref{CO2}),  
 using the  pseudo-spectral method described in \citet{Alias2013}, 
for the four different cases, corresponding to the parameters of the linear dispersion curves described in section III.   
We note again that in these equations $X,T$ are scaled variables, see (\ref{scale}), and have dimensions of $C^{-1/2}, C^{-3/2}$ respectively, where $C$ is the velocity scale. The dependent variables $u$ and $v$ have the dimension of $C$. The coefficients $n,m,\alpha,\delta,p,q,\lambda$ are dimensionless, while $\beta, \gamma, \mu,\nu$ have dimensions of $C^{2}$, and $\Delta$ has the dimension of $C$. 
For all cases considered here we have
$$
n=m=p=q=\gamma=\nu=0.
$$
For the initial conditions we use either an approximation to a solitary wave solution of the 
corresponding coupled KdV  system, which is mainly suitable for Case A,
or an approximation to a nonlinear wave packet, which is more suitable for Cases B,C,D.  
The former initial condition is described by  \citet{Alias2013}, is denoted as ``weak coupling KdV solitary waves'', and given by,
\begin{eqnarray}\label{weak}
& u =  a\,\hbox{sech}^2 (\gamma_1 X ) \,, \quad  \frac{a}{3} =4(1 + \alpha )\gamma_{1}^2 \,, \\
& v = \, b\,\hbox{sech}^2 (\gamma_2 X) \,, \quad  \frac{b}{3} =4(\delta + \lambda) \gamma_{2}^2 \,. 
\end{eqnarray}
 This was  mostly implemented with the constraint that $\gamma_1 = \gamma_2 $. Note that here the nonlinear terms $(u^2 /2)_{XX}$, $(v^2 /2)_{XX}$ have maximum absolute values of 
$2a^2 \gamma^{2}_1 = a^3 /6(1+ \alpha)$ and $2b^2 \gamma^{2}_2 = b^3 /6(\delta + \lambda )$ respectively.

The nonlinear wave packet initial condition is based on either a maximum point in the group velocity curve
where   $\partial c_g /\partial k = 0$ and $k=k_m $, or a maximum point in the phase velocity curve where
$c_p = c_g $   and $k=k_s $.  The former corresponds to the unsteady nonlinear wave packet travelling
 at a speed close to the maximum group velocity, and is relevant for both modes in Case A, but  only for 
 mode $1$  in Cases B and C.  The latter corresponds to a steady wave packet and is relevant 
for mode $2$ in Cases B and C, and both modes in Case D.

To obtain a suitable wave packet initial condition, the procedure is to choose 
$k$, either $k_m $ or $k_s$, and  then find the ratio $r = u_0 /v_0 $ from (\ref{disp1}) or (\ref{disp2}) 
in the form $u_0 = U_0 a_0 , v_0 = V_0 a_0 $
where $a_0 $ is an arbitrary function of $X$, but $U_0 , V_0 $ are known functions of $k$.
Based on the expected outcome that the nonlinear wave packet will be governed by an evolution equation
such as the nonlinear Schr\"odinger equation, we choose $a_0 (X) = A_0 \, \hbox{sech}(K_0 X )$.
Note that  the underlying theory suggests  that the shape should be $\hbox{sech}$, and that 
$K_0 $ depends on the amplitude $A_0$  (e.g.,  \citet{Grimshaw08}).  
Here  instead we choose a value of $K_0 << k$. 
 Then the wave packet initial condition is 
\begin{equation}\label{wpic}
u(X,0)= r V_0 A_0 \, \hbox{sech}(K_0 X) \, \hbox{cos}(k X)\,, \quad 
v(X,0) = V_0 A_0 \,  \hbox{sech}(K_0 X) \, \hbox{cos}(k X) \,,
\end{equation}
where $r =U_0 /V_0 $ is a known function of $k$, and we can choose $V_0 $ arbitrarily, say $V_0 =1$.

Our main aim is to understand and interpret  the observed dynamical behaviour by relating it to  
the main features of the relevant dispersion curves, comparing especially the theoretically predicted group speeds and 
$r = u_0/v_0$ amplitude ratios with those found in the numerical simulations. For the latter, we adopt the following methodology;
the speed is measured at the maximum of the dominant wave packet, and the numerical ratio is measured as
$R = \max |u|/\max |v|$ in the interval between the two nearest peaks, containing the maximum value of the dominant wave packet.
Note that $R$ is necessarily positive, unlike $r$, since phase determination numerically is quite difficult. 
 In some cases wave packets generated in 
the numerical simulations are either contaminated by radiation, or show signs of more than one carrier wavelength. 
In these cases the ratio is not so instructive, and instead we choose the relevant points on the dispersion curves primarily by the speed of the wave  packet, ruling out some points if the corresponding wavelength is too long or too short.

\subsection{Numerical results}

\begin{figure}[htbp]
\begin{center}
\includegraphics[height=5cm,width=8cm]{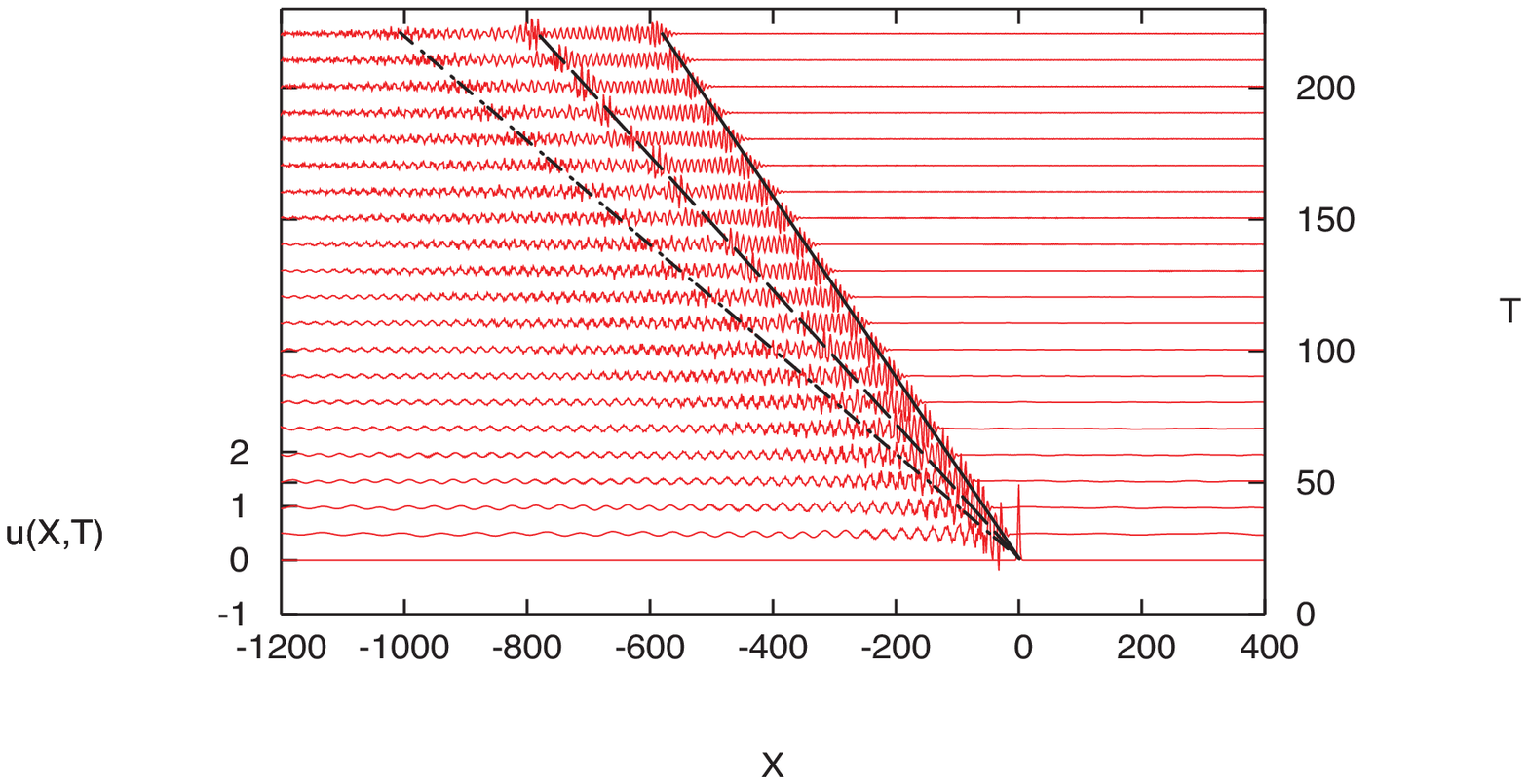} 
\includegraphics[height=5cm,width=8cm]{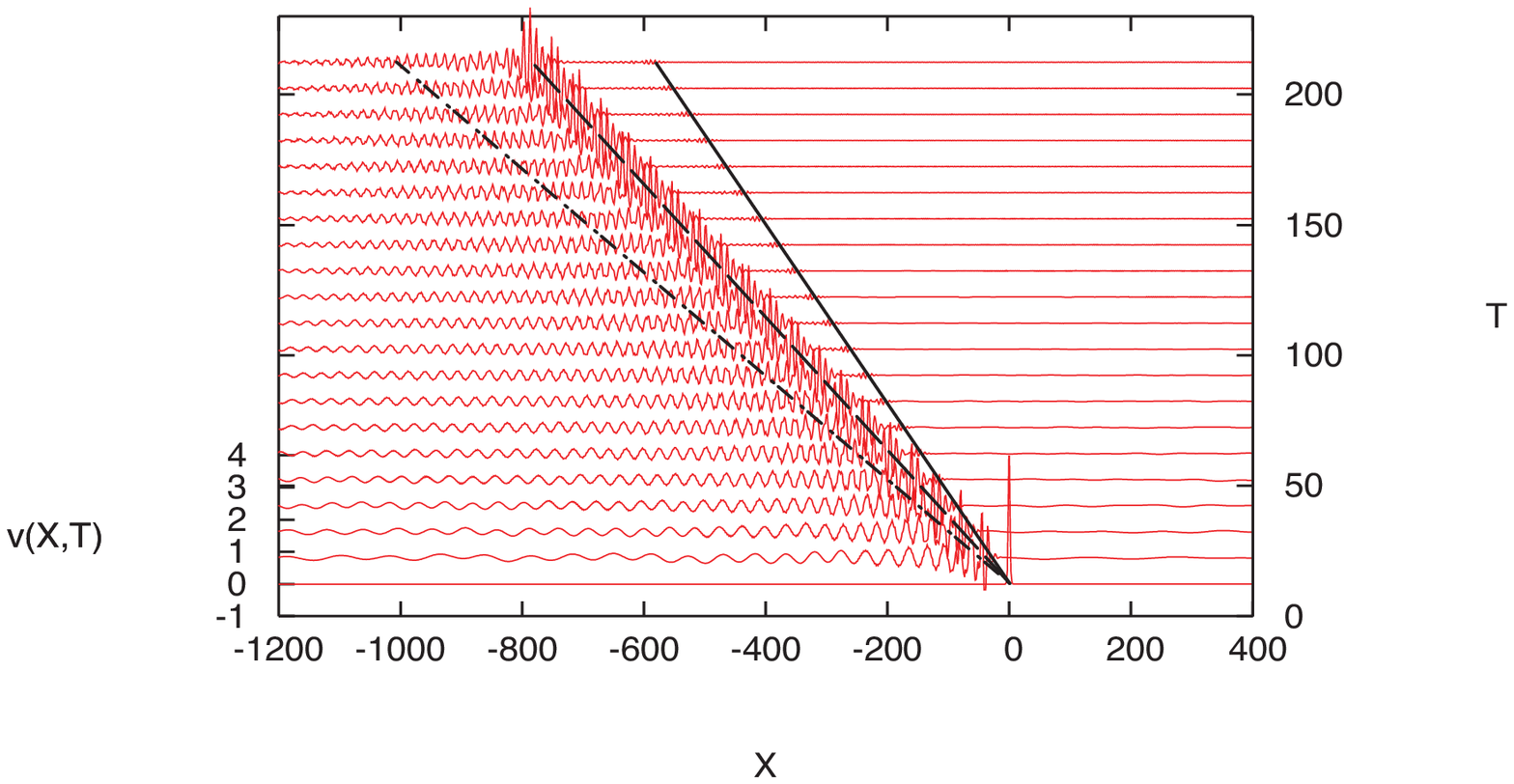}
\caption{Numerical simulations for Case A using a KdV initial condition of weak coupling with $a=1.4$ and $b=4.38$ in (\ref{weak}). The solid, dashed and dash-dot lines in both plots refer to the points $A$, $B$ and $G_1$ in Figure \ref{fig:1}.}
\label{fig:1a}
\end{center}
\end{figure}

\begin{figure}[htbp]
\begin{center}
\includegraphics[height=4cm,width=8cm]{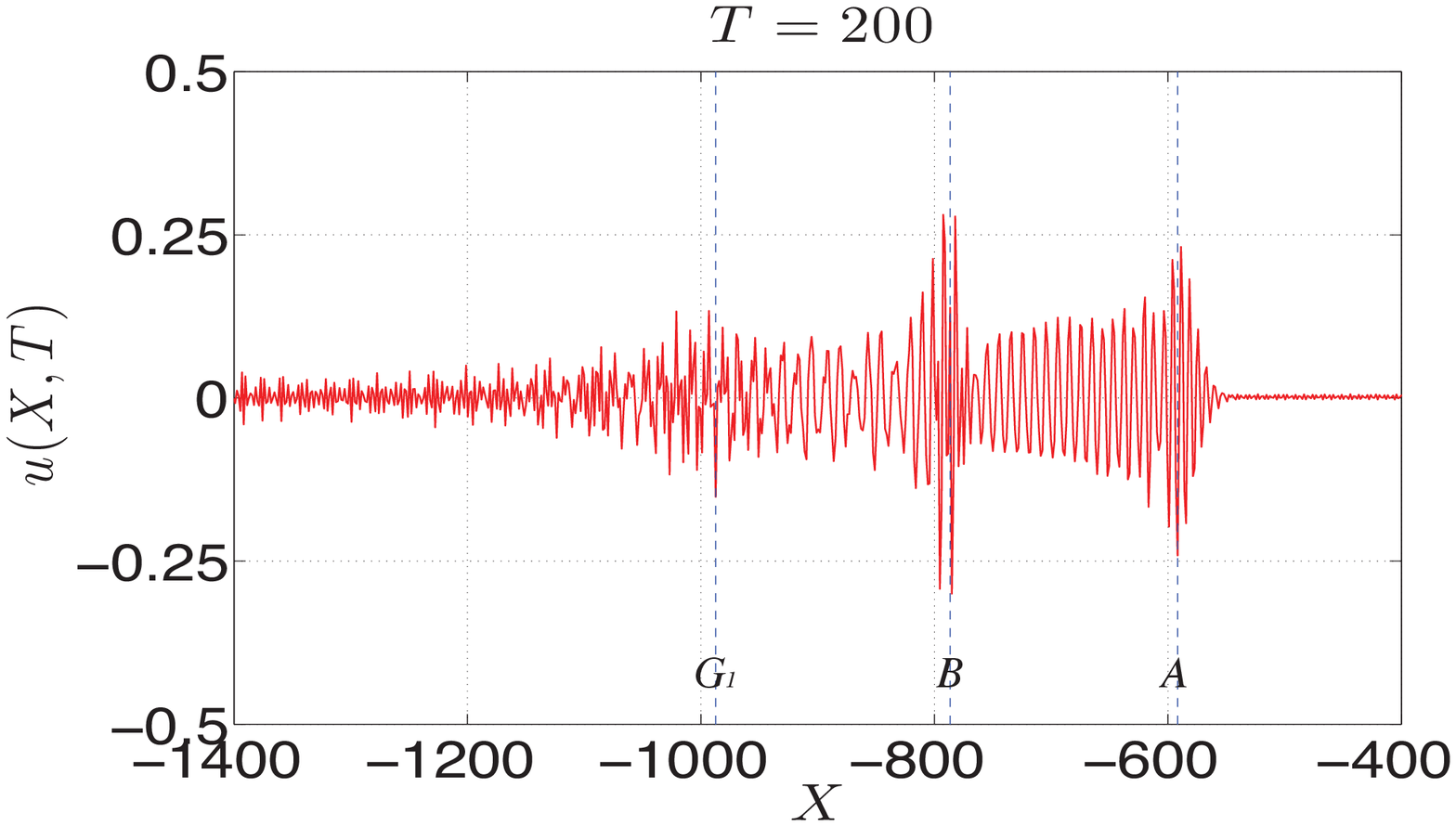} 
\includegraphics[height=4cm,width=8cm]{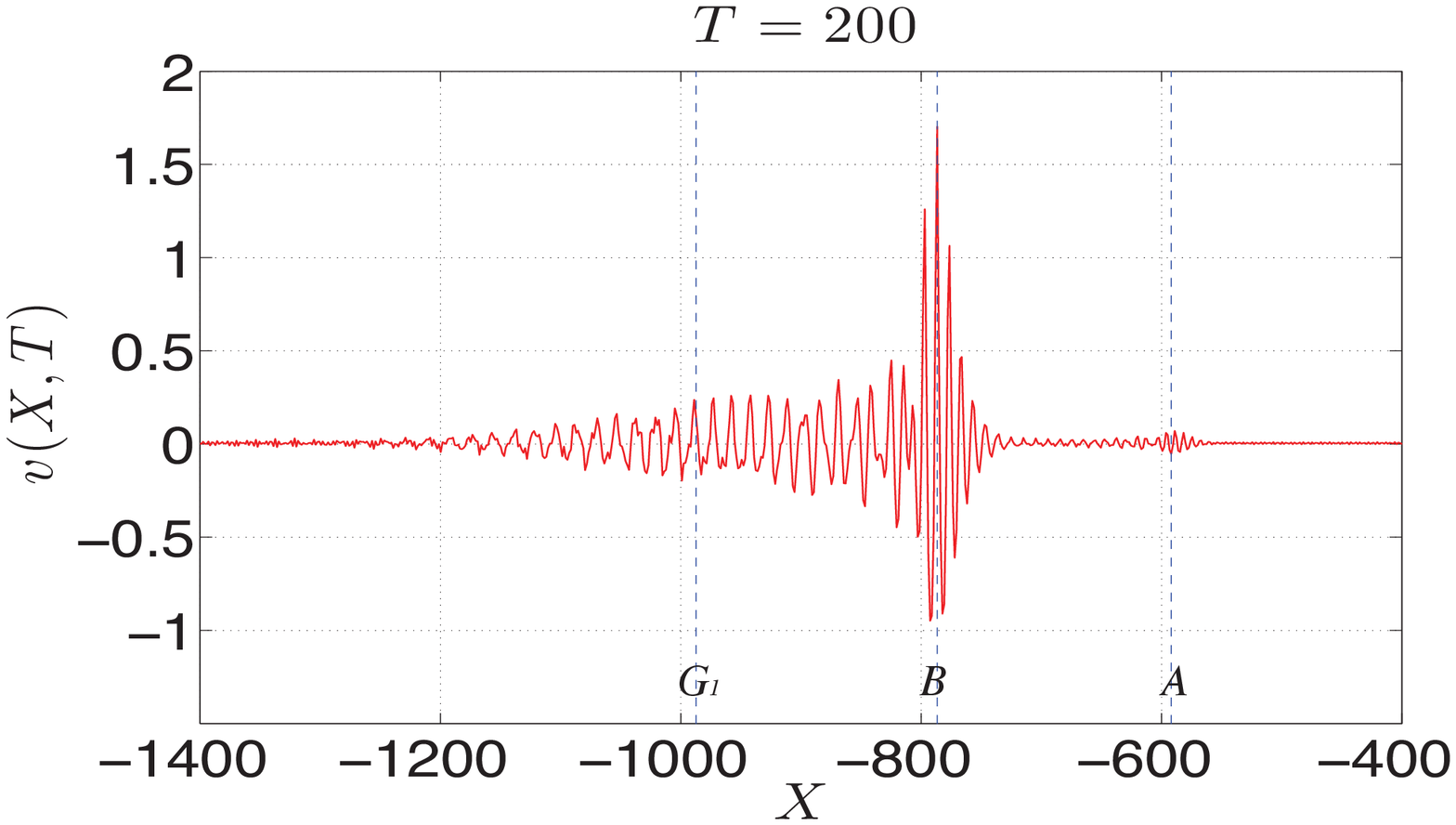}
\caption{Same as Figure \ref{fig:1a}, but a cross-section at $T=200$. }
\label{fig:1b}
\end{center}
\end{figure}

\noindent
{\bf Case A}:

 \noindent
 A typical numerical result is  shown in Figures \ref{fig:1a} and \ref{fig:1b} 
using the KdV solitary wave initial condition (\ref{weak}).
The generation of two wave packets can be seen in the $u$-component, 
but one of them is too small to be seen in the $v$-component. 
The comparison of the numerical modal ratio, 
 $R$ determined as described above,  shows  very good agreement with 
the theoretical prediction from the dispersion relation, see Table \ref{Table1}.
The  theoretical modal ratio is $r= 3.692$ for mode $1$ and $r =-0.132$ for mode $2$, while the 
 speeds  are $c_{g1}= -2.912, c_{g2} =  -3.854$ and $k_{m1}=0.895, k_{m2}=0.584$.
 The ratios of the numerically found wave packets  obtained from the  vertical dashed lines $A$ and $B$ in 
 Figure \ref{fig:1b} are given, respectively by $R=3.433$ for mode 1 and $R=0.176$ 
 for mode 2,  which are in agreement with the  theoretical predictions, and the numerically found speeds 
 $-2.960, -3.933$ are also in good agreement.
 However, we see that there is also some significant radiation to the left of these wave
 packets, and in particular some focussing possibly associated with the point $G_1$ in  Figure \ref{fig:1}. This is
 a resonance between the group velocity of mode $1$ and the phase velocity  of mode $2$. 
 The numerical speed and ratio at this point are given by, respectively, $-4.937$ and $0.639$ 
 in reasonable agreement with the theoretical prediction. 
 However, this resonance is perhaps contaminated here because the resonance points 
 $C, D, E, F$ on the dispersion curves  near $G_{1,2}$ are quite  close for a wide range of wavenumber $k$.

Next Figure \ref{fig:1c} shows the numerical results initiated using the wave packet initial conditions (\ref{wpic}) with 
$k=k_{m1}=0.895$ and ratio $r= 3.692$ for mode $1$, while we set $A_0=0.1$. 
These parameters correspond to mode $1$, see point $A$ in Figure \ref{fig:1}.
In qualitative agreement with the analogous results for a single Ostrovsky equation, 
we see the  emergence  of a nonlinear wave packet propagating  to the left with  speed $-2.940$ and ratio $3.685 $, 
which are both close to the theoretical prediction for point $A$, see Table \ref{Table1}. 
Here we also can detect a mode $2$ wave packet, corresponding to point $B$ in Figure \ref{fig:1}, 
as well as some radiation due to modal energy exchange associated with the resonance point $G_1$. 
The numerically found speeds  are, respectively, $-4.805$ for point B and $-5.996$ for point $G_1$,
with ratios $R=0.411$ and $R=1.162$.
In this simulation, we do not see any evidence of  waves associated with the  points $C,D,E,F$.

\begin{figure}[htbp]
\begin{center}
\includegraphics[height=5cm,width=8cm]{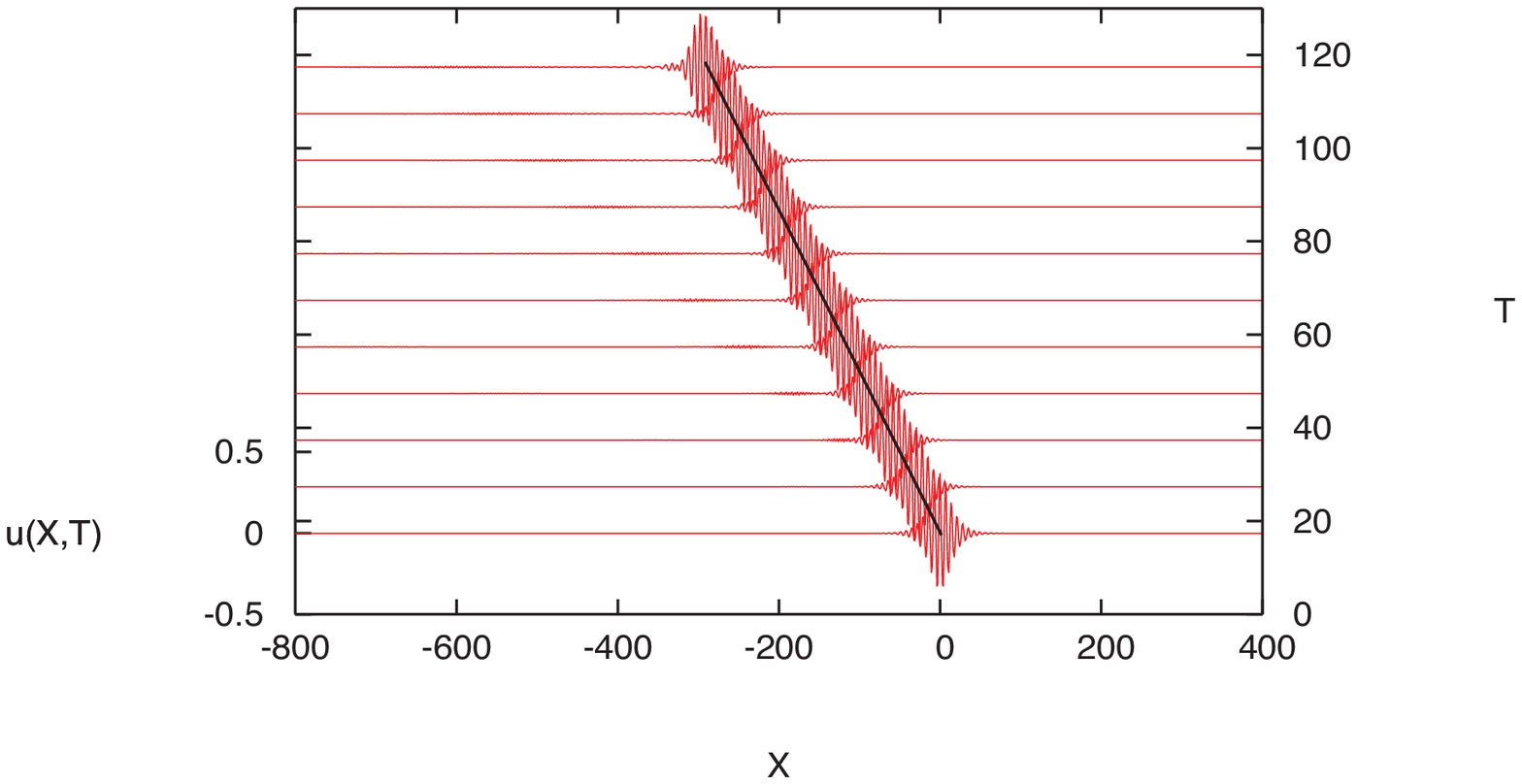} 
\includegraphics[height=5cm,width=8cm]{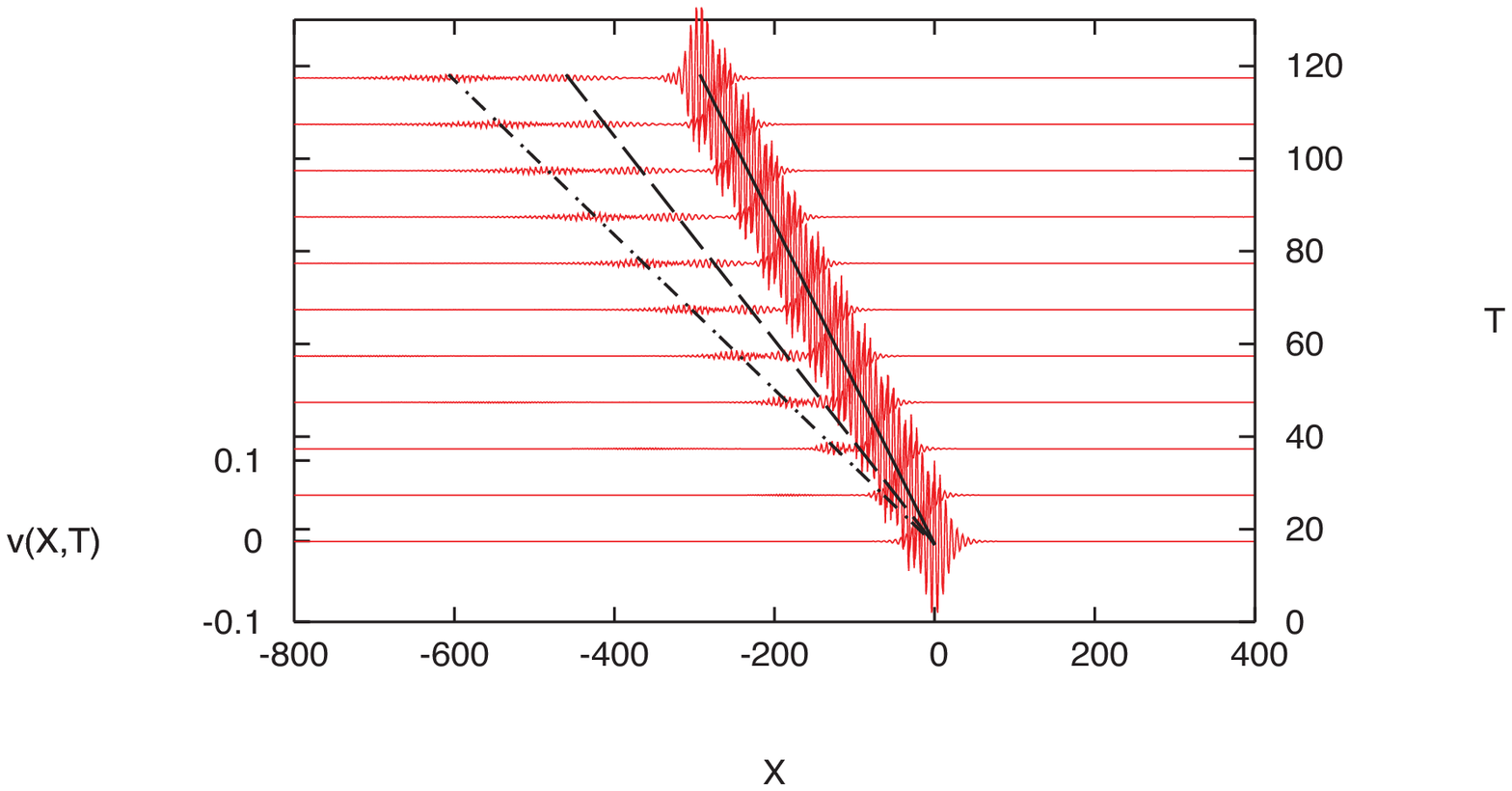}
\caption{Numerical simulations for Case A using the wave packet initial condition (\ref{wpic}) with 
 $k=k_{m1} = 0.895$ for mode $1$, and $A_0 =0.1, K_0=0.1\, k.$
 The solid, dashed and dash-dot lines refer to the points $A$, $B$ and $G_1$ in Figure \ref{fig:1}.}
\label{fig:1c}
\end{center}
\end{figure}

\begin{figure}[htbp]
\begin{center}
\includegraphics[height=3.5cm,width=8cm]{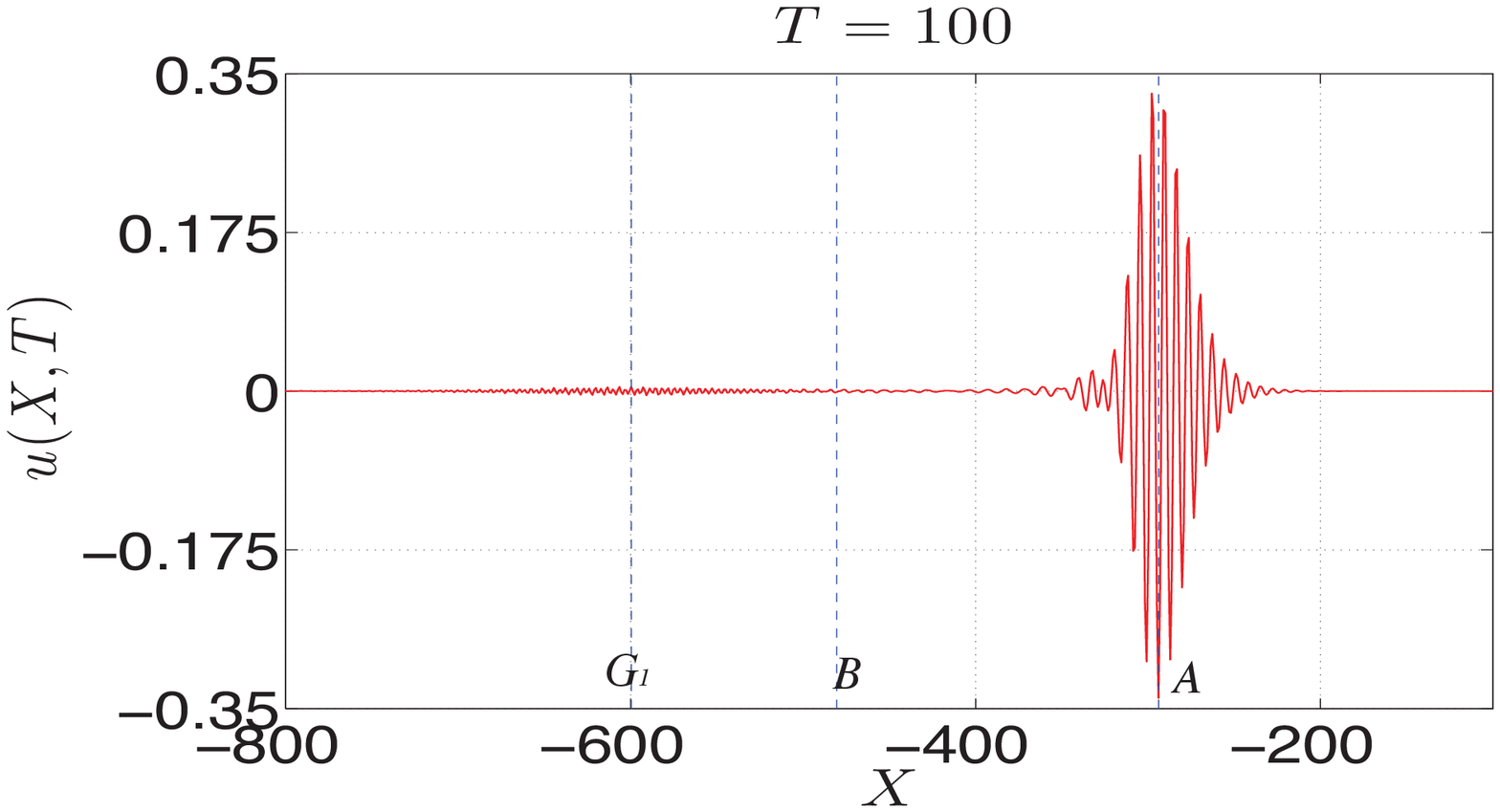} 
\includegraphics[height=3.5cm,width=8cm]{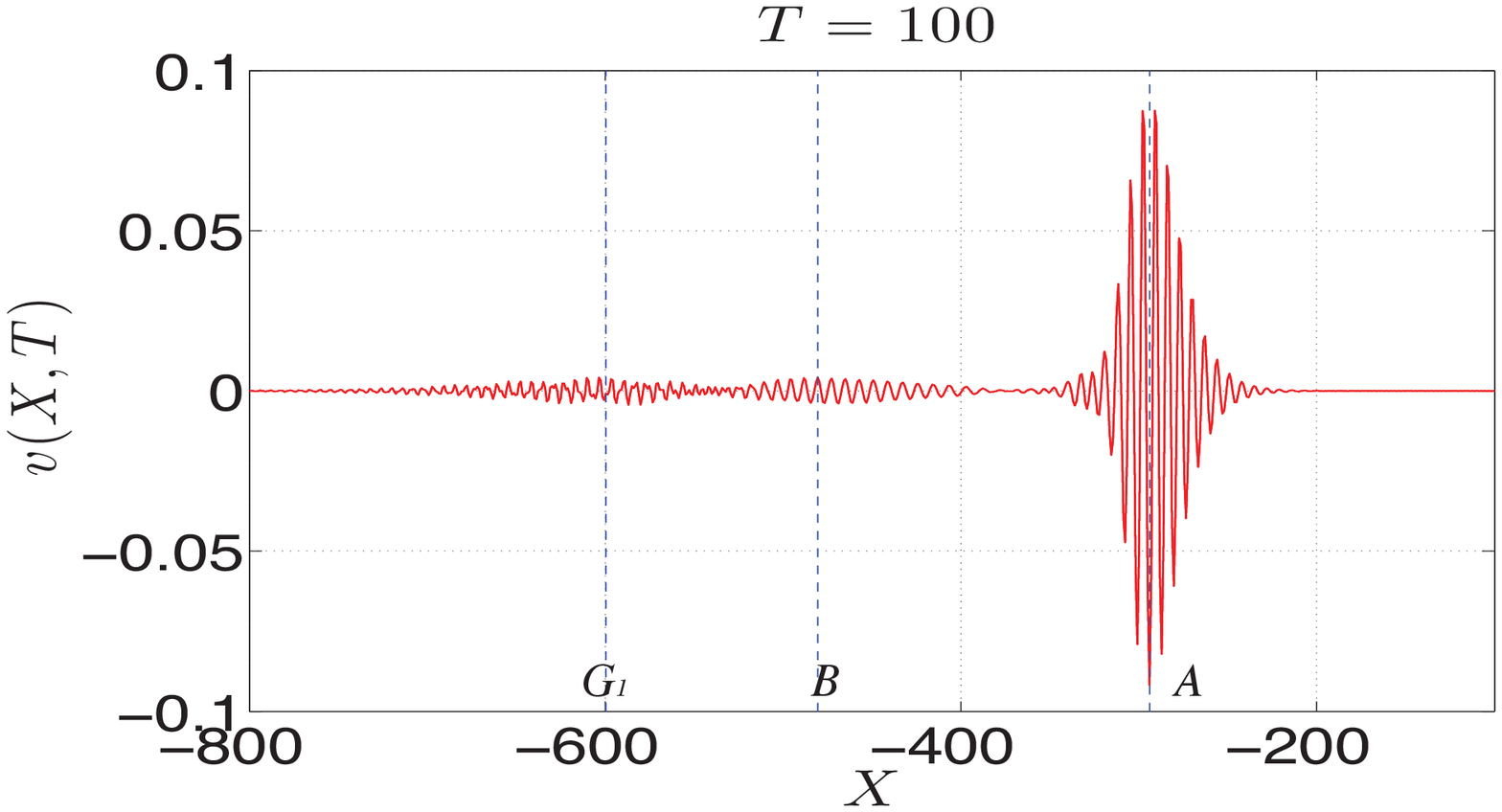}
\caption{Same as Figure \ref{fig:1c}, but a cross-section at $T=100$.}
\label{fig:1c_t100}
\end{center}
\end{figure}

\begin{figure}[htbp]
\begin{center}
\includegraphics[height=5cm,width=8cm]{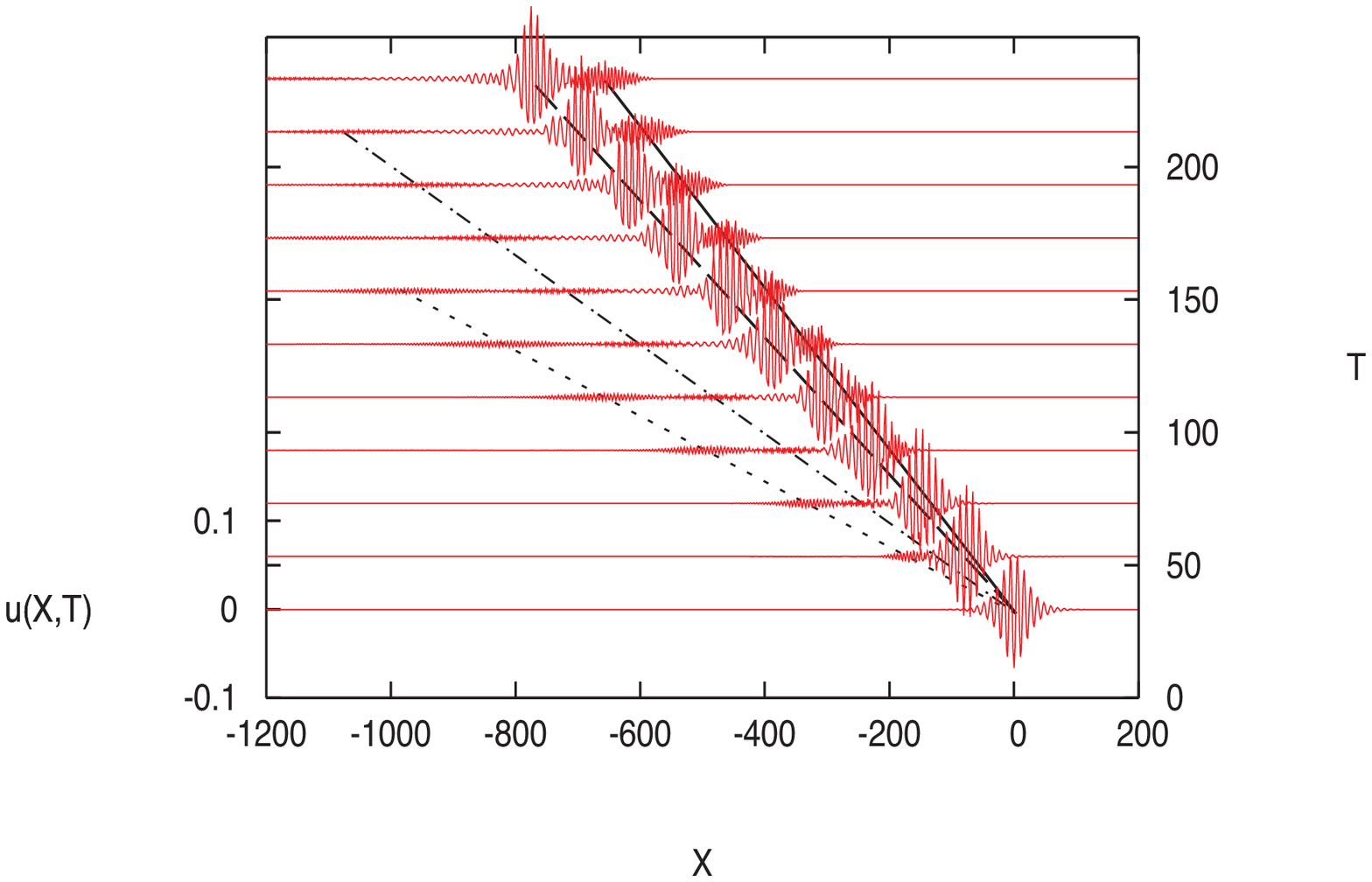} 
\includegraphics[height=5cm,width=8cm]{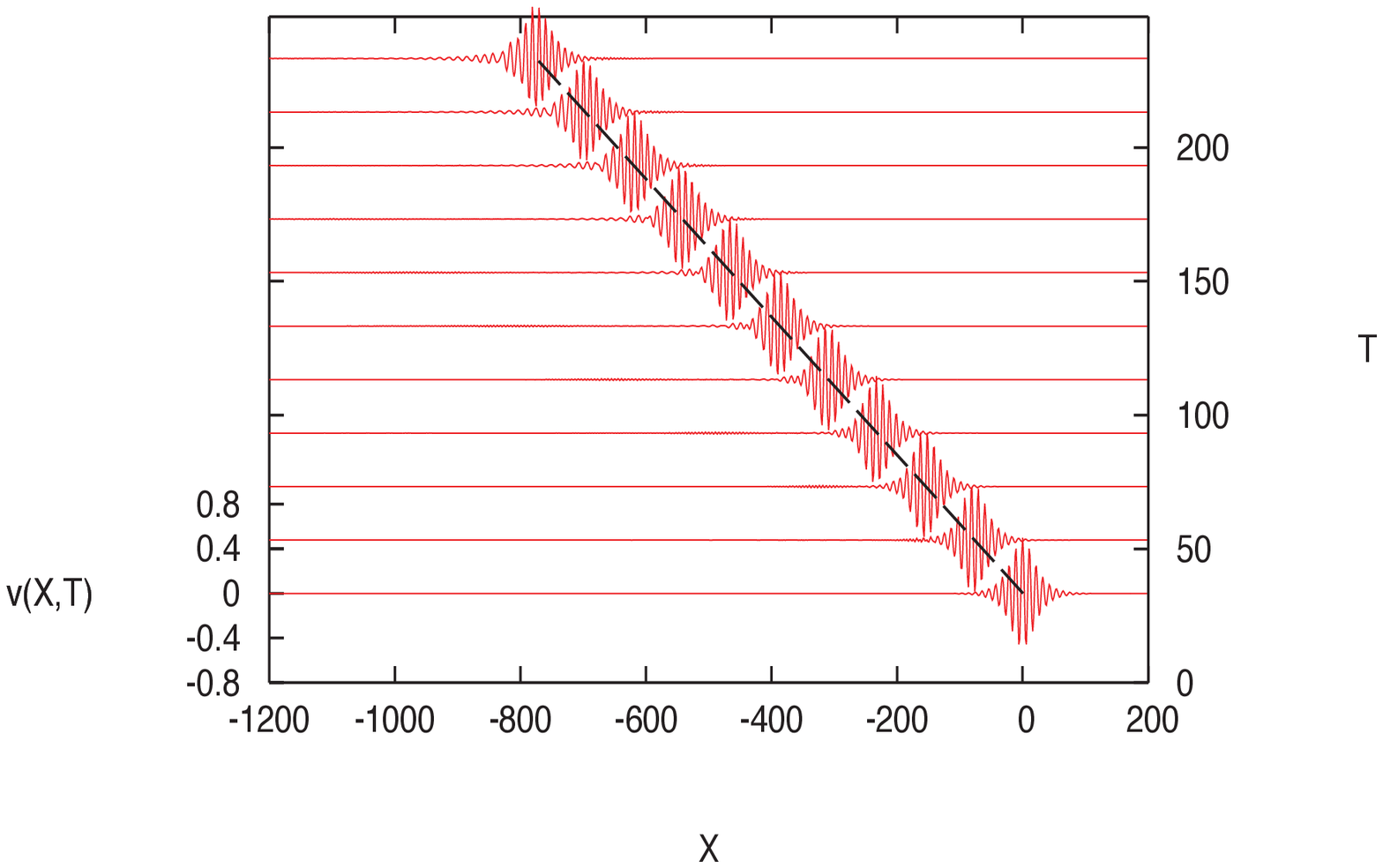}
\caption{Numerical simulations for Case A using the wave packet initial condition (\ref{wpic}) with  $k=k_{m2} = 0.584$ for mode $2$, 
and $A_0 =0.5, K_0=0.1\, k, V_0 =1$. The solid, dashed, dash-dot and dotted lines refer to the points $A$, $B$, $G_1$ and ($C, E$) in Figure \ref{fig:1}. Note that the scales for the $u$ and $v$ components are different.}
\label{fig:1d}
\end{center}
\end{figure}

\begin{figure}[htbp]
\begin{center}
\includegraphics[height=3.5cm,width=8cm]{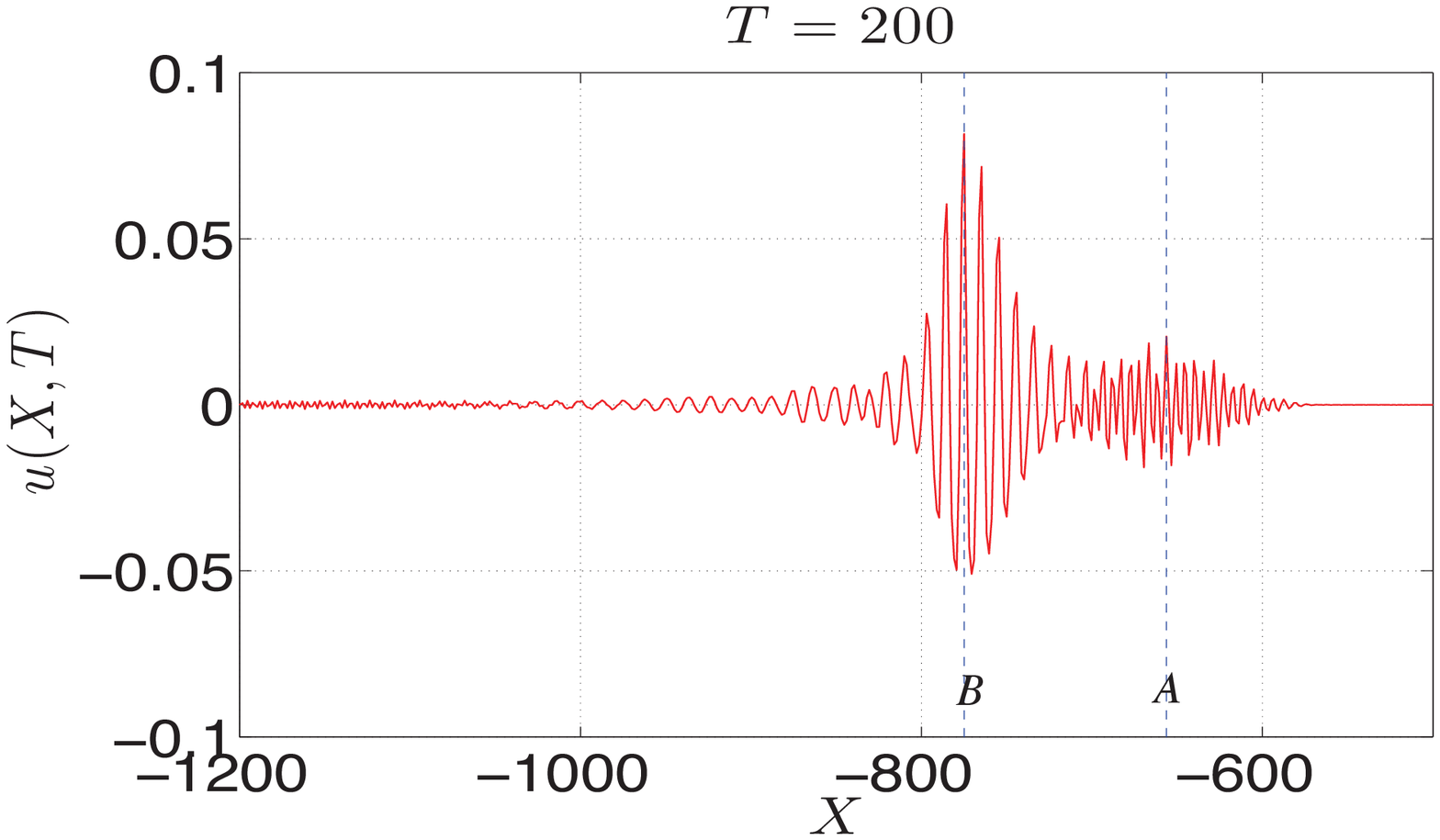} 
\includegraphics[height=3.5cm,width=8cm]{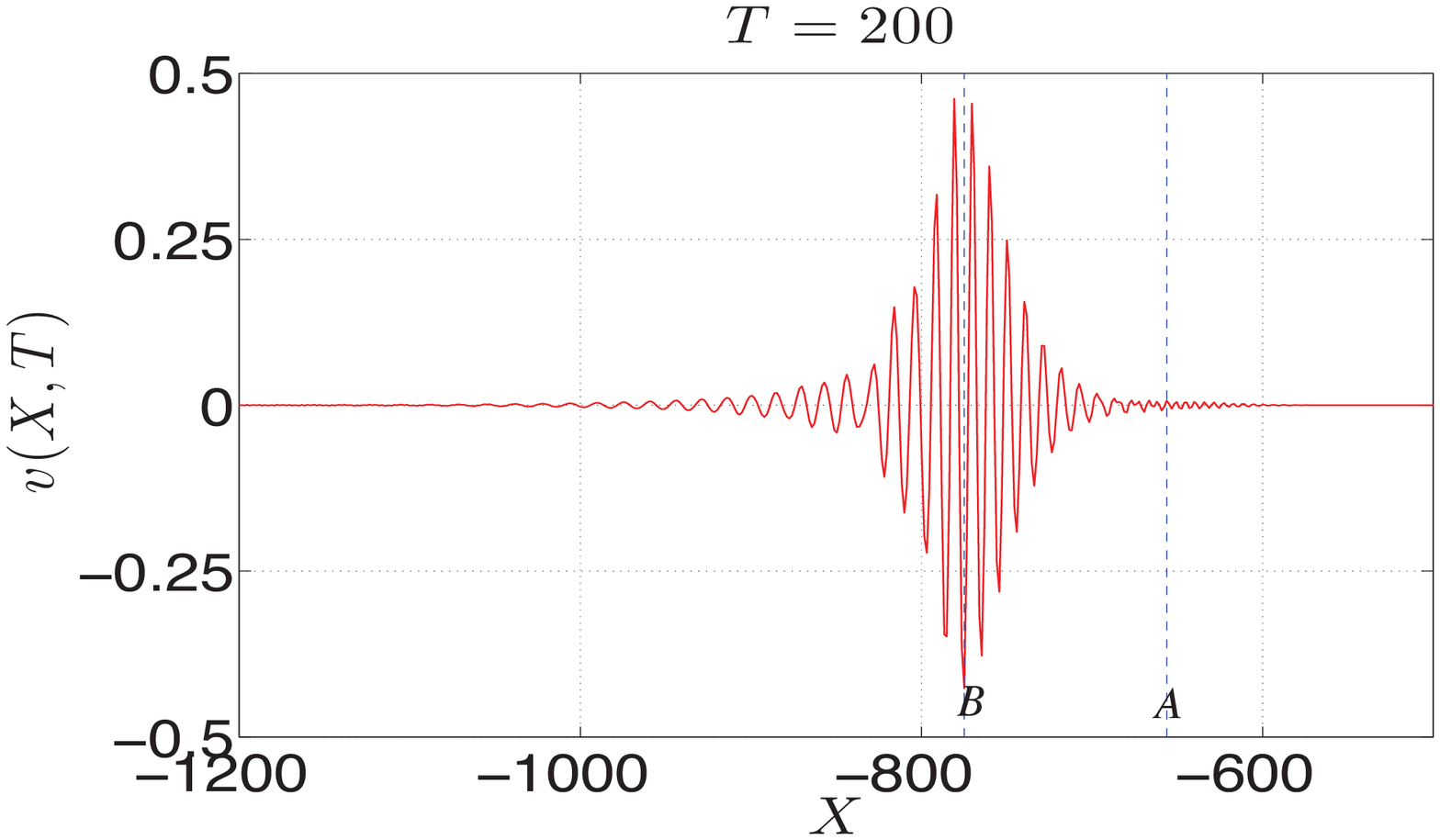}
\caption{Same as Figure \ref{fig:1d}, but a cross-section at $T=200$.}
\label{fig:1f_t200}
\end{center}
\end{figure}

\begin{figure}[htbp]
\begin{center}
\includegraphics[height=3.5cm,width=8cm]{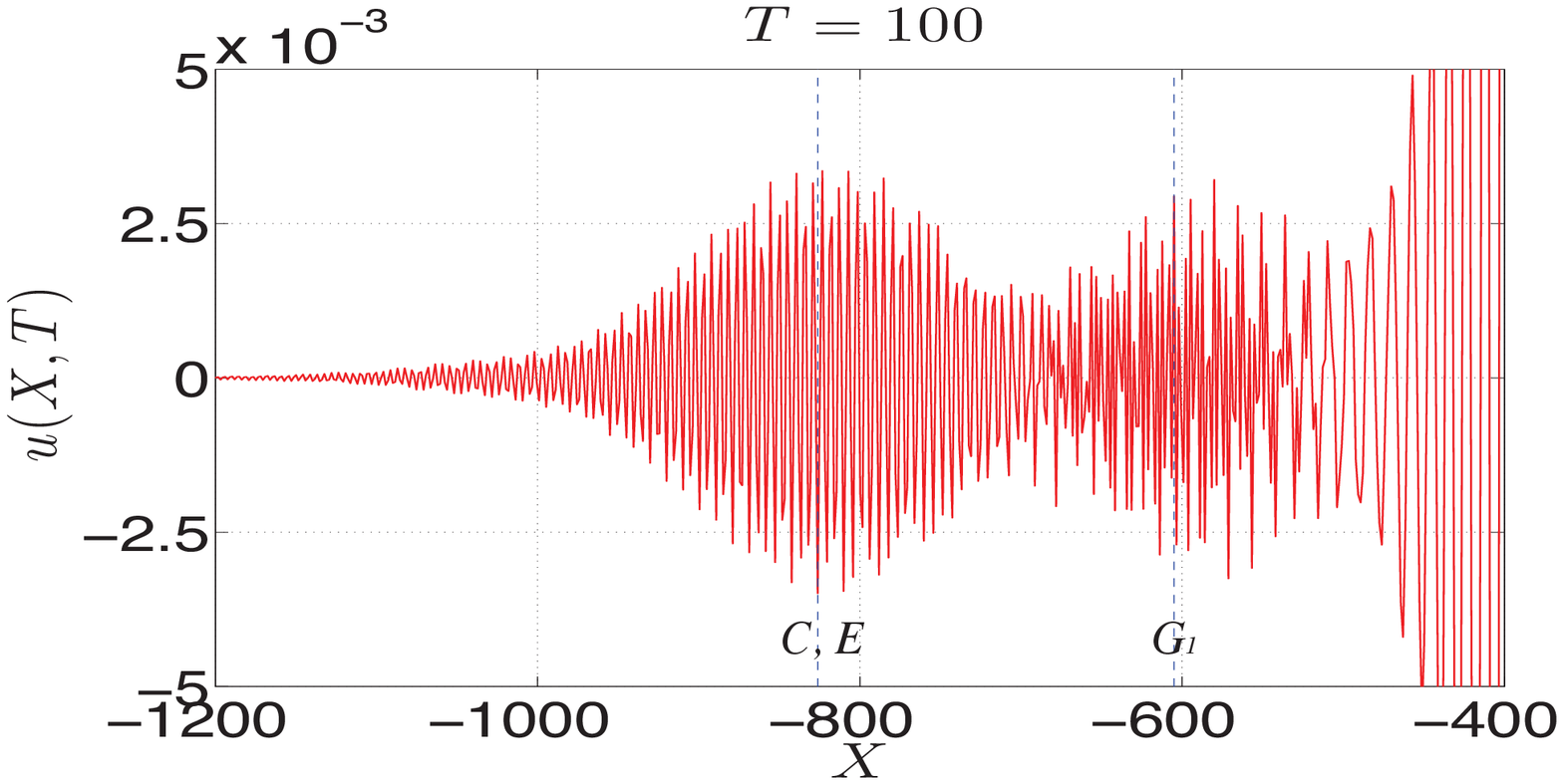} 
\includegraphics[height=3.5cm,width=8cm]{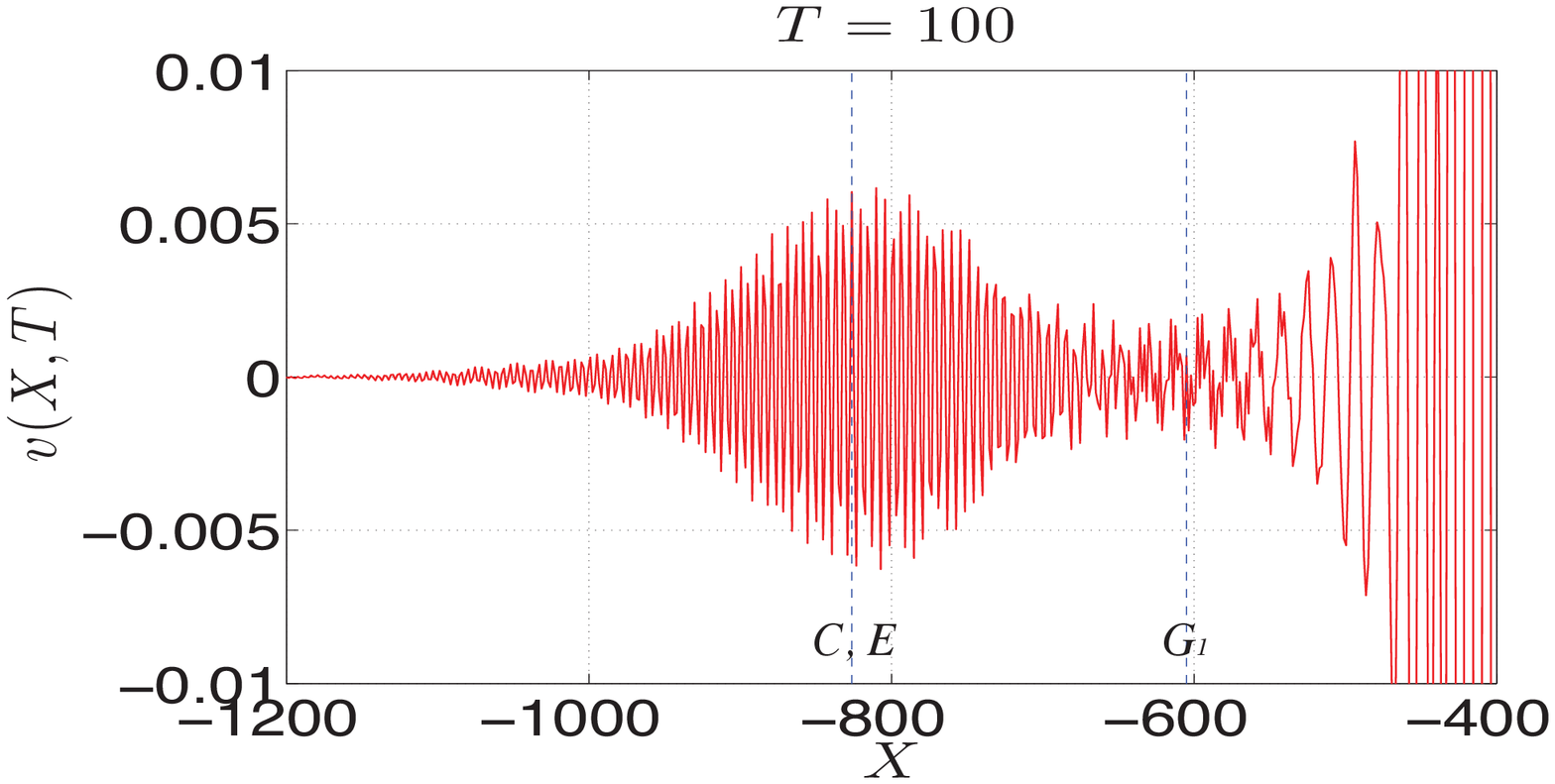}
\caption{Same as Figure \ref{fig:1d}, but a cross-section at $T=100$ of the domain $-1200< X < -400$
and with rescaled $u,v$.}
\label{fig:1f_t100}
\end{center}
\end{figure}

Figures  \ref{fig:1d}, \ref{fig:1f_t200} and \ref{fig:1f_t100}  
show  the numerical results commenced with wave packet initial conditions (\ref{wpic}) 
 with $k=k_{m2}=0.584$ and ratio $r= -0.132$ for mode $2$.
 These parameters correspond to point $B$ in Figure \ref{fig:1}.
 Again, we  can clearly see one wave packet emerging and propagating with
 a speed  $-3.904$ and ratio $0.177$, both close to the theoretical prediction for point $B$, see Table \ref{Table1}. 
 But here there  is also a small unsteady wave packet, seen in the $u$-component, 
 moving with the speed  $-3.281$ close to the
 theoretical prediction of $c_{g1}=-2.912$ and ratio $R = 2.555$ for a mode $1$ wave packet, corresponding to point $A$
 in Figure \ref{fig:1} and Table \ref{Table1}. Here  we also can see the formation of wave packets to the left, 
 corresponding to points $G_1$ and $C,E$ with the numerically found speeds $-6.050, -8.262$ and ratios 
 $1.436, 0.567$ also in reasonable agreement with the theoretical prediction.

\begin{figure}[htbp]
\begin{center}
\includegraphics[height=5cm,width=8cm]{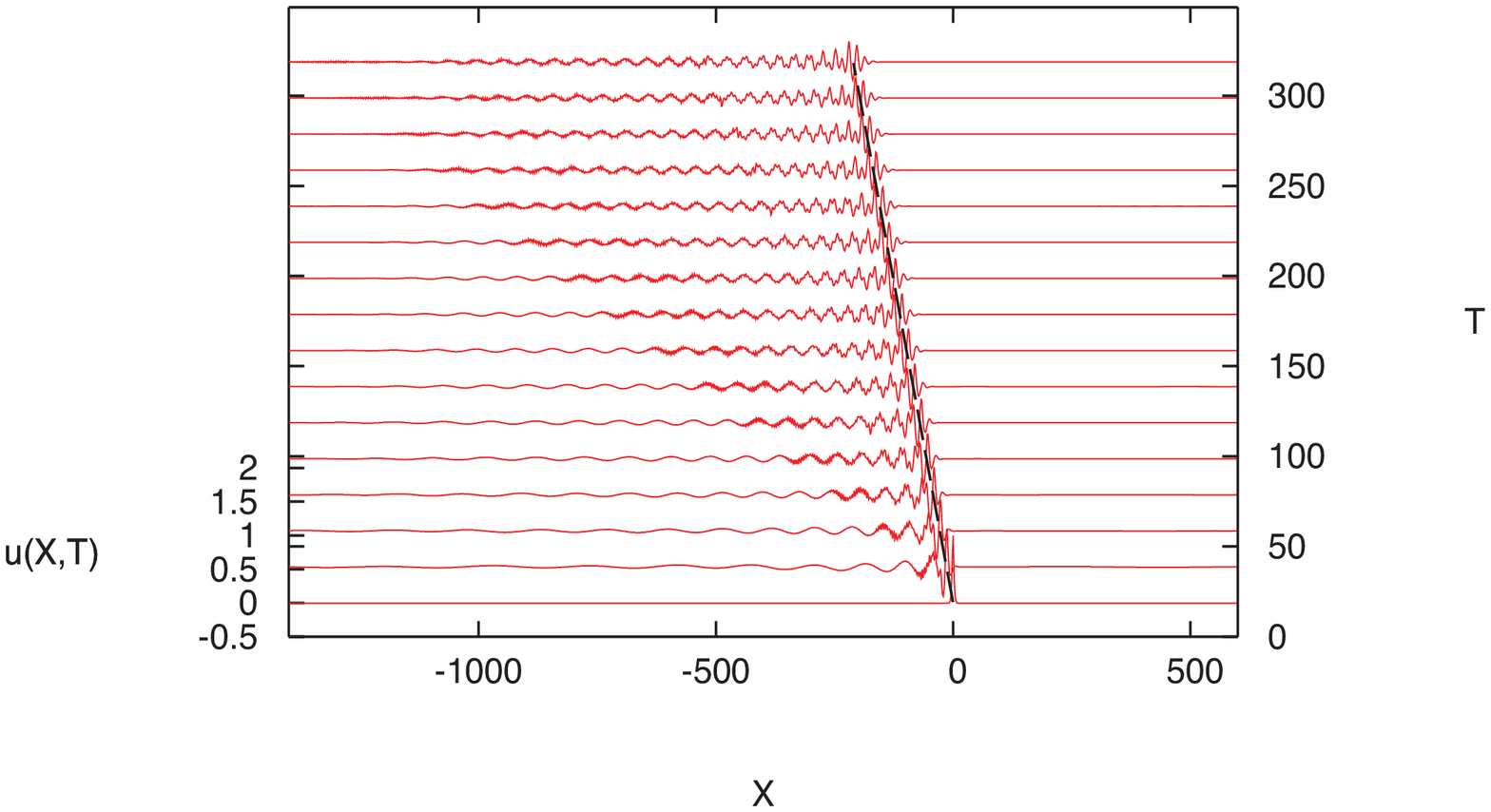}
\includegraphics[height=5cm,width=8cm]{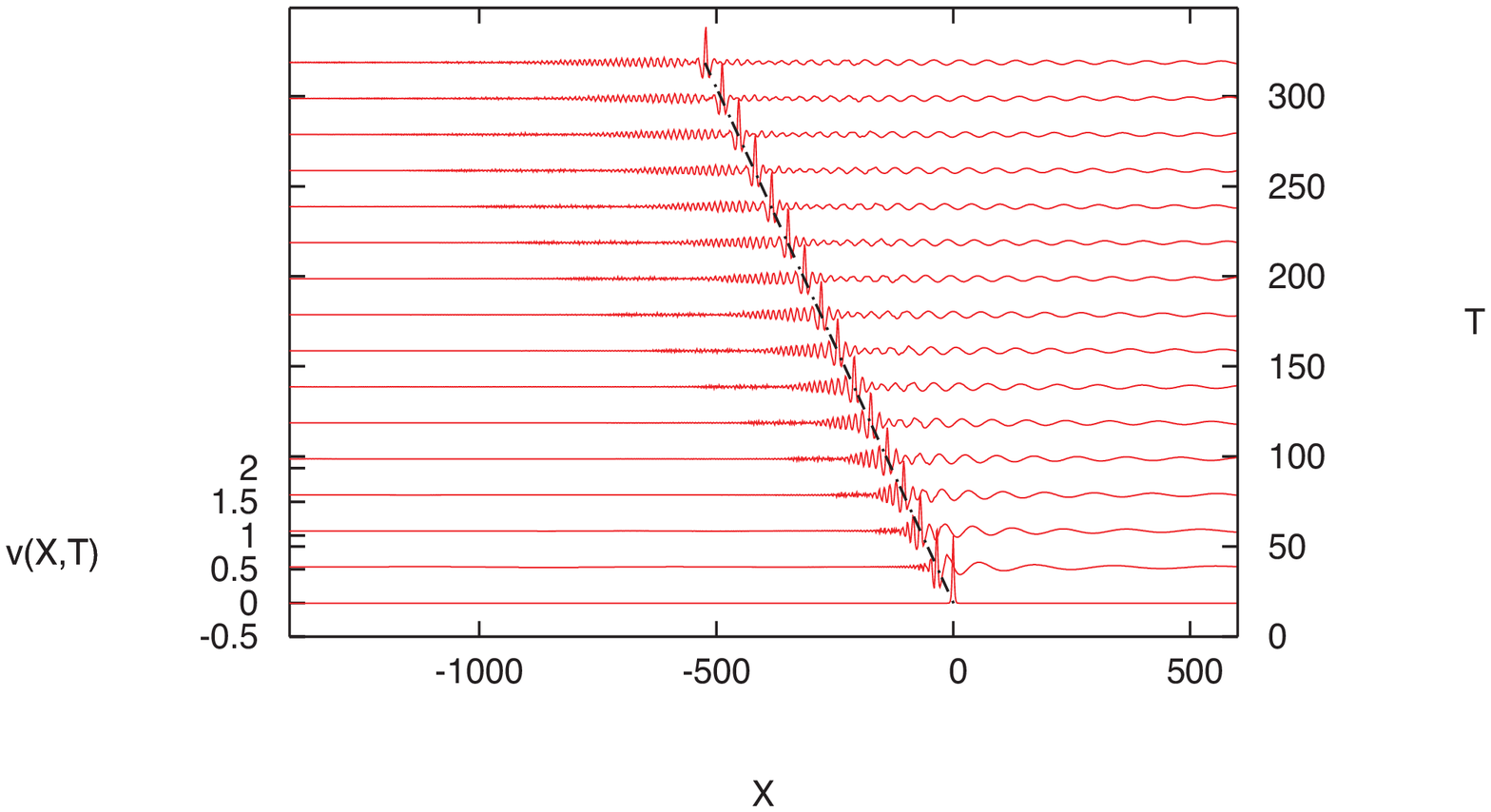}
 \caption{Numerical simulations for Case B using a KdV initial condition of weak coupling with the parameter $a=b=1$. The dashed line in $u$-plot refers to point $A$ and the dash-dot line in $v$-plot refers to point $B$.}
\label{fig:2a}
\end{center}
\end{figure}

\begin{figure}[htbp]
\begin{center}
\includegraphics[height=3.5cm,width=8cm]{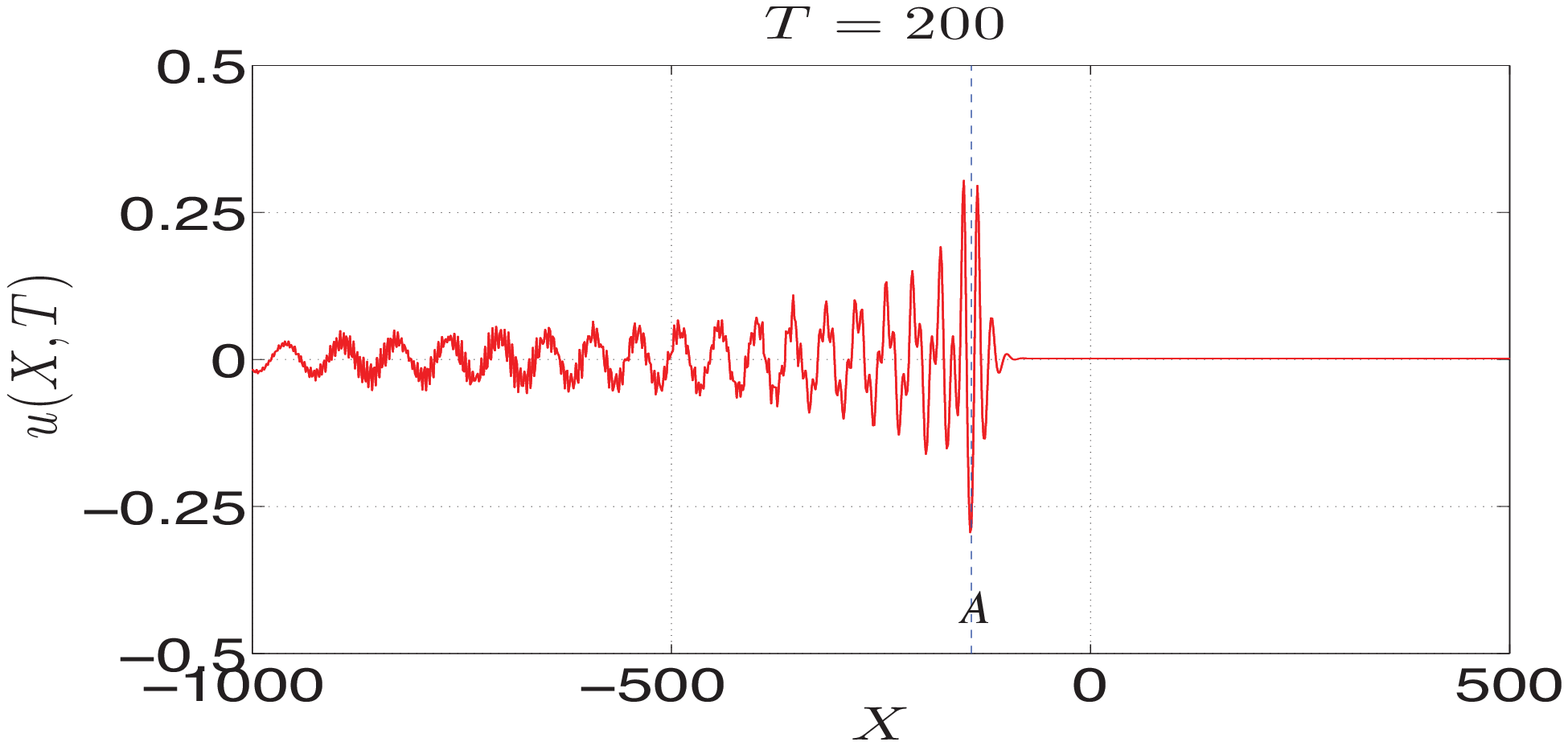}
\includegraphics[height=3.5cm,width=8cm]{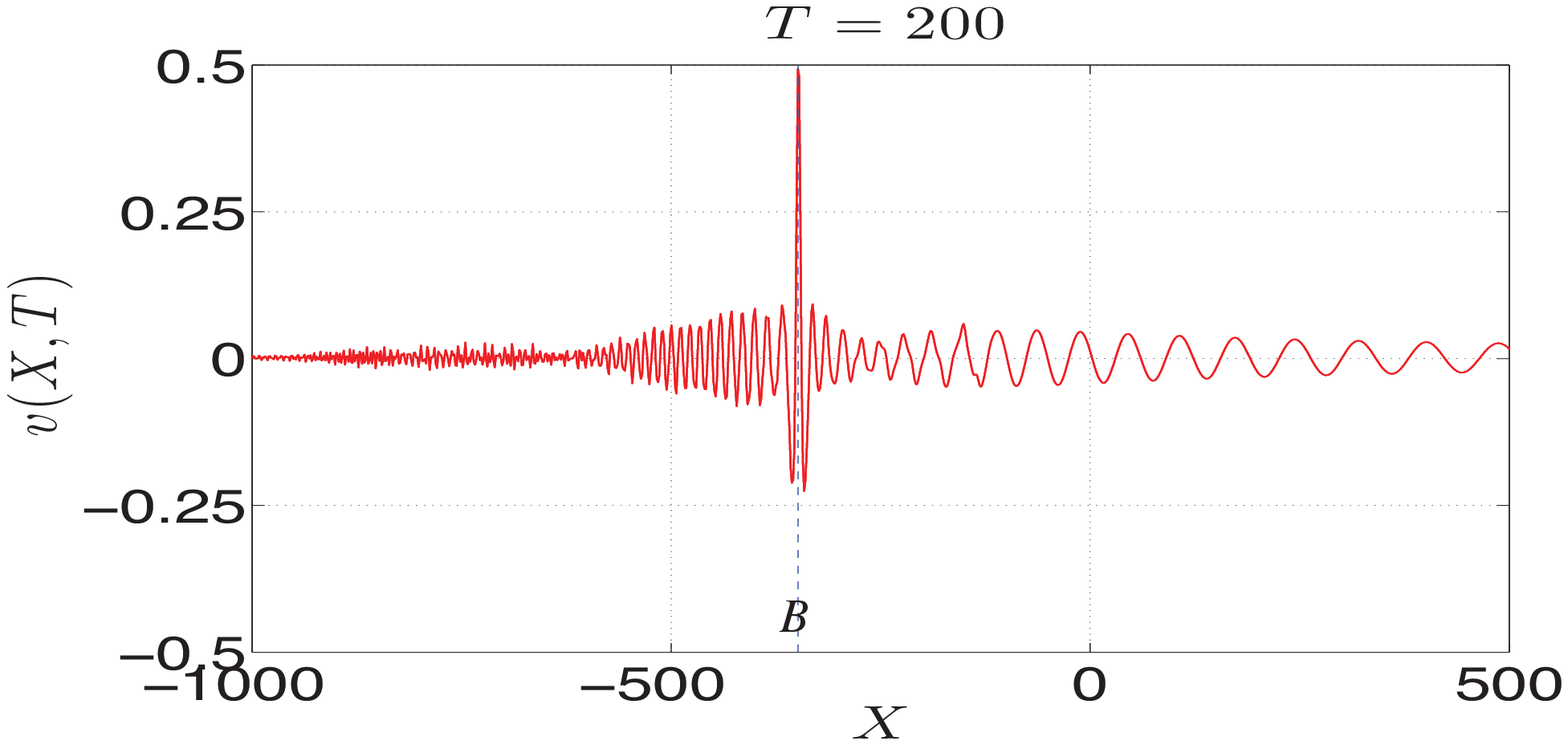}
 \caption{Same as Figure \ref{fig:2a}, but a cross-section at $T=200$ for both modes.}
\label{fig:2b}
\end{center}
\end{figure}

\medskip
\noindent
{\bf Case B}:

 \noindent
A typical numerical result is  shown in Figures \ref{fig:2a}, \ref{fig:2b}  using the KdV solitary wave initial condition (\ref{weak}).
We can clearly see a wave packet in the $u$-component identified by the vertical dashed line $A$, 
 with  speed $-0.710$ and ratio $4.815$. The corresponding 
  theoretical predictions  are a speed $c_{g1} = -0.6834$  and ratio $r= 33.696$, 
  corresponding to point $A$ in Figure  \ref{fig:2}, see Table \ref{Table2}.  
  However, here the wave packet is strongly nonlinear, and we note that if $v$ is measured at the point where
  $|u|$ is a maximum, then the numerical ratio is $25$, closer to the theoretical value.
Another wave packet can be clearly seen in $v$-component with  speed $-1.743$ and  ratio $-0.221$.
Here the corresponding theoretical predictions are a speed $-1.785$ and ratio $r=-0.036$, 
corresponding to point $B$ in Figure  \ref{fig:2}, see Table \ref{Table2}. Again, this
wave packet is strongly nonlinear, and if $u$ is measured at the point where $|v|$ is a 
maximum, then the numerical ratio is $-0.0229$, closer to the theoretical value.
Also note that since there is considerable radiation in the plot, we cannot  
detect the wave packet associated to point $A$ in the $v$-plot, and similarly for the 
point $B$ in the $u$-plot.

In Figures \ref{fig:2c} and \ref{fig:2d} we use the  wave packet initial condition
({\ref{wpic}), with $k=k_{m1} = 0.345$ and the ratio $R= 33.696 $ corresponding to a maximum group velocity 
$c_{g1}=-0.683$ in mode $1$
corresponding to point $A$ in Figure \ref{fig:2}, see Table \ref{Table2}.
As expected, an unsteady wave packet emerges, clearly seen in both the $u$ and $v$ plots in the first solid line, propagating with speed $-0.610$ and ratio $21.261$ in reasonable agreement with the theoretical predictions. 
The dashed line in the $v$-plot shows a wave packet propagating with  speed $-1.343$, but the ratio 
cannot be measured here as in the $u$-plot, this location is the tail of the larger wave packet associated
with point $A$.  Based on the speed and wavenumber, we suggest this is associated with 
point $B$ in  Figure \ref{fig:2}, see Table \ref{Table2}.
A third small wave packet can be observed in the $v$ -mode represented by the dash-dot line with  speed $-2.446$ 
and ratio $3.160$, which we associate with the resonance point $C$ for mode $1$  in Figure \ref{fig:2}, 
see Table \ref{Table2},  generated by a mode $1$ unsteady wave packet associated with the point $A$. 
Then, a fourth small wave packet can also be observed in the $v$-mode represented by the dotted line 
with  speed $-3.057$ and ratio $0.148$, which we associate with the point $E$, based on ratio and wavenumber considerations. Both these third and fourth wave packets have speeds which might be associated with
the point $D_{1}$, but we have ruled out this connection due to a large disparity between the predicted and observed ratio and wavenumber.

\begin{figure}[htbp]
\begin{center}
\includegraphics[height=5cm,width=8cm]{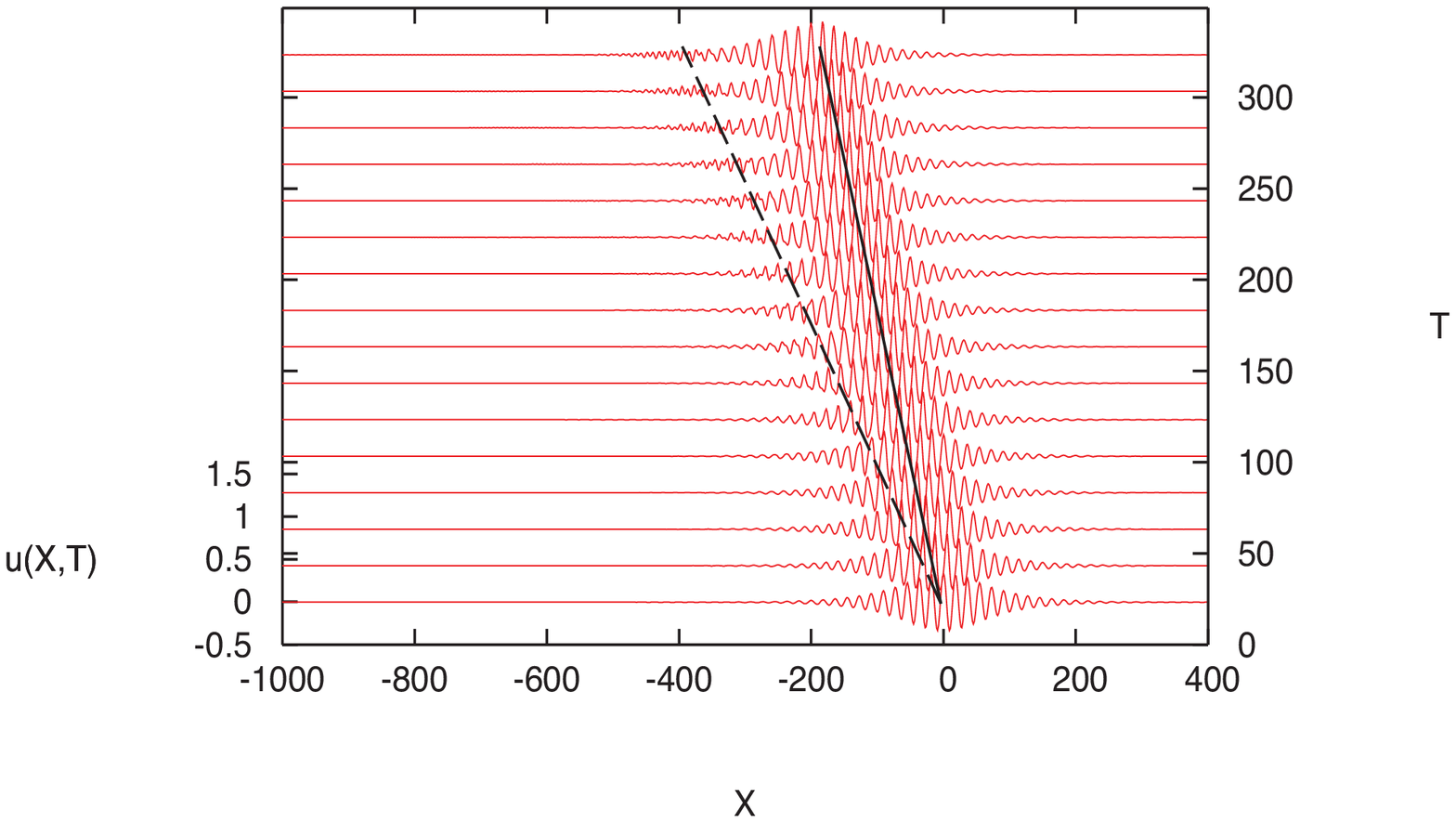}
\includegraphics[height=5cm,width=8cm]{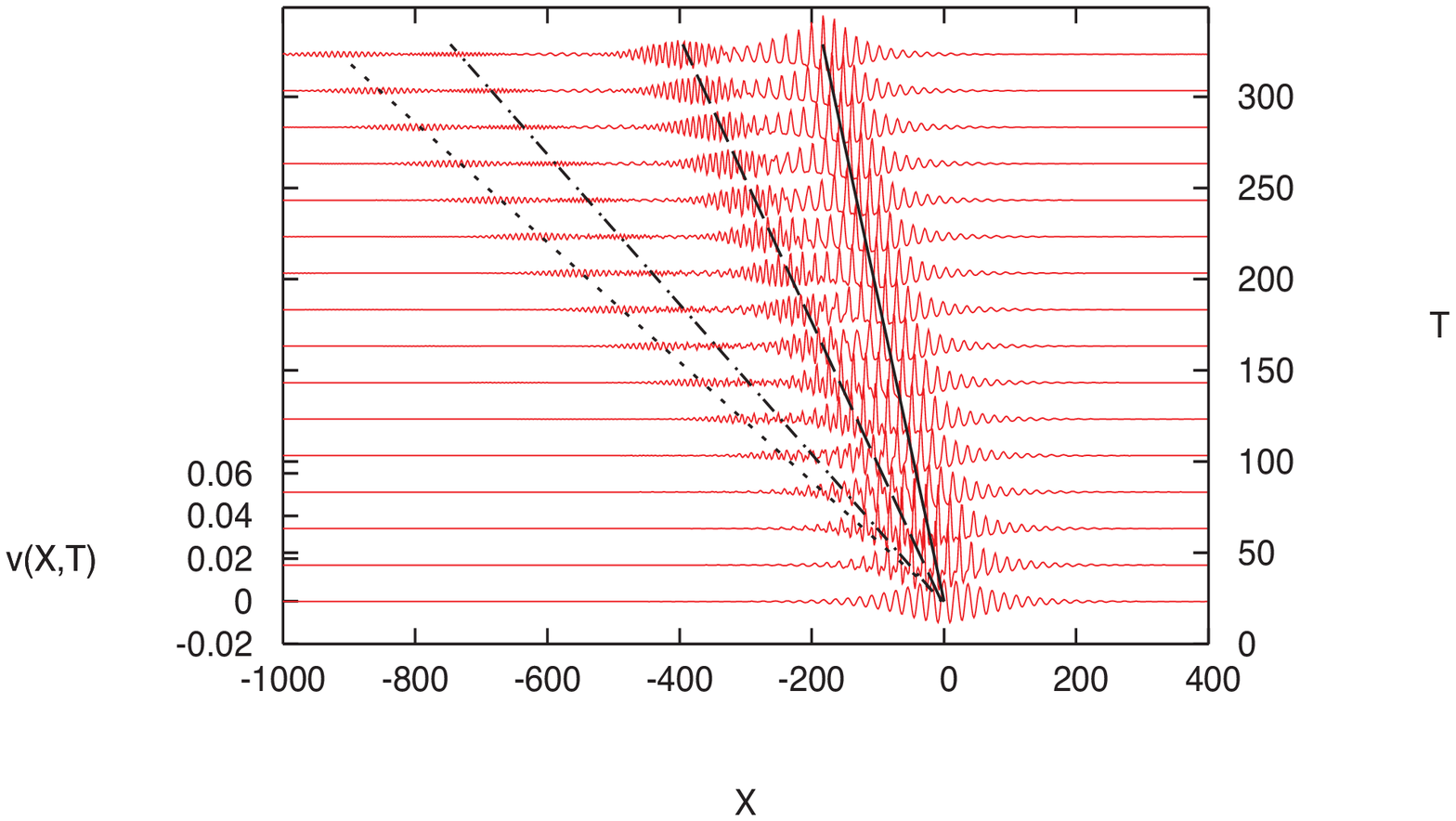}
 \caption{Numerical simulations for Case B using a nonlinear wave packet initial condition corresponding to the value $k=k_{m1}=0.345$ with $A_0=0.01 \, , K_0=0.05 \, k$ and $V_0=1$. The solid, dashed, dash-dot and dotted lines respectively refer to points $A$, ($B,C$), ($C,D_1$) and ($D_1,E$) in the dispersion relation.}
\label{fig:2c}
\end{center}
\end{figure}

\begin{figure}[htbp]
\begin{center}
\includegraphics[height=3.5cm,width=8cm]{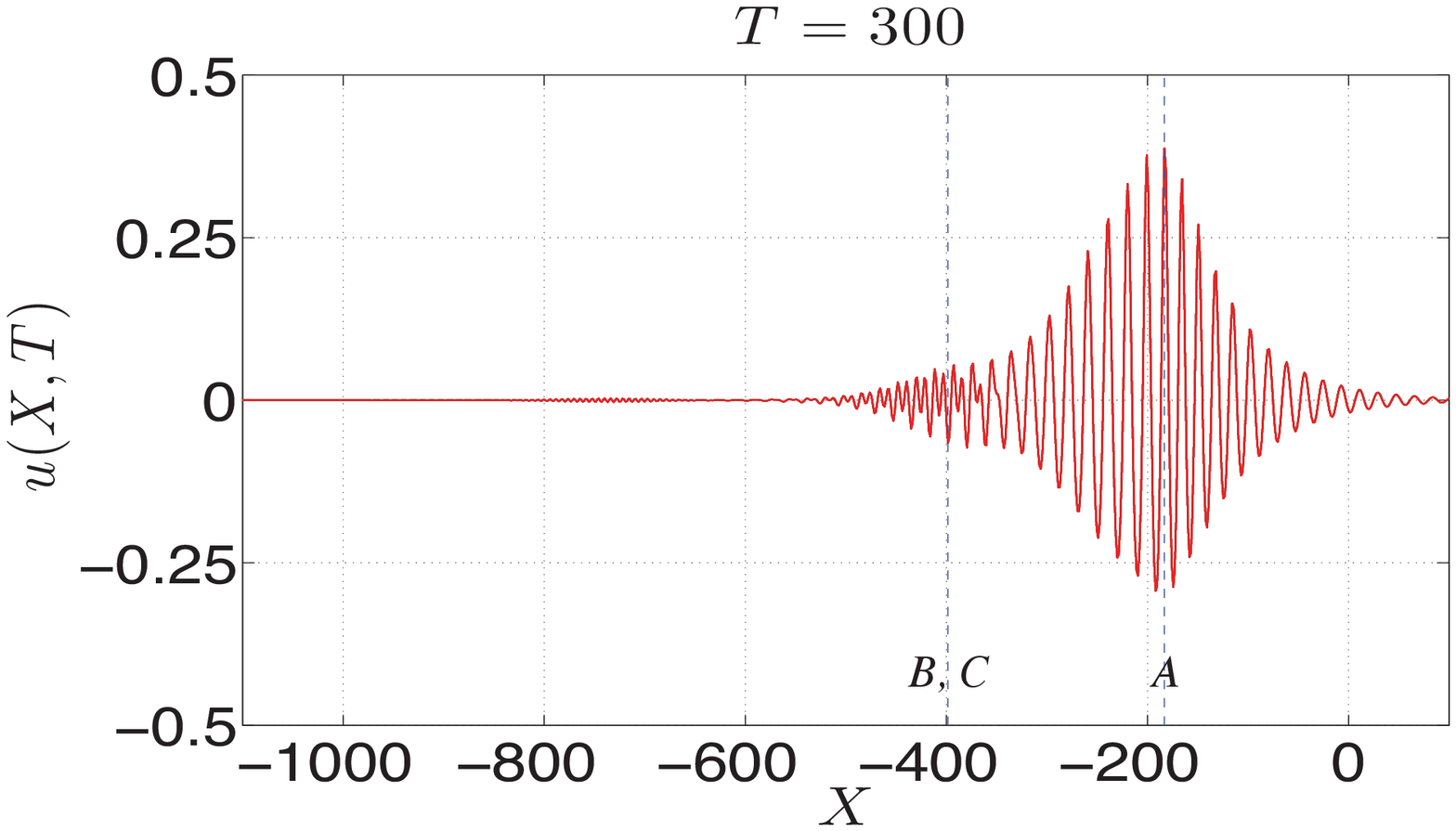}
\includegraphics[height=3.5cm,width=8cm]{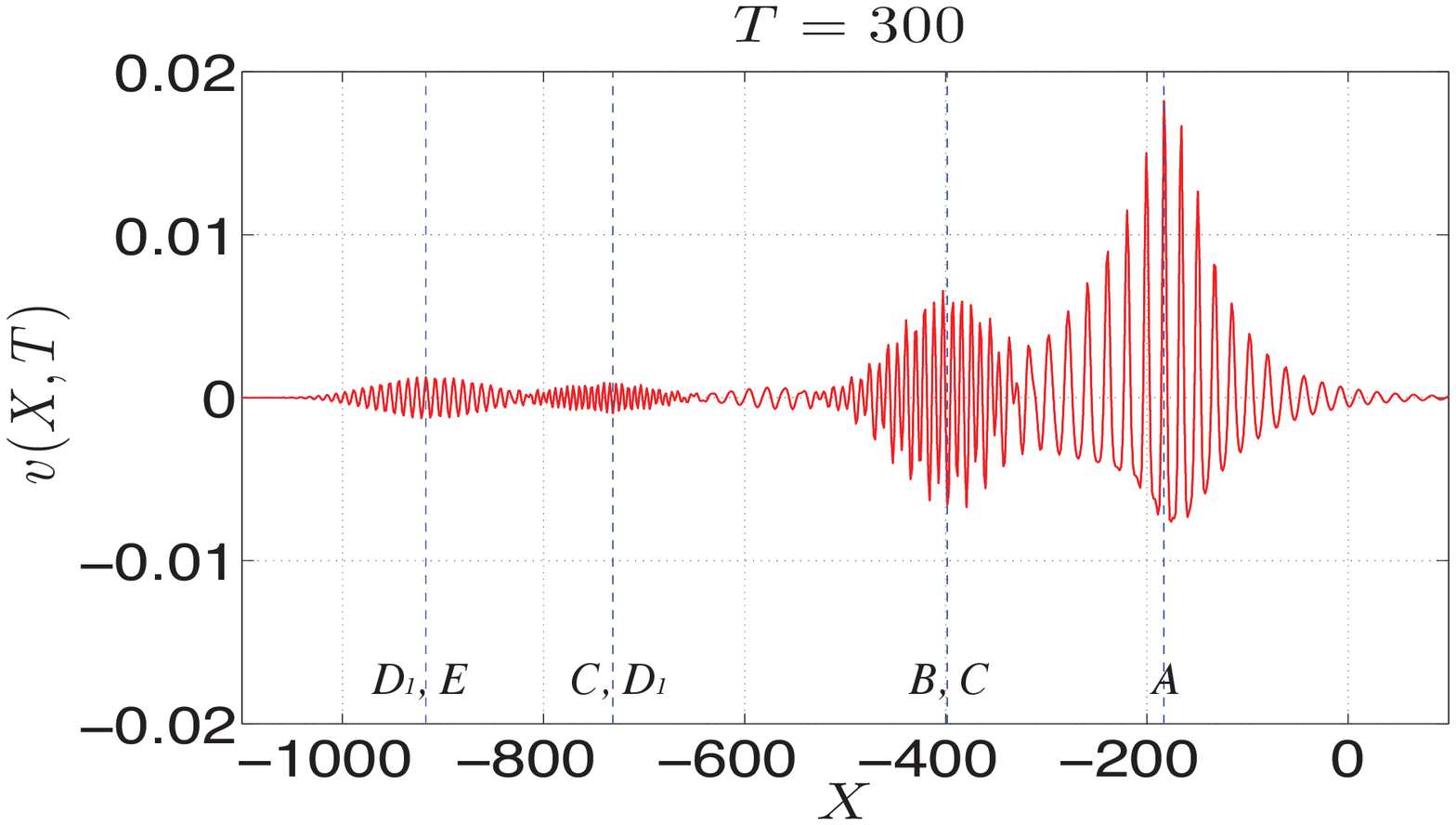}
 \caption{Same as Figure \ref{fig:2c}, but a cross-section at  $T=300$ for both modes.}
\label{fig:2d}
\end{center}
\end{figure}

 Figures \ref{fig:2e} and \ref{fig:2f} show the case when the  wave packet initial condition (\ref{wpic}) has 
 $k=k_{s2}=0.372$ with  ratio $R=-0.036$ corresponding  to a  maximum phase speed in mode $2$, represented
 by the point $B$ in Figure {\ref{fig:2}}, see Table \ref{Table2}.  
  In the both modes, the main feature is a steady wave packet with  speed  $-1.787$ and 
 ratio $0.042$,  see the dashed line, in good agreement with  the predicted values  from the dispersion relation, 
 see Table \ref{Table2}.  There is a very small wave packet indicated by the solid line with a speed
 $-0.461$ which we associate with point $A$ based on the speed.  Here the ratio 
 cannot be measured as this location lies in the tail of the larger wave packet associated
with point $B$.  There is a third wave packet shown by the blue line  with speed $-3.362$ and ratio $0.144$,
 which we associate with the point $E$, based on the consideration of the speed and wavenumber, as the ratio cannot 
 be measured accurately since in the $v$-plot this location lies in the tail of the main wave packet. 
Wave packets have speeds which might be associated with the point $D_{1}$, but we have ruled 
out this connection due to a large disparity between the predicted and observed ratio and wavenumber.

\begin{figure}[htbp]
\begin{center}
\includegraphics[height=5cm,width=8cm]{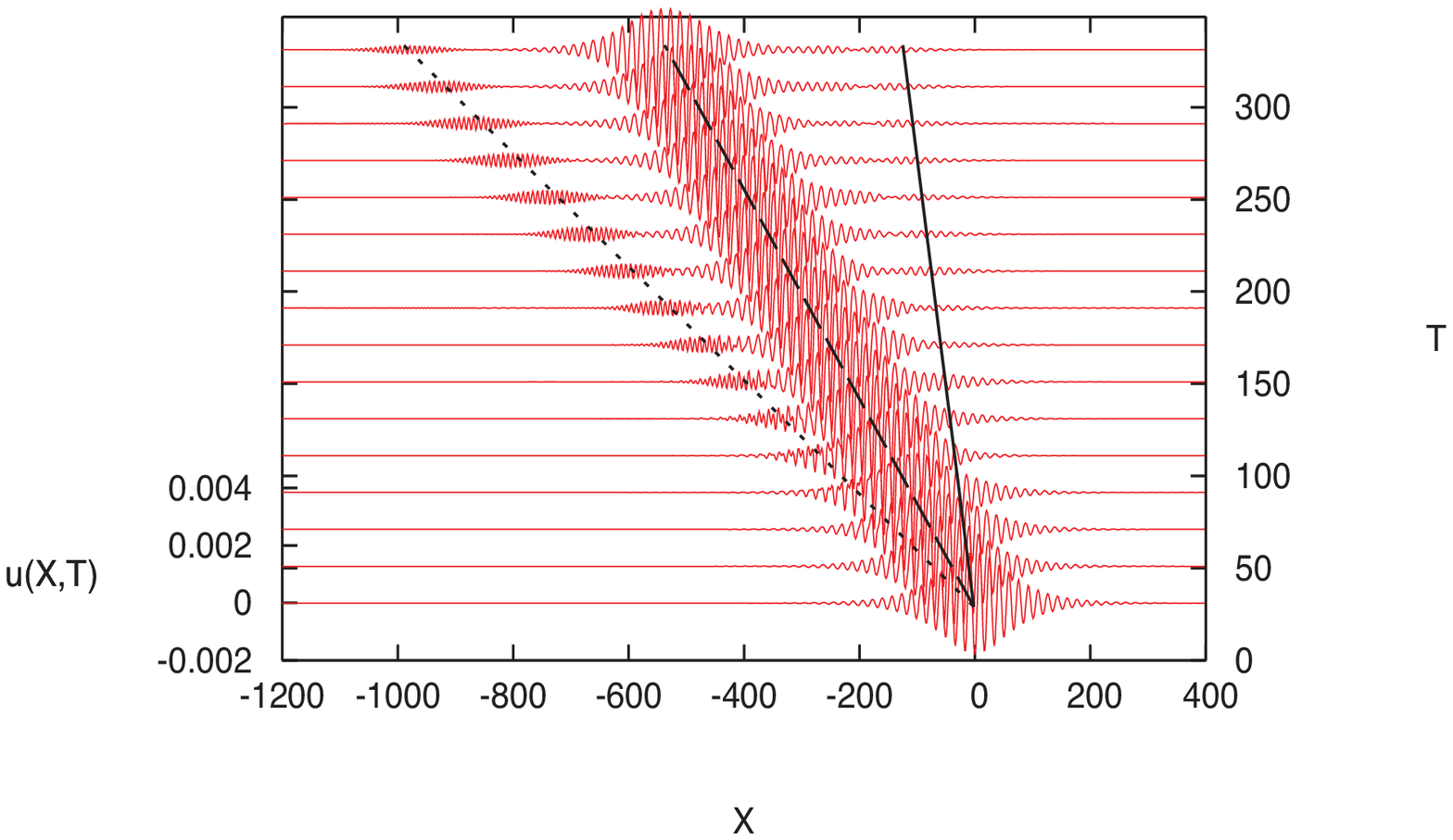}
\includegraphics[height=5cm,width=8cm]{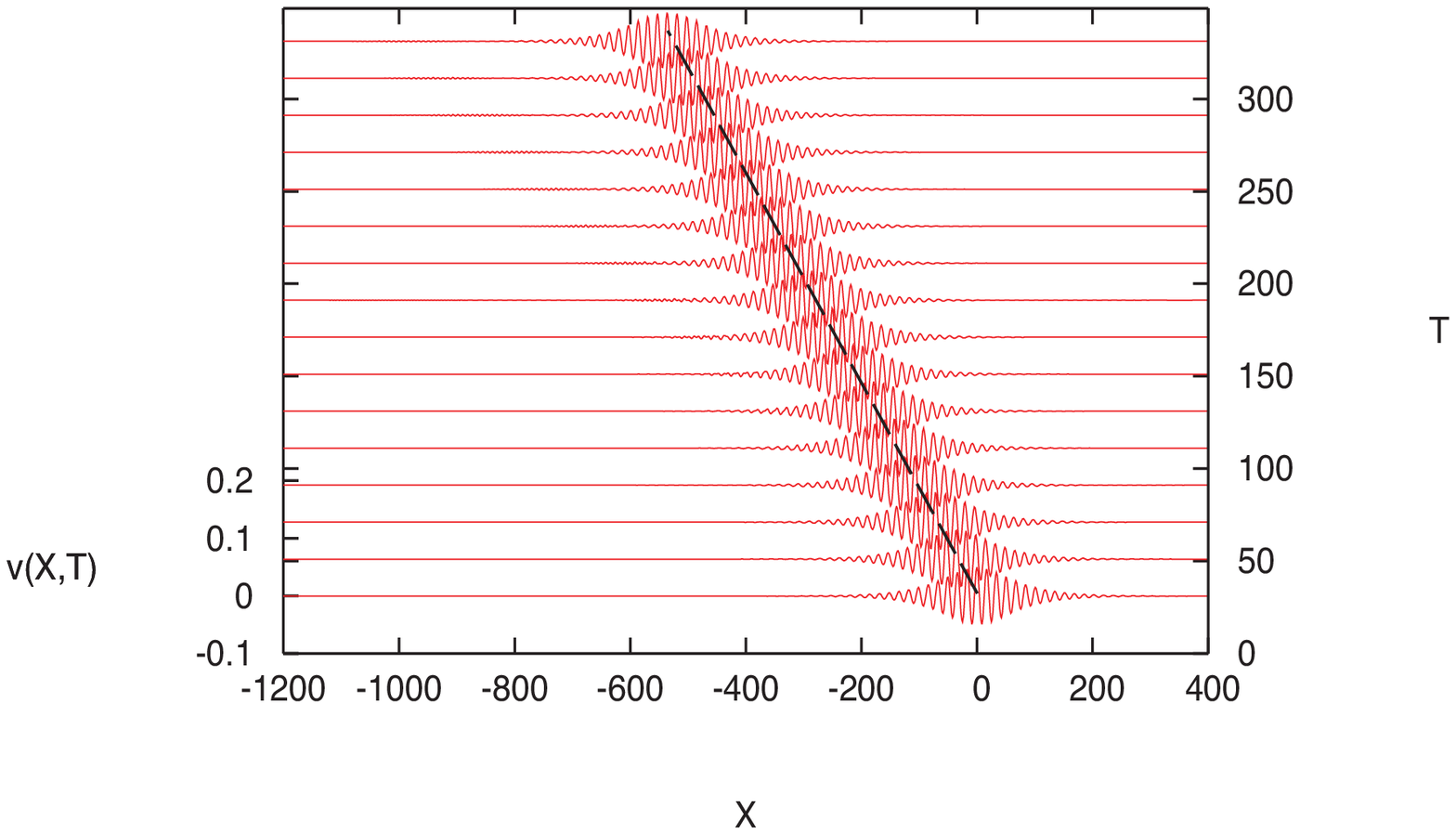}
 \caption{Numerical simulations for Case B using a nonlinear wave packet initial condition corresponding to the value $k=k_{s2}=0.372$ with $A_0=0.05, K_0=0.05 \,k, V_0=1$. The solid, dashed and dash-dot lines respectively refer to points  $A$, $B$ and ($D_1,E$) in the dispersion relation.}
\label{fig:2e}
\end{center}
\end{figure}

\begin{figure}[htbp]
\begin{center}
\includegraphics[height=3.5cm,width=8cm]{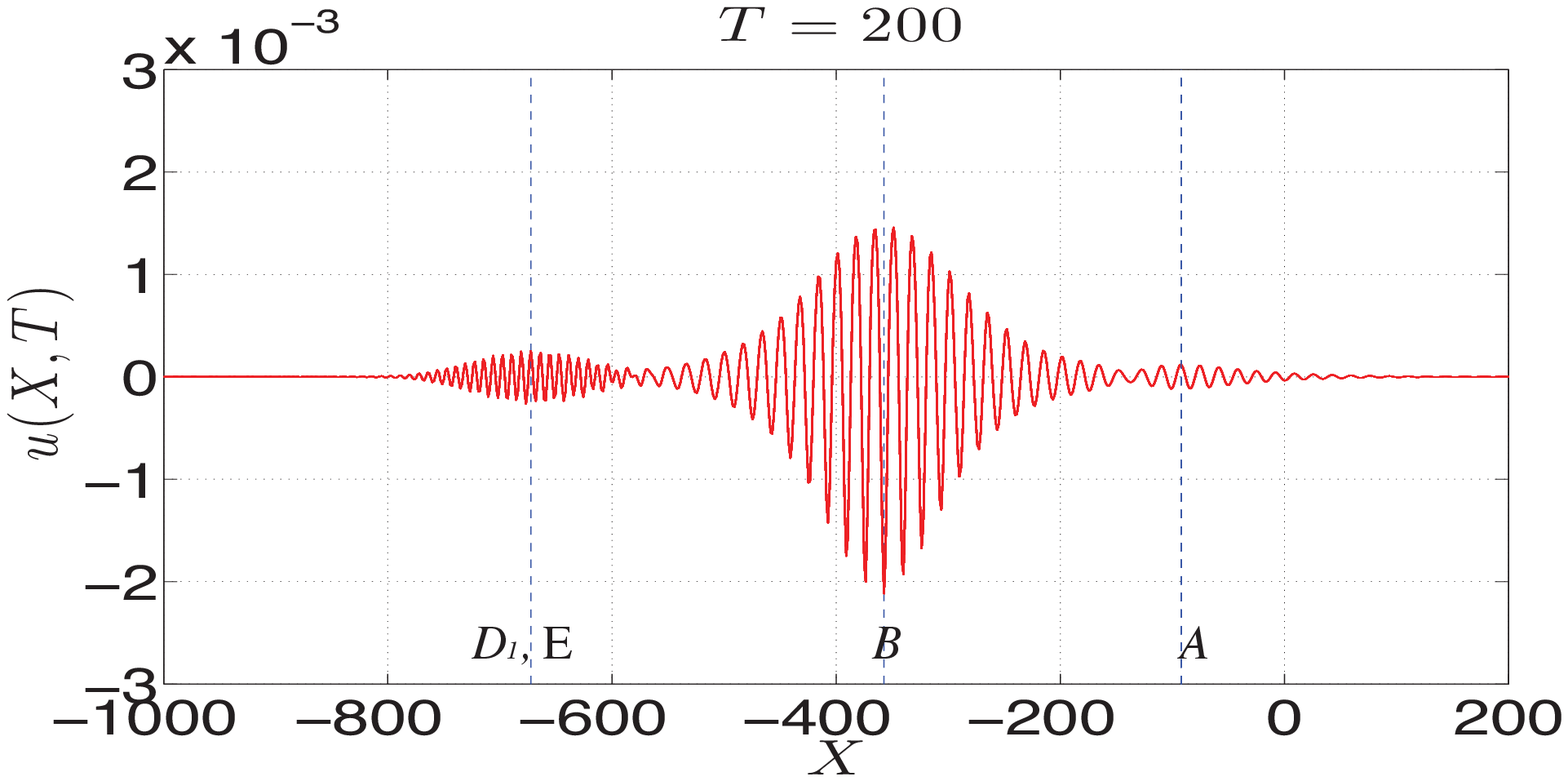}
\includegraphics[height=3.5cm,width=8cm]{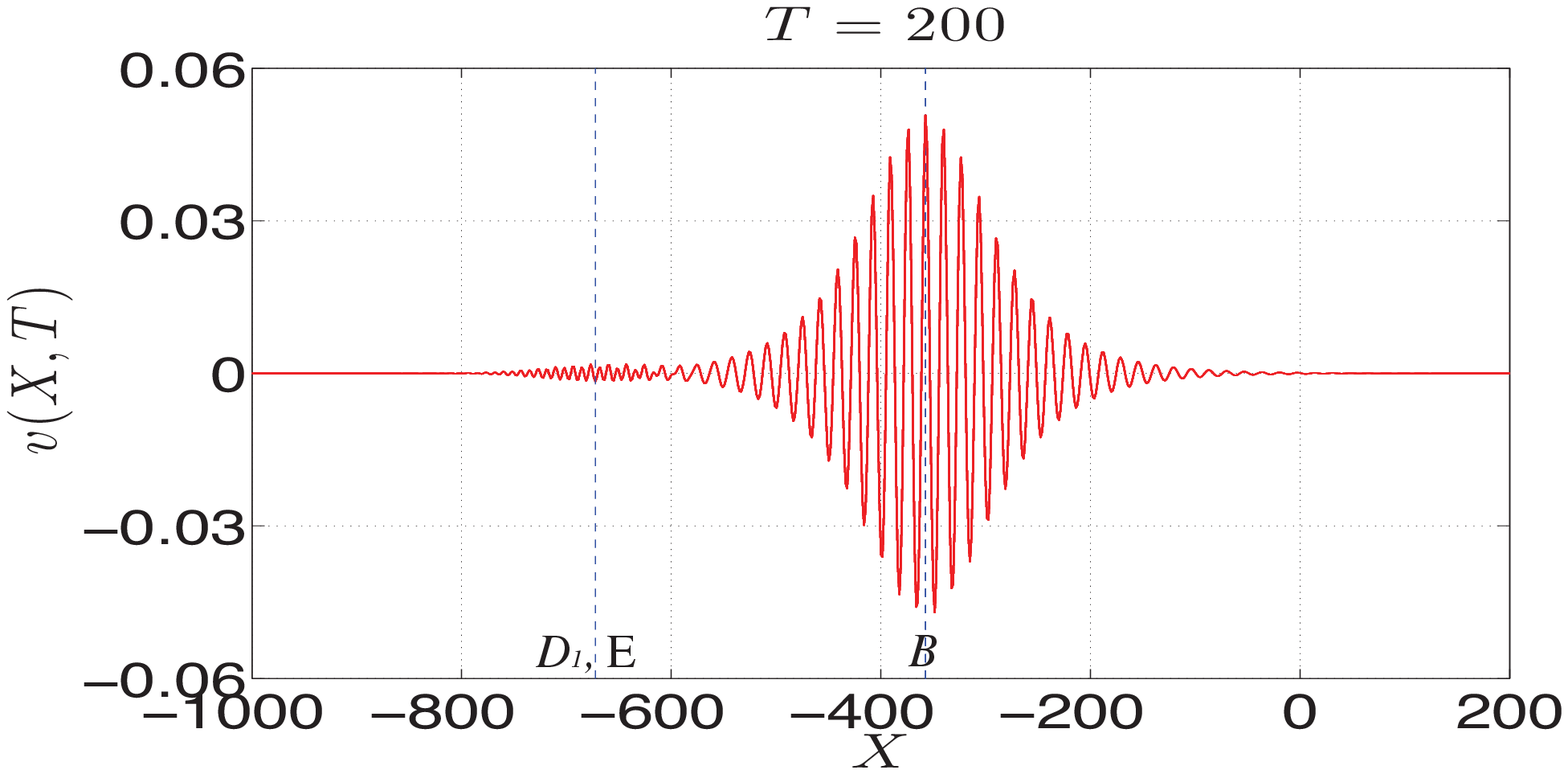}
 \caption{Same as Figure \ref{fig:2e}, but a cross-section at  $T=200$ for both modes.}
\label{fig:2f}
\end{center}
\end{figure}

\medskip
\noindent
{\bf Case C}:

\noindent
Case C is analogous to  Case B.  A typical numerical result is  shown in Figures \ref{fig:C1_2} and \ref{fig:C2_2} using the KdV solitary wave initial condition (\ref{weak}).  But here we chose $\gamma_{1} \ne \gamma_{2}$ in order that the ratio
 $a/b$ should coincide with the predicted ratio $1.3$ corresponding to the point $A$ in Figure \ref{fig:3}.
 A strongly nonlinear unsteady wave packet emerges,
  denoted by the vertical line $A$ in Figure \ref{fig:C2_2}, with speed  $-0.156$ and ratio $0.5496$,
   in agreement for the speed with the theoretical predictions from the point $A$ in the dispersion plots of 
 Figure \ref{fig:3} and Table \ref{table3}.  This wave packet has a phase speed which is very close to the 
 group velocity over the range of wave numbers from the point $D$ to $E$,  leading to
 strongly nonlinear effects and difficulty in numerically determining a ratio. 
 In Figures \ref{fig:C1_2} and \ref{fig:C2_2}  there is also evidence of significant radiation 
 both to the right and to the left
 of  the main wave packet. The waves to the right with positive speed can be  associated with the 
points $F_2$ and/or $N$ as these have a positive group velocity for mode $2$ and a ratio  of  nearly $-10$, which 
 means that the amplitude in the $v$-plot is  too small to be seen.   
 Although the points  $F$ and $N$ are very close,  they have  a different interpretation. 
 The point $F_2$ is a resonance between $c_{g1}$ and $c_{p2}$, 
 while  the point $N$ is a resonance between the speed at the
 minimum point of $c_{g1}$ with $c_{p2}$. 
 Moreover, this wave to the right has the appearance of
 a  linear dispersive wave,  and hence there is no very clear identifiable speed or wavenumber.  
 The waves to the left show both small-scale and large scale features in both  $u$ and $v$, 
 with the small-scale features more prominent in $u$ and the large-scale features more prominent in $v$.
 The large-scale feature may be associated with either $B$ or $K$ and the small-scale with either $J$ or $M$. 
 That is, these are mode $1$ waves associated with turning points in the group velocity, 
 and a resonance with the phase velocity. 
 Also note that for both $B$ and $K$  the ratio is such that $v$ dominates, while for $J$ and $M$ it is $u$ that dominates, features consistent 
 with the numerical simulation.   Thus, overall all the features in the numerical simulation can be associated with  the turning
 points in the group velocity curve $c_{g1}$ for mode $1$.

 \begin{figure}[htbp]
\begin{center}
\includegraphics[height=5cm,width=8cm]{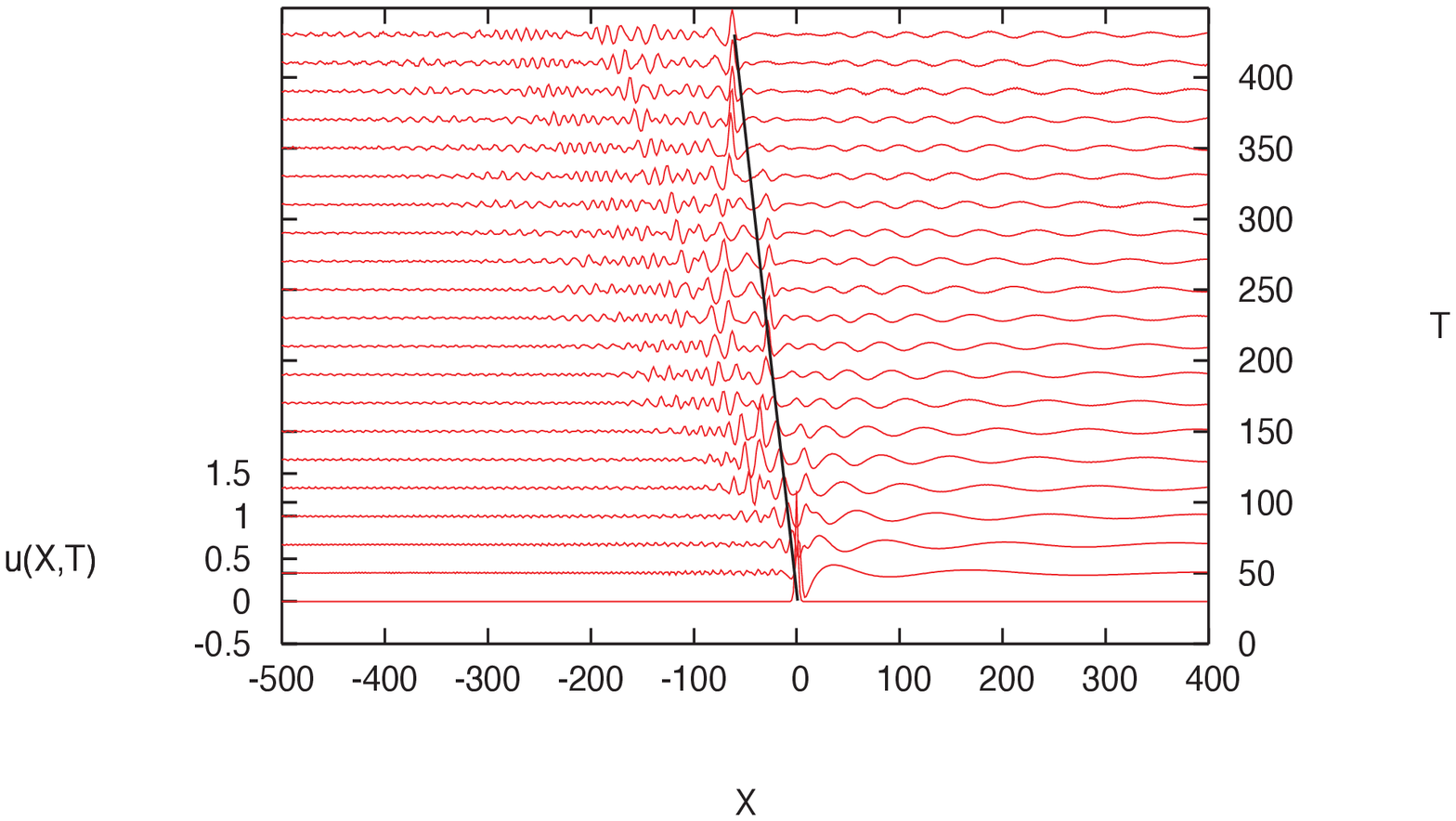} 
\includegraphics[height=5cm,width=8cm]{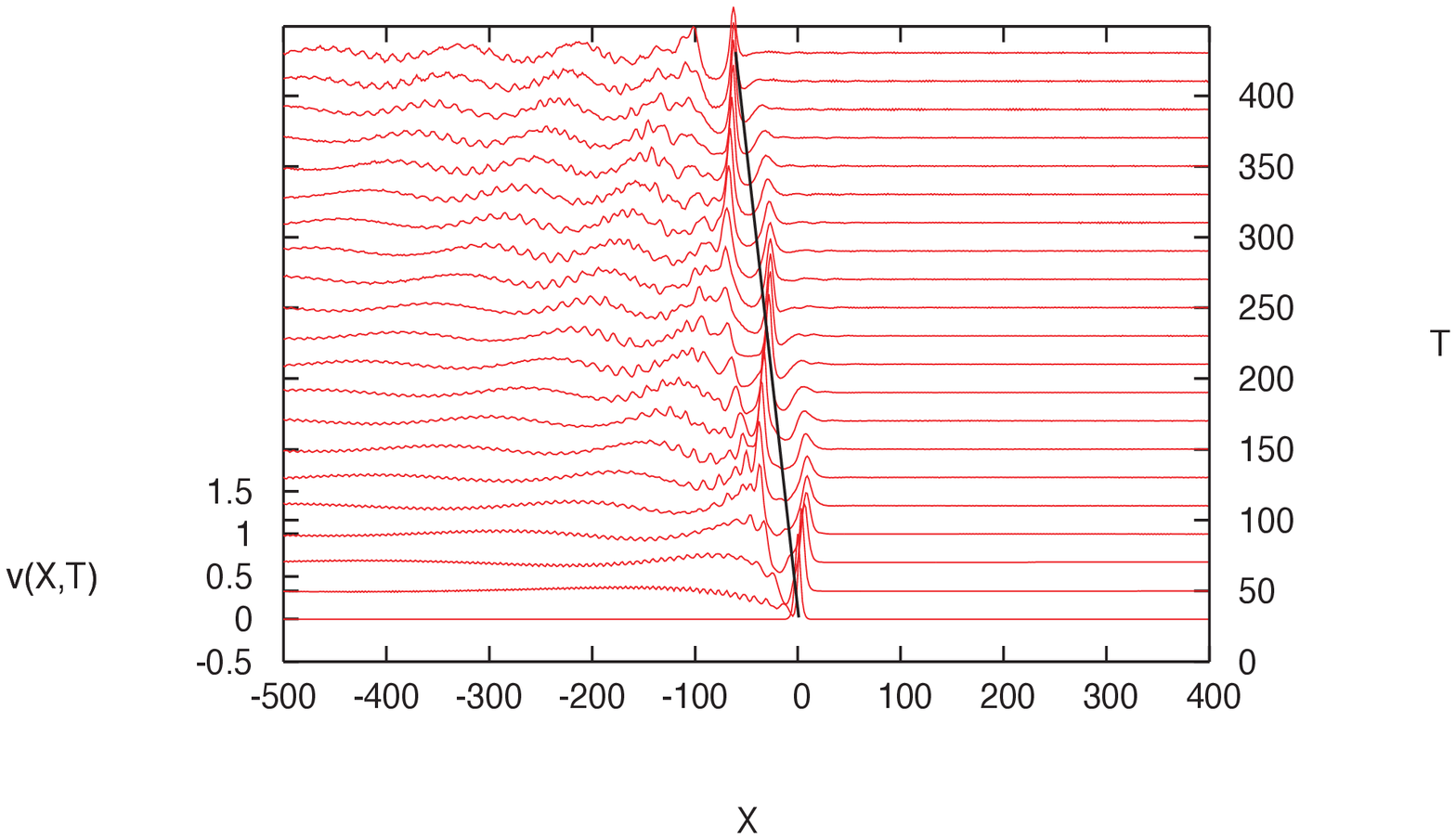}
\caption{Numerical simulations for Case C using a KdV initial condition of weak coupling (\ref{weak}) with $a=1.3$ and $b=1$.}
\label{fig:C1_2}
\end{center}
\end{figure}

\begin{figure}[htbp]
\begin{center}
\includegraphics[height=3.5cm,width=8cm]{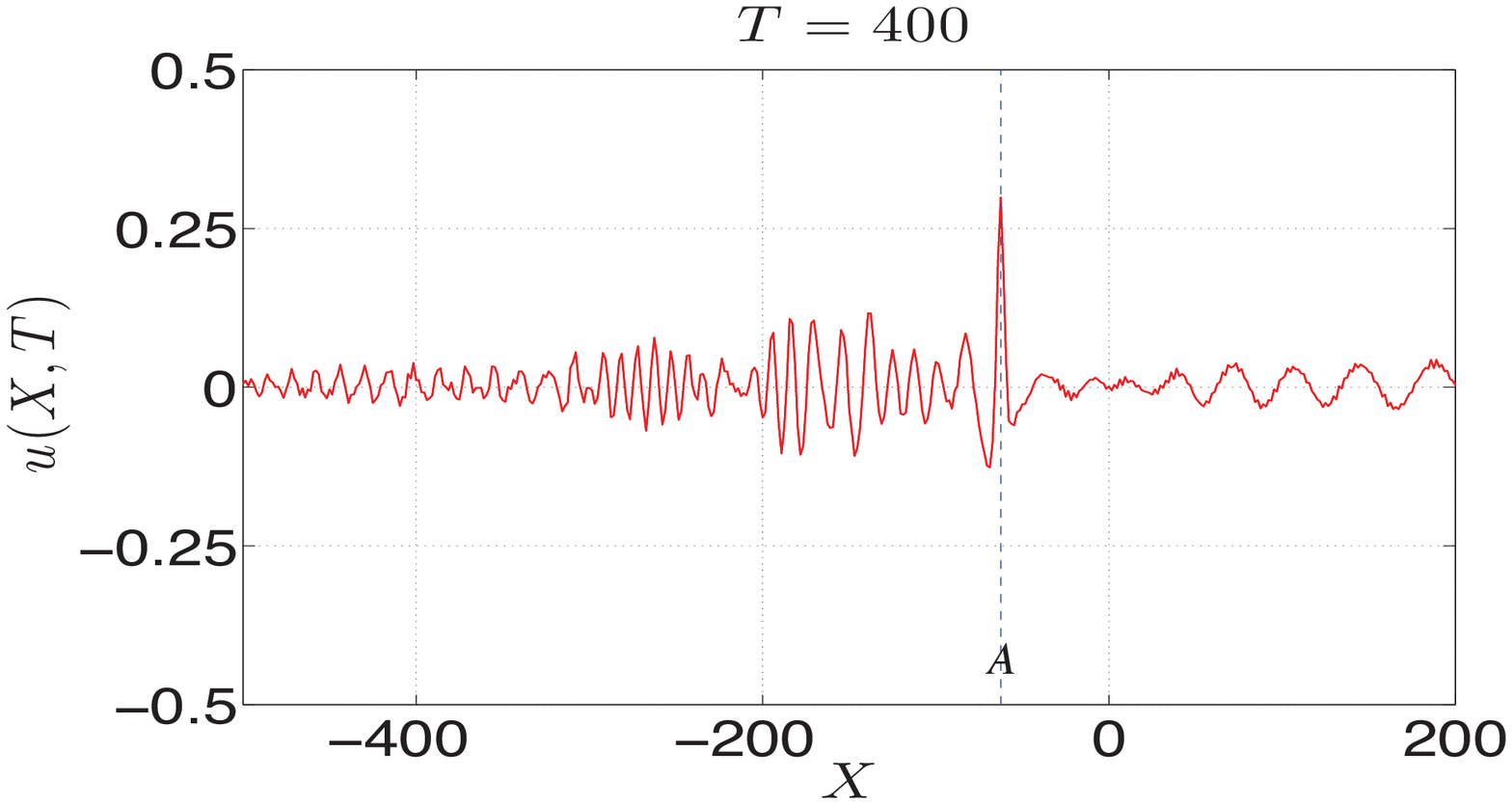}
\includegraphics[height=3.5cm,width=8cm]{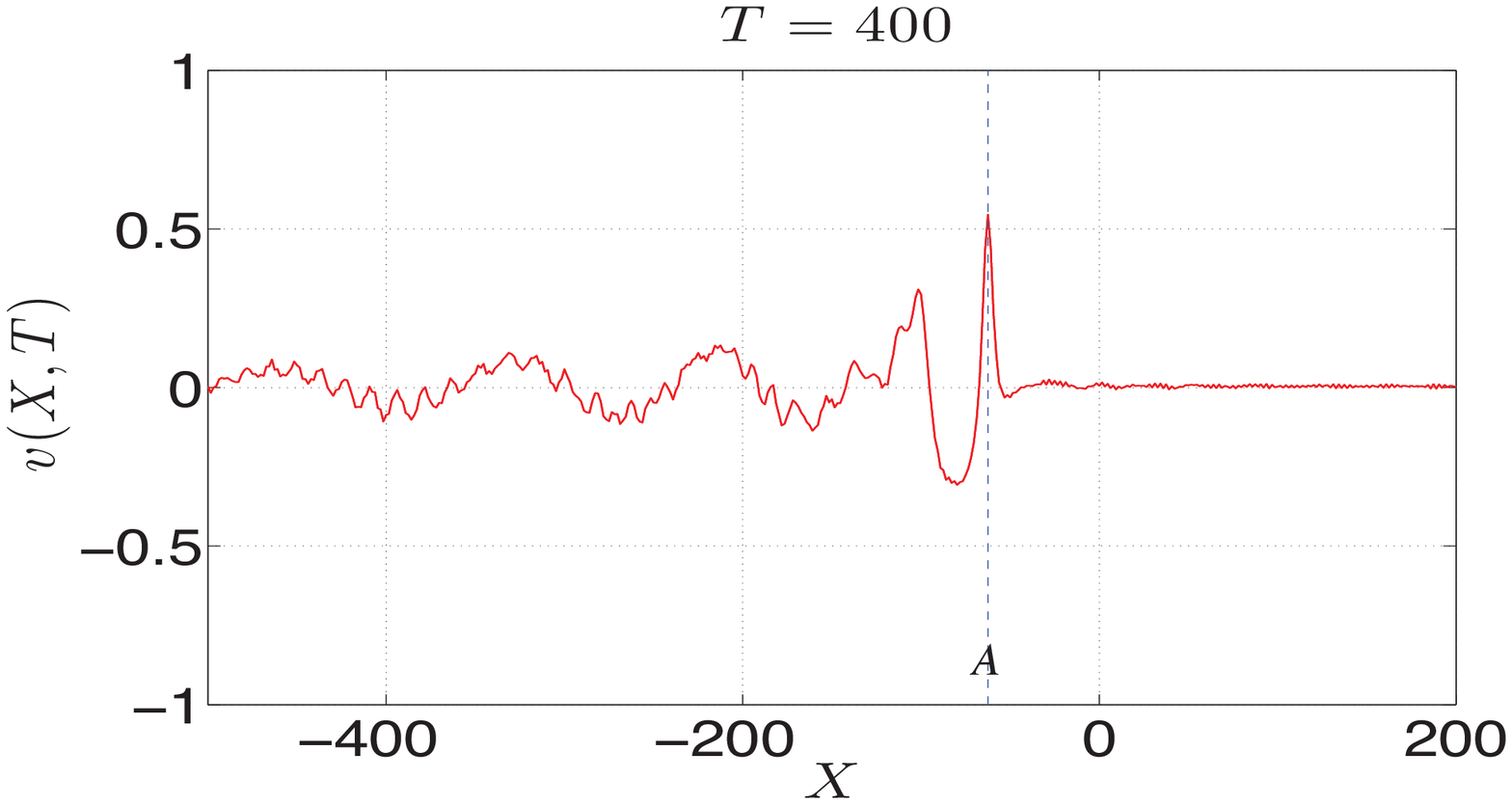}
 \caption{Same as Figure \ref{fig:C1_2}, but a cross-section at  $T=400$ for both modes.}
\label{fig:C2_2}
\end{center}
\end{figure} 
 
 As noted above, the group velocity curve $c_{g1}$ for mode $1$ has three turning points,
 while there are no such turning points for $c_{g2}$.
To examine each of these, we first  examined the
turning point $A$ in Figure \ref{fig:3} and Table \ref{table3},  and  used the wave packet initial condition 
 (\ref{wpic}) with wavenumber   $k=k_{m1} =0.306$ and ratio $R = 1.309$. 
 The numerical results are shown in Figures \ref{fig:C3a}, \ref{fig:C4a} and the emergence
 of a  nonlinear wave packet is clearly seen. At the vertical line  $A$, the speed is $-0.146$ with ratio 
 $1.387$, in agreement with  the theoretical prediction.
 There is a secondary wave packet now discernible on the vertical line $I$,  moving with speed $-0.625$ 
 and ratio $1.910$, which from the dispersion relation in Figure \ref{fig:3} is identified with 
 the point  $I$, which is  a  resonance between 
the maximum value of the phase speed of mode $2$ (point $C$) with mode $1$.
However, we note that the resonance points $J, M$ are close by with similar values, and so may also be relevant.

\begin{figure}[htbp]
\begin{center}
\includegraphics[height=5cm,width=8cm]{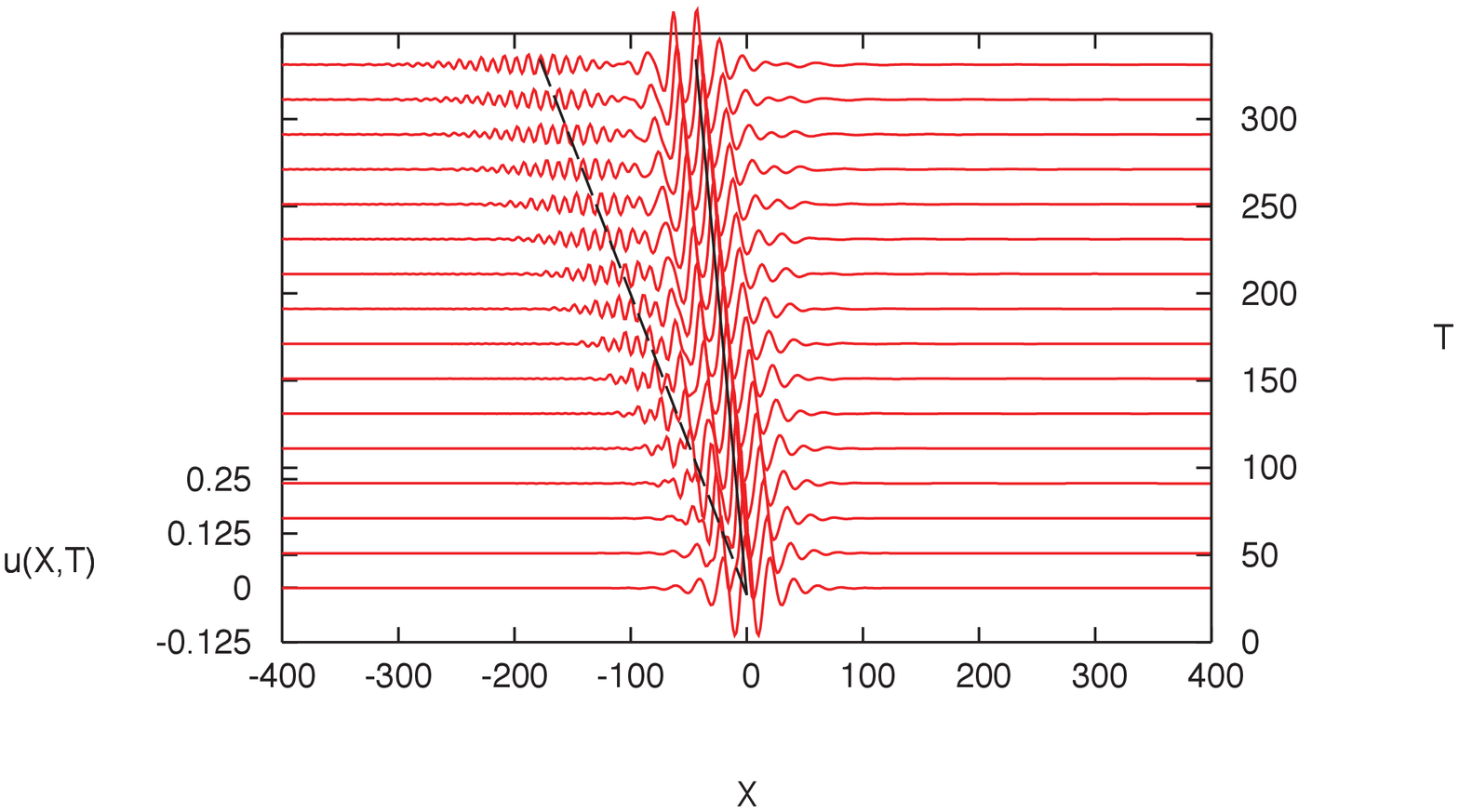}
\includegraphics[height=5cm,width=8cm]{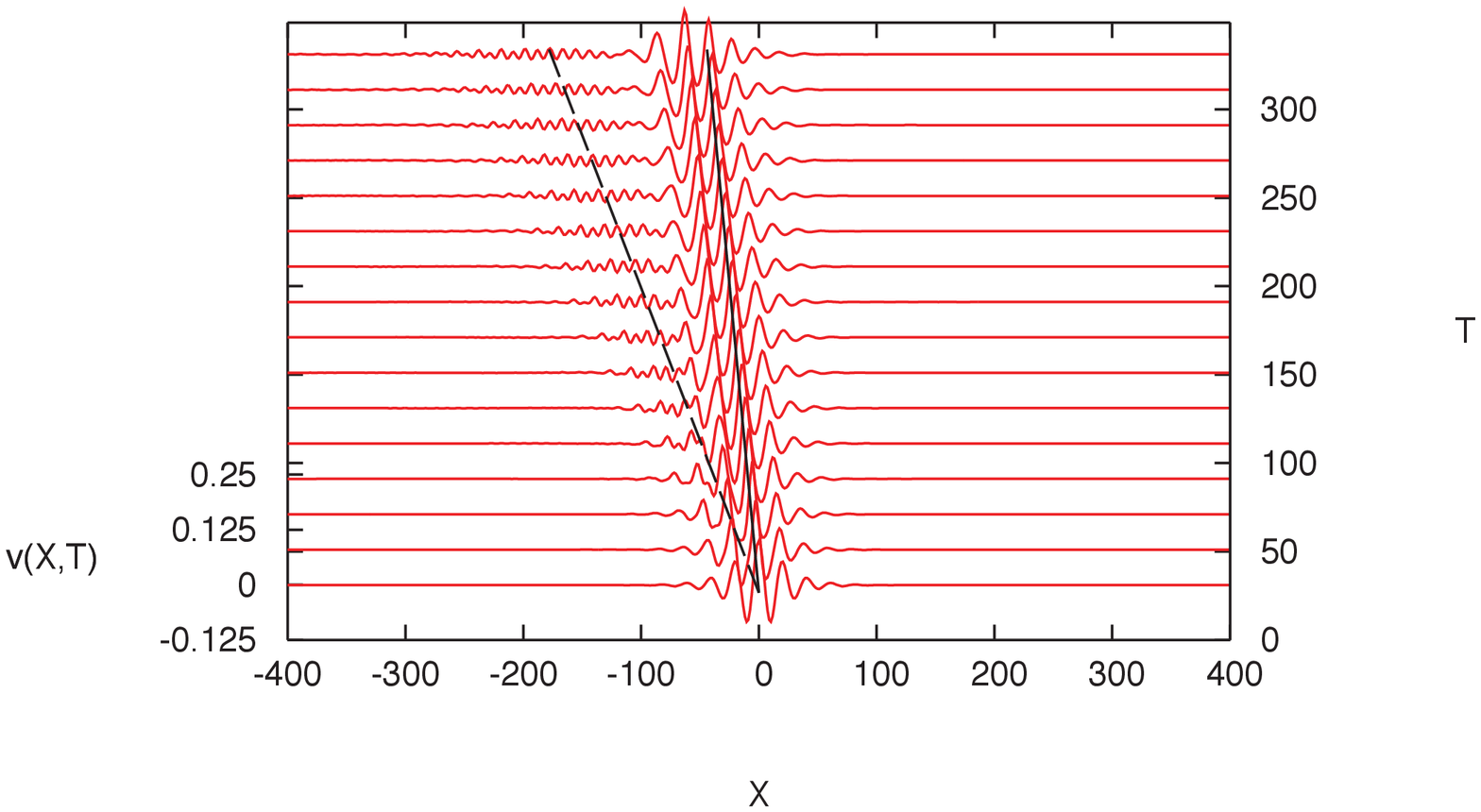}
 \caption{Numerical simulations for Case C using the wave packet initial condition (\ref{wpic}) with $k=k_{m1} = 0.306$ 
 corresponding to  point $A$ with $A_0 =0.1, K_0=0.2 \,k, V_0=1$. The {solid and dashed lines} respectively refer to points $A$ and ($I, J, M$) indicated in Figure \ref{fig:C4a}. }
\label{fig:C3a}
\end{center}
\end{figure}

\begin{figure}[htbp]
\begin{center}
\includegraphics[height=3.5cm,width=8cm]{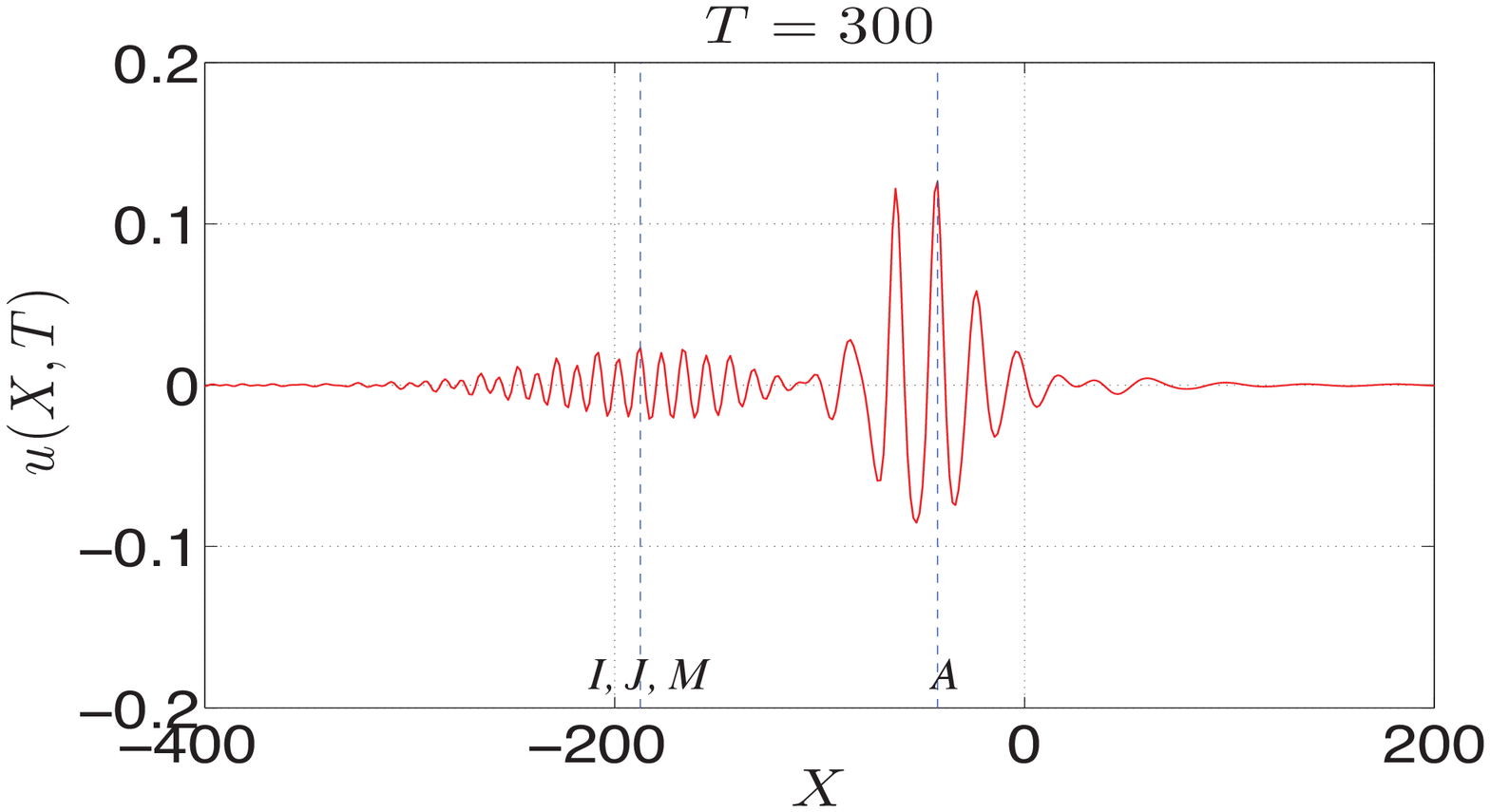}
\includegraphics[height=3.5cm,width=8cm]{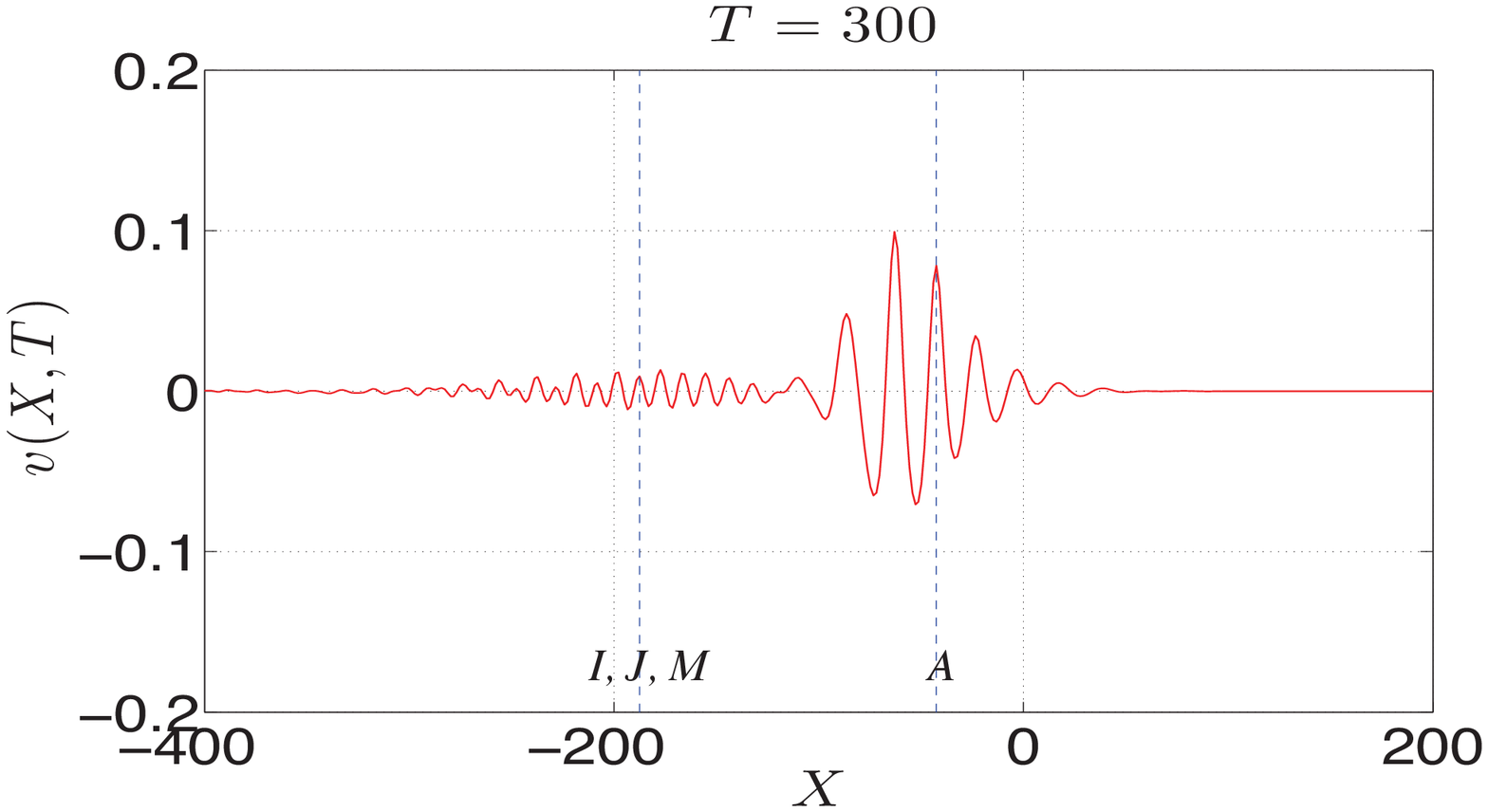}
 \caption{Same as Figure \ref{fig:C3a}, but a cross-section at  $T=300$ for both modes.}
\label{fig:C4a}
\end{center}
\end{figure}
 
  Next we used the wave packet initial condition associated with the 
turning point $B$ in Figure \ref{fig:3}, with $k=0.152$ and $A_0 =0.25, K_0 = 0.2k, V_0 =1$. 
The numerical result is shown in Figures \ref{fig:C5} and \ref{fig:C5_t200}.
A  nonlinear wave packet emerges with speed $-0.205$ and ratio $0.460$, 
 whereas the predicted values  are $-0.281$ and $0.04$ in Table \ref{table3}. 
 The speed is approximately  consistent with the theoretical prediction for point $B$ 
 but the ratio is not. However we note here that due to the variability in the emerging wave packets
 in the $u$-variable, the ratio  is quite hard to determine here. 
This  may be due to contamination with 
waves associated with the points $A$ or $D_{2}$. 

The corresponding numerical result 
for an initial condition associated with the turning point $K$ are shown in Figures \ref{fig:C6} and \ref{fig:C7}.
A strongly nonlinear wave packet emerges,  with speed $-0.303$ and  ratio $0.492$, 
can be seen in both the $u$ and $v$ plots,
and is in reasonable  agreement with the theoretical prediction. However, the resonance points 
$D_{1}, F_{1}$ have similar speeds and the strong nonlinearity suggests there may be some interaction here,
leading to difficulty in determining a numeral ratio.
There is also a small wave propagating  to the right, seen in the $u$-plot,  with the speed $0.234$ and the ratio $4.939$, indicated by the vertical line $N$, which can  be associated with one or more  of the resonance points $N, F_{2}, G$ in Figure \ref{fig:3}.

\begin{figure}[htbp]
\begin{center}
\includegraphics[height=5cm,width=8cm]{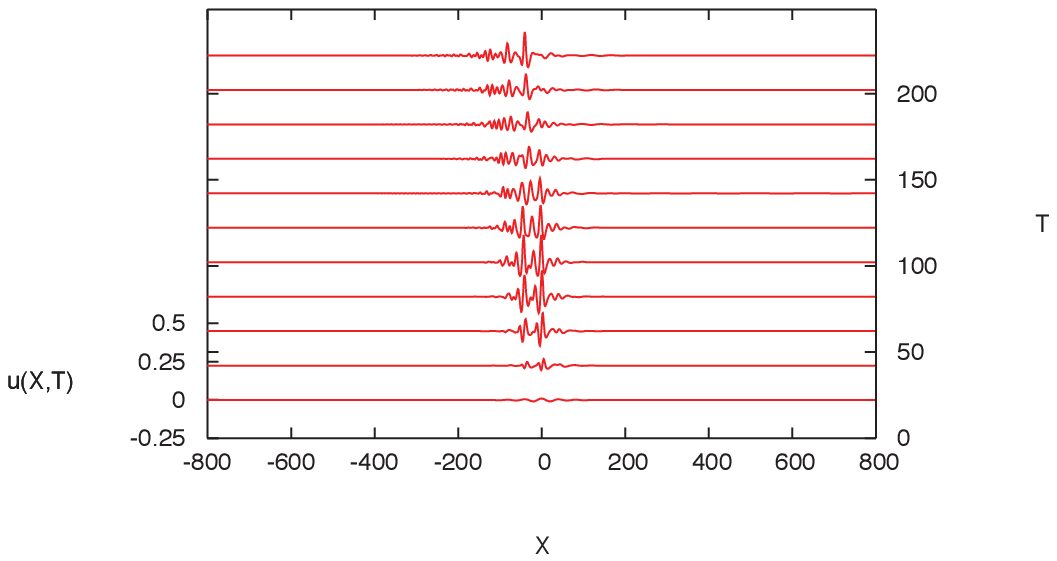}
\includegraphics[height=5cm,width=8cm]{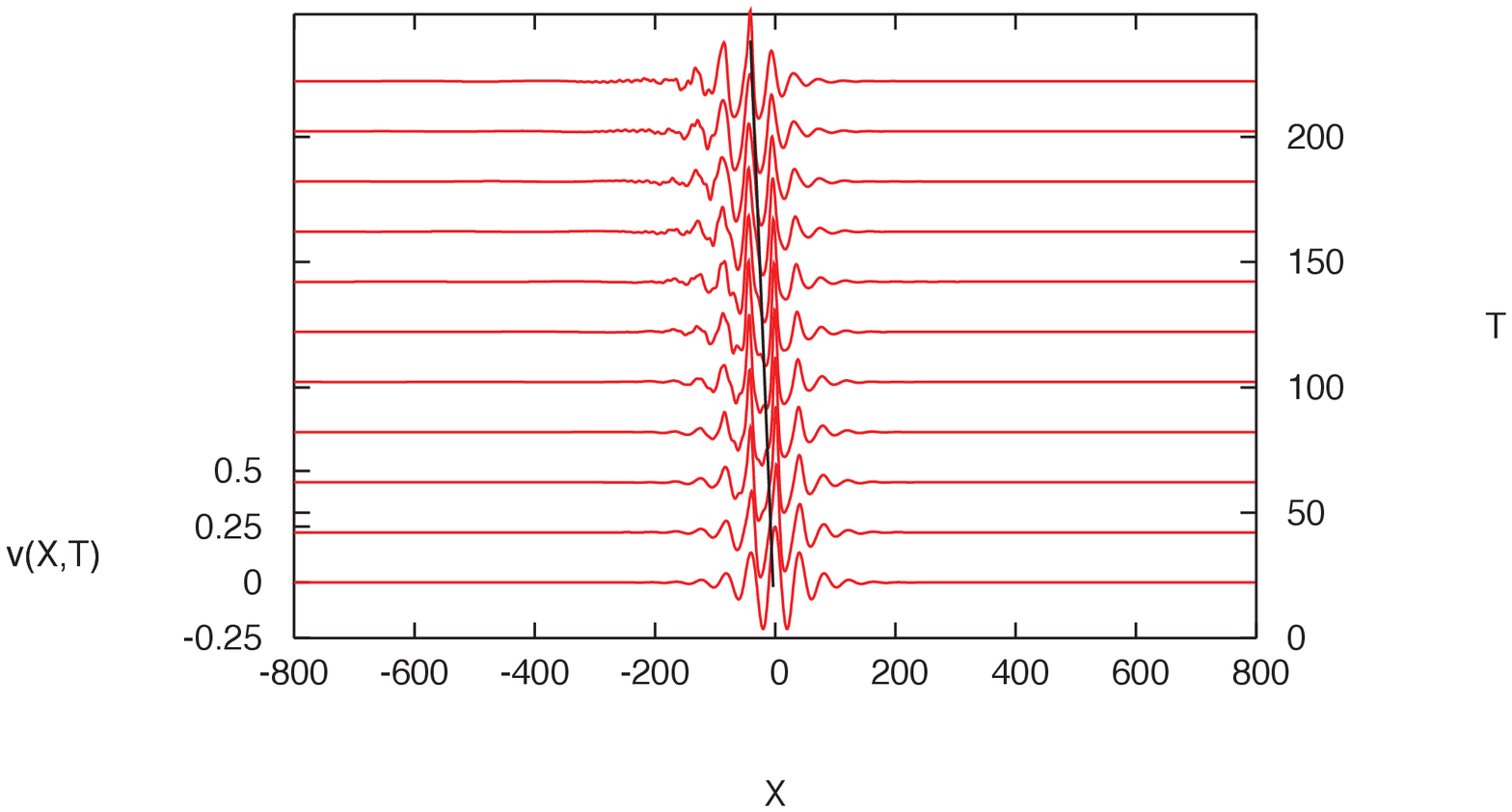}
 \caption{Numerical simulations for Case C using the wave packet initial condition (\ref{wpic}) with $k= 0.152$ 
 corresponding  to point $B$ with  $A_0 =0.25, K_0=0.2\, k, V_0=1$.}
\label{fig:C5}
\end{center}
\end{figure}

\begin{figure}[htbp]
\begin{center}
\includegraphics[height=3.5cm,width=8cm]{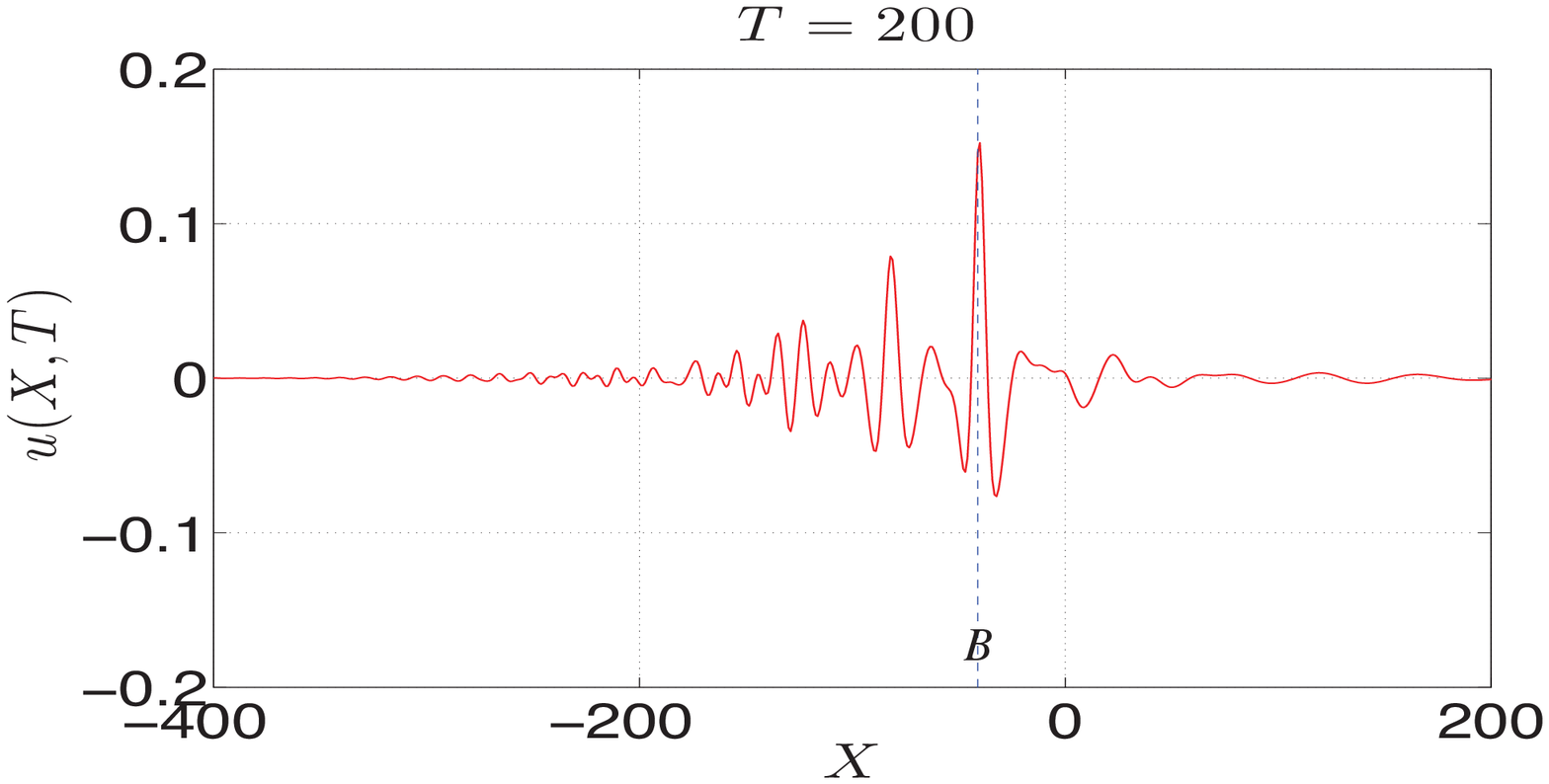}
\includegraphics[height=3.5cm,width=8cm]{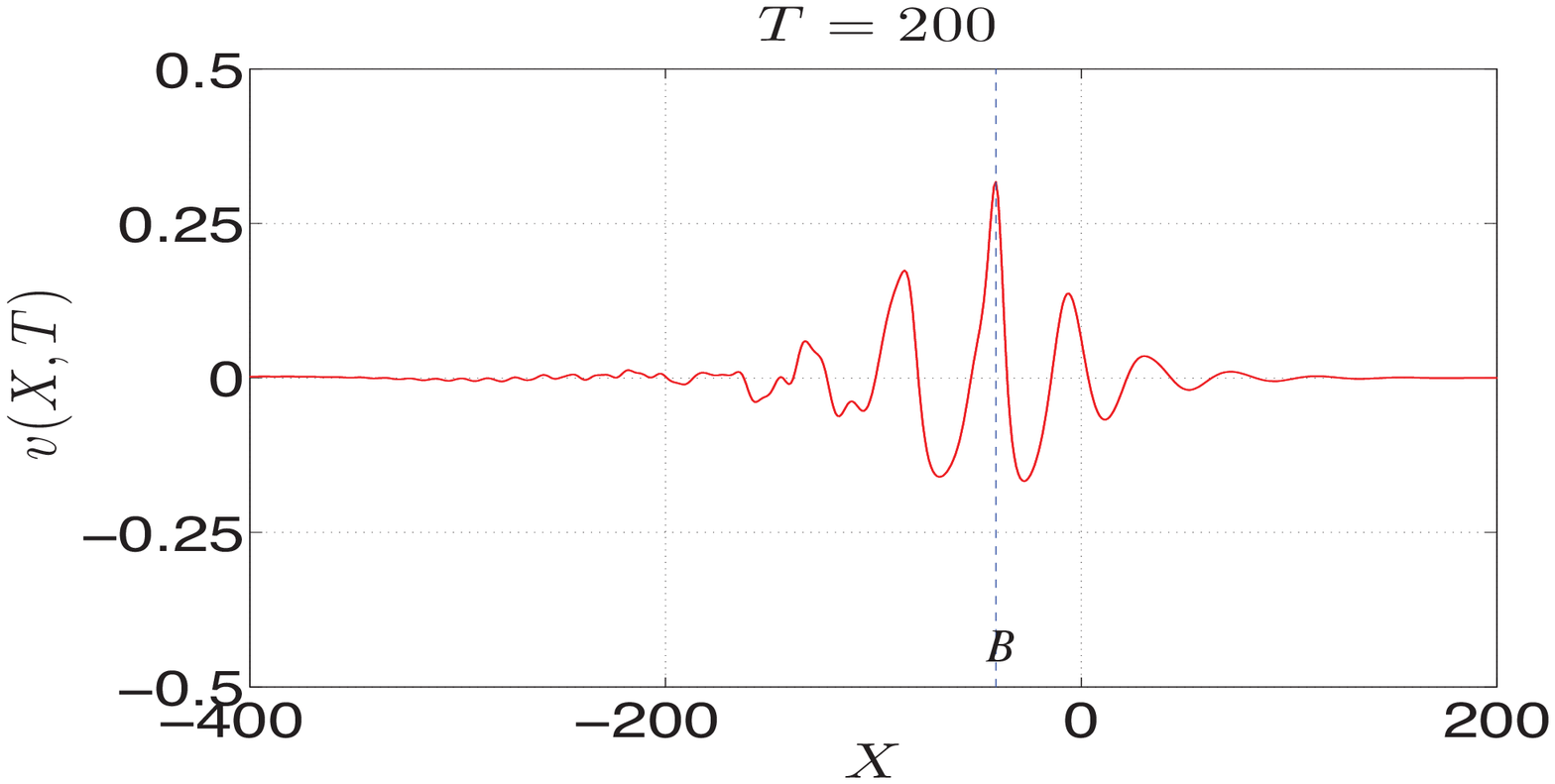}
 \caption{Same as Figure \ref{fig:C5}, but a cross-section at  $T=200$ for both modes.}
\label{fig:C5_t200}
\end{center}
\end{figure}

\begin{figure}[htbp]
\begin{center}
\includegraphics[height=5cm,width=8cm]{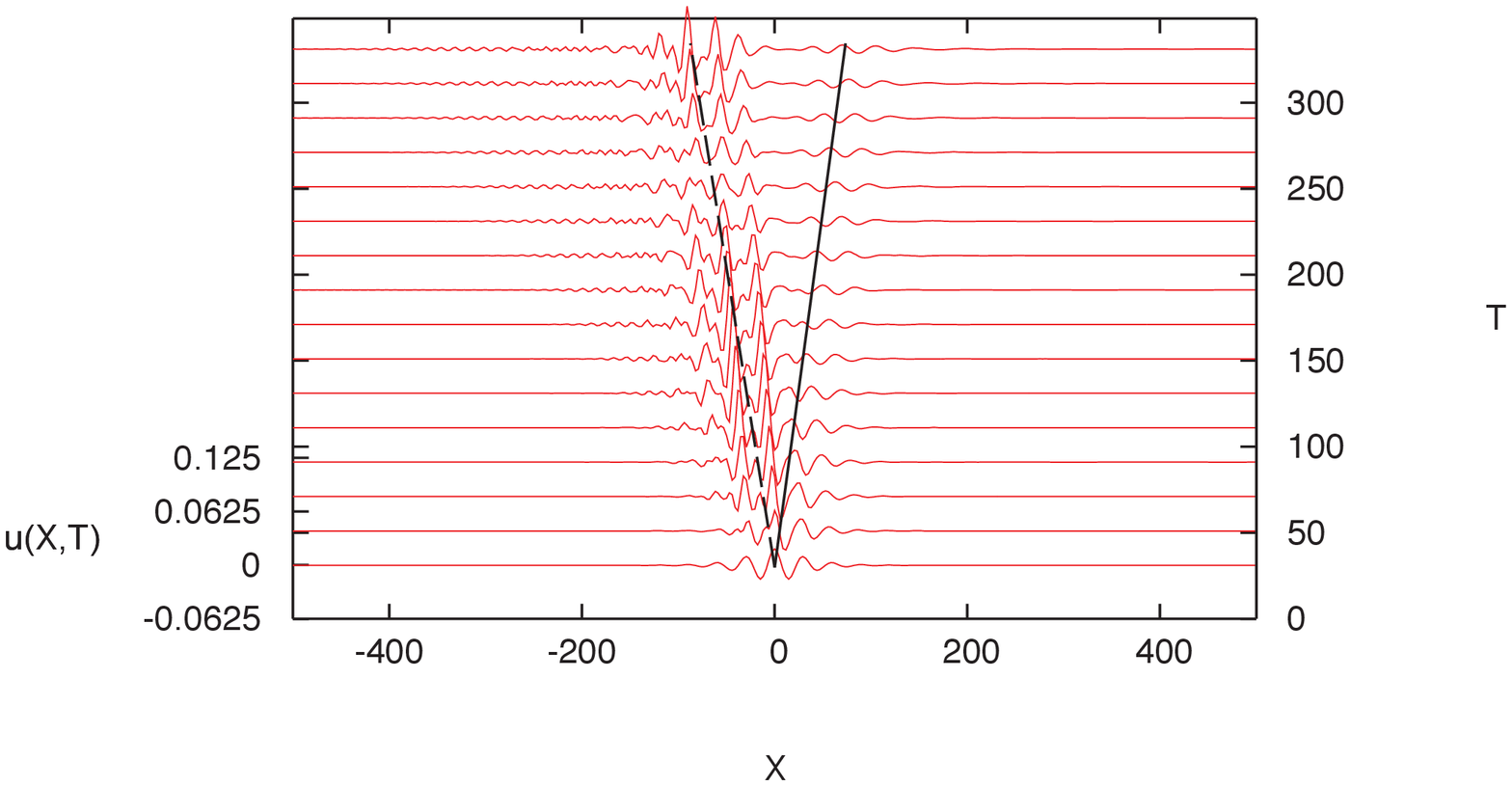}
\includegraphics[height=5cm,width=8cm]{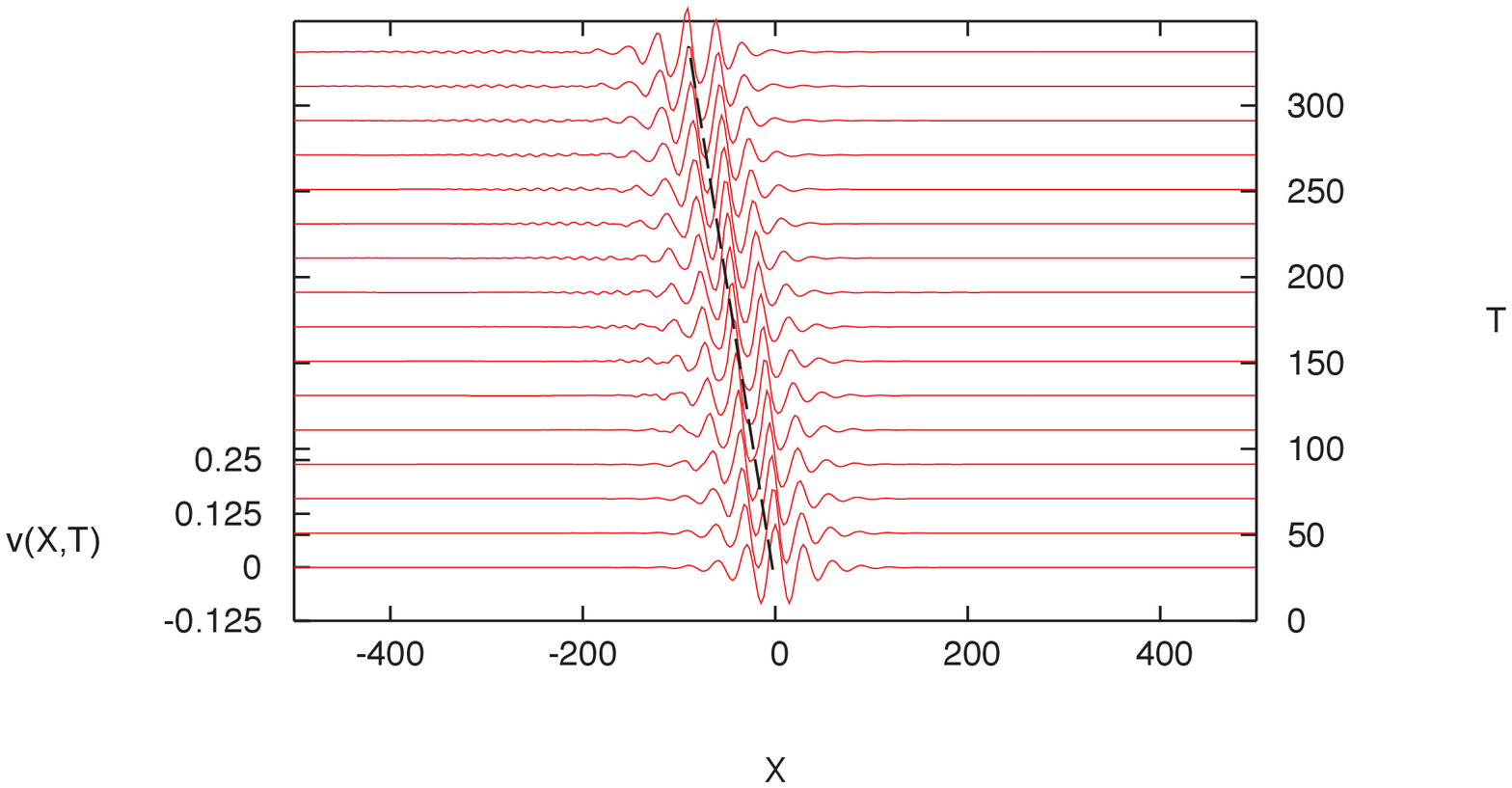}
 \caption{Numerical simulations for Case C using the wave packet initial condition (\ref{wpic}) with $k= 0.209$ corresponding to point $K$ with $A_0 =0.1, K_0 = 0.2 \,k, V_0 =1$. The {solid and dashed lines} respectively refer to points ($F_2, G, N$) and ($D_1, F_1, K$) in Figure \ref{fig:3}. }
\label{fig:C6}
\end{center}
\end{figure}

\begin{figure}[htbp]
\begin{center}
\includegraphics[height=3cm,width=8cm]{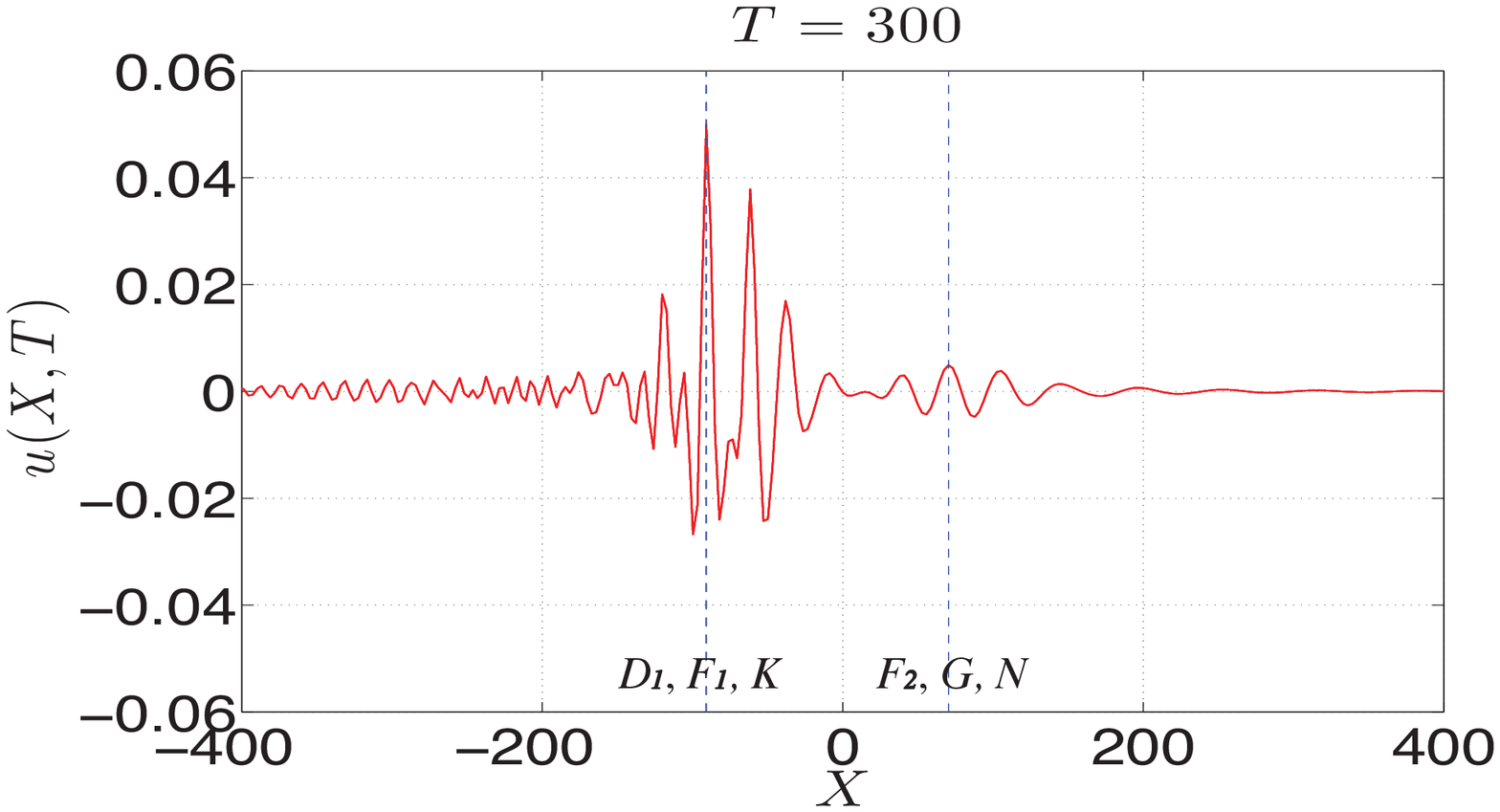}
\includegraphics[height=3cm,width=8cm]{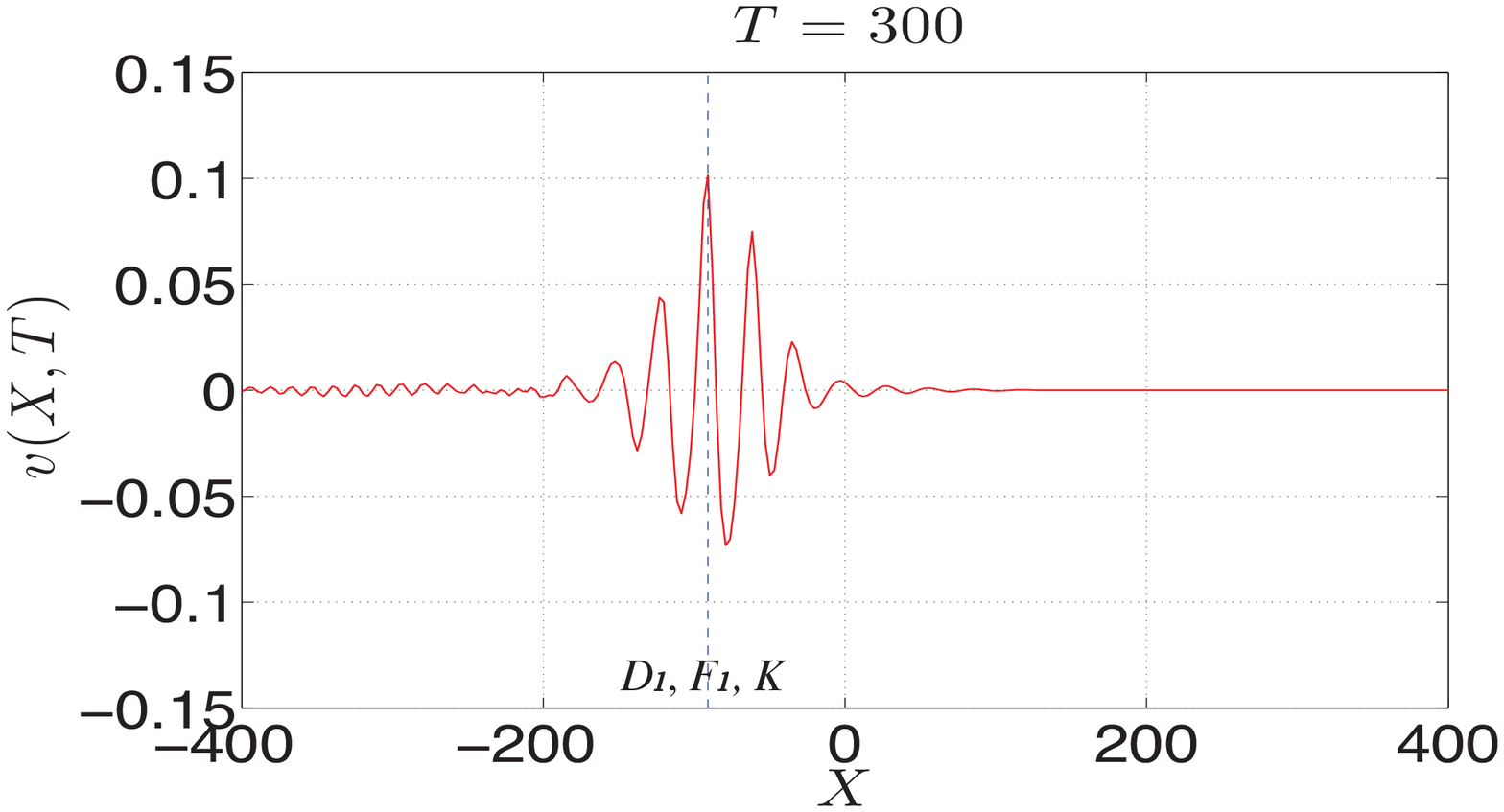}
 \caption{Same as Figure \ref{fig:C6}, but a cross-section at  $T=300$ for both modes. }
\label{fig:C7}
\end{center}
\end{figure}

Finally, we turn to the simulation associated with the turning point $C$ in Figure \ref{fig:3},
using the wave packet initial condition (\ref{wpic}) with $A_0 =0.025, K_0=0.05 k, V_0=1$. 
The numerical result is shown in Figures \ref{fig:C9} and \ref{fig:C9_t300}. 
In this case a steady wave packet  clearly emerges, indicated by the solid line, with 
speed  $-0.244$ and  ratio $1.874$, in good agreement with the predicted theoretical values.
Note that the resonance point $E$ has a similar speed, but quite different wavenumber, and indeed
we do not see that wave forms associated with this point.   
  
\begin{figure}[htbp]
\begin{center}
\includegraphics[height=5cm,width=8cm]{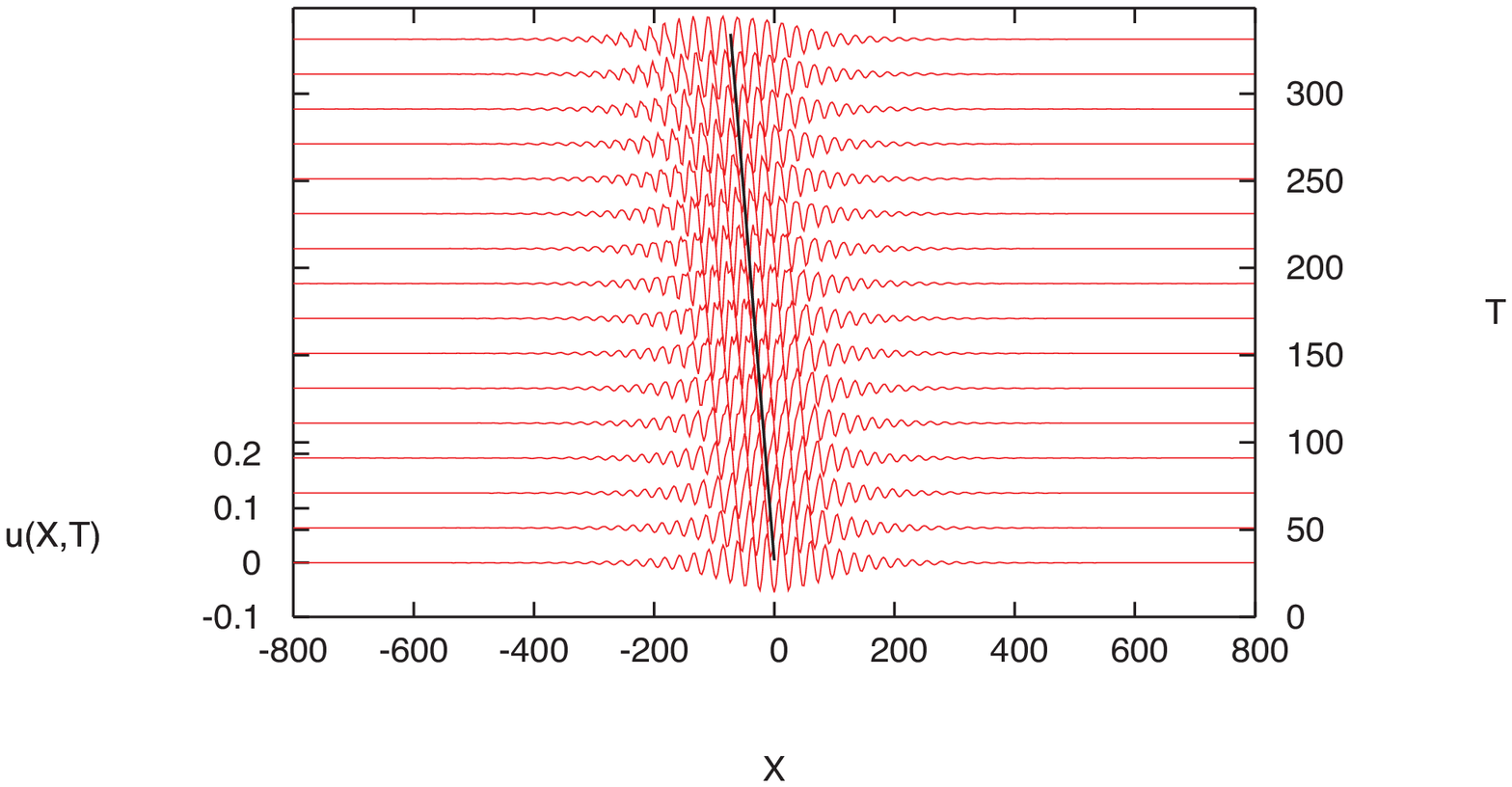}
\includegraphics[height=5cm,width=8cm]{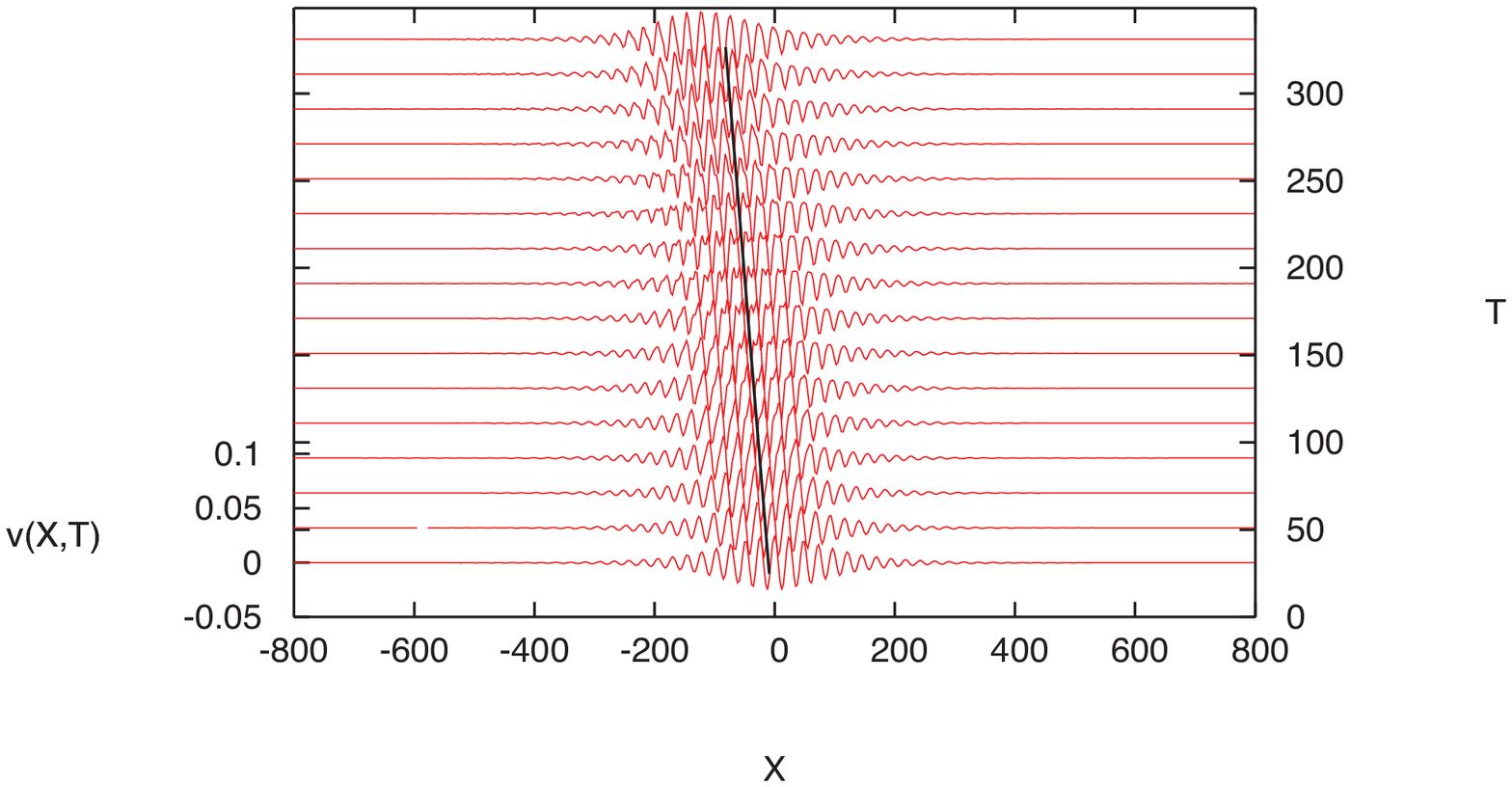}
 \caption{Numerical simulations for Case C using the wave packet initial condition (\ref{wpic}) with $k=k_{s2} = 0.259$ corresponding to point $C$ with $A_0 =0.025, K_0=0.05\, k, V_0=1$.}
\label{fig:C9}
\end{center}
\end{figure}

\begin{figure}[htbp]
\begin{center}
\includegraphics[height=3cm,width=8cm]{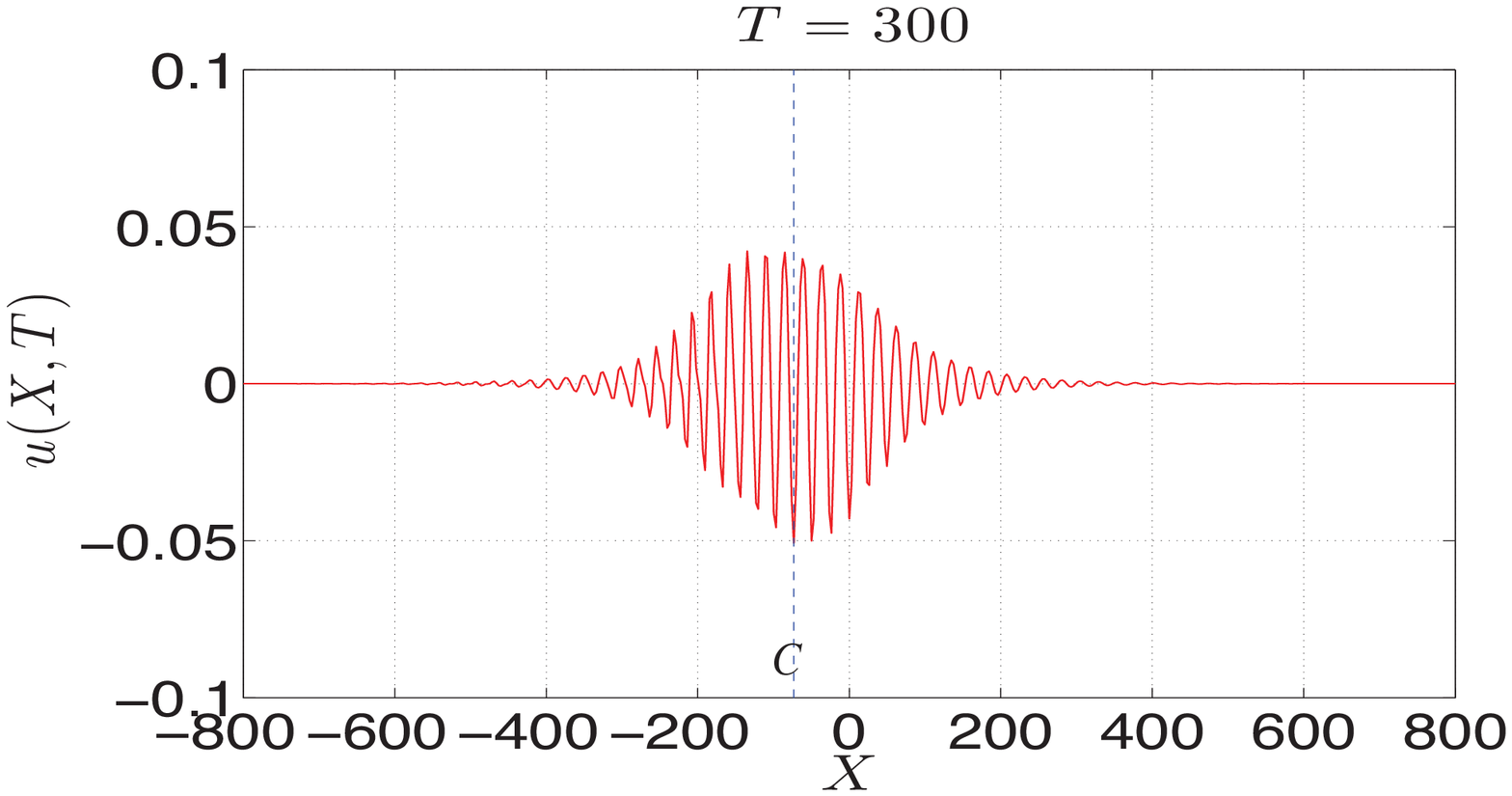}
\includegraphics[height=3cm,width=8cm]{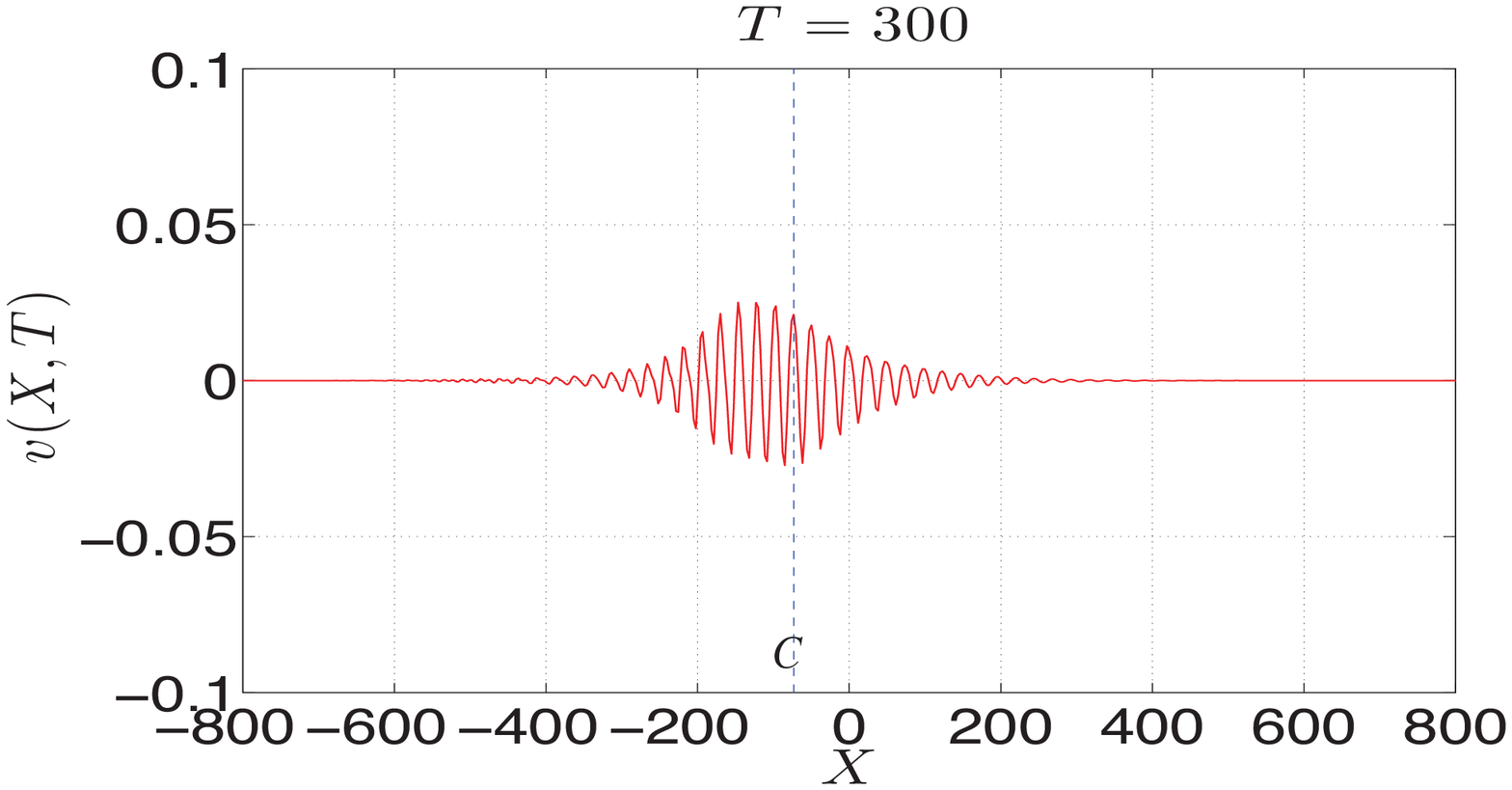}
 \caption{Same as Figure \ref{fig:C9}, but a cross-section at  $T=300$ for both modes. }
\label{fig:C9_t300}
\end{center}
\end{figure}

\newpage

\medskip
\noindent
{\bf Case D}:

\noindent
A typical numerical result is  shown in Figures \ref{fig:D1} and \ref{fig:D2}
 using the KdV solitary wave initial condition (\ref{weak}).
The numerical results show two steady wave packets emerging, as expected, 
 with  speeds $-0.146, -0.586$ and ratios $10.136, 1.753$ associated with the vertical lines $A$ and $B$ 
 respectively in Figure \ref{fig:D2},  in reasonable  agreement with the theoretical values.
 These wave packets are strongly nonlinear and there is considerable evidence of resonances and radiation. 
In particular, the vertical line $F$ in  Figure \ref{fig:D2}  is interpreted as an interaction between  
the  points $B$ and $F_{1}$, the latter being a resonance between the  
group velocity of mode $1$ and phase speed of mode $2$, see Figure \ref{fig:4} and Table \ref{Table4}. 
There is also a transient wave propagating to the right, probably due to 
fact that the  negative signs of both $\beta$ and $\mu$ allow both modes to have positive group velocities for low wavenumbers.

\begin{figure}[htbp]
\begin{center}
\includegraphics[height=5cm,width=8cm]{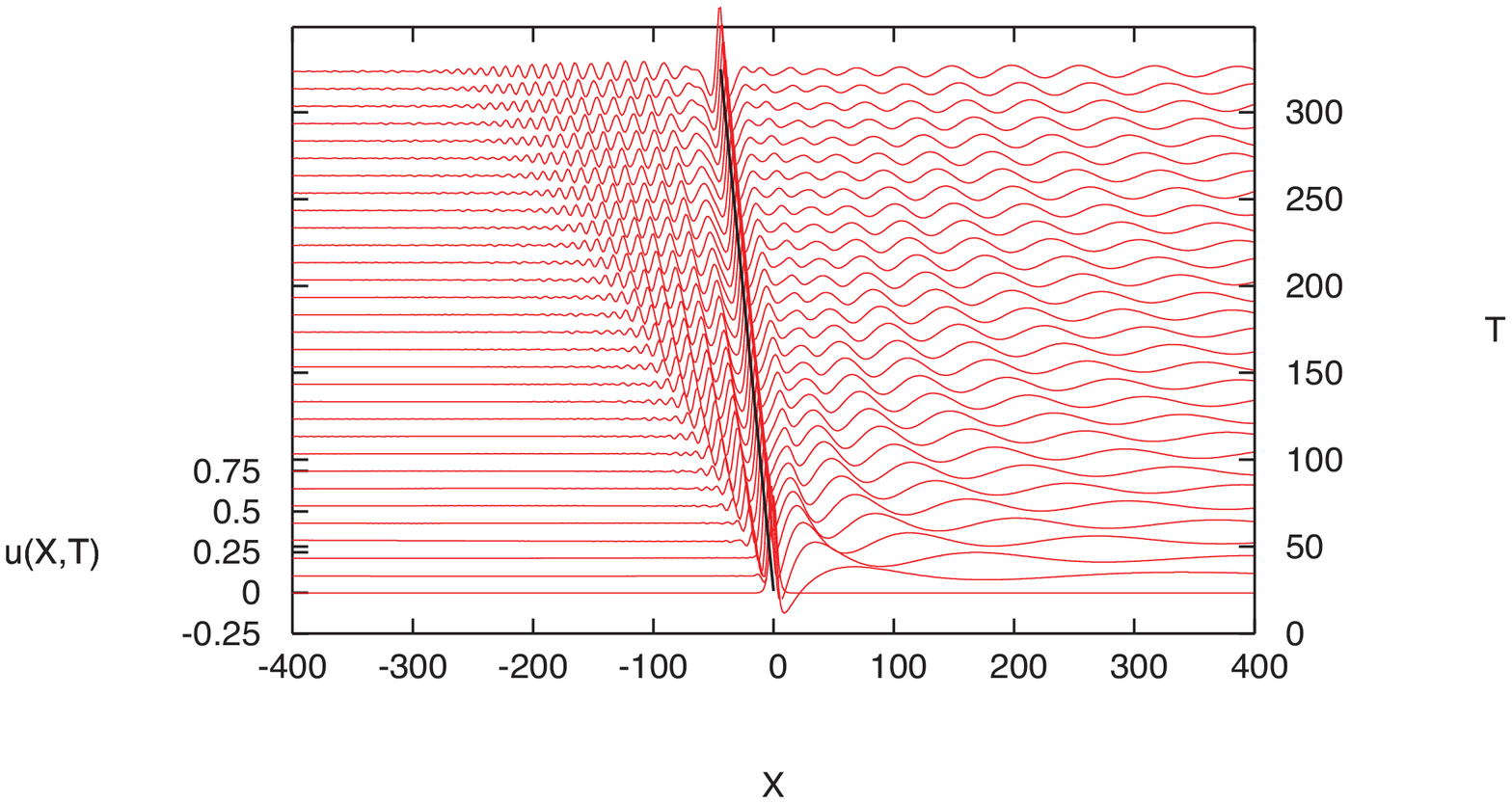} 
\includegraphics[height=5cm,width=8cm]{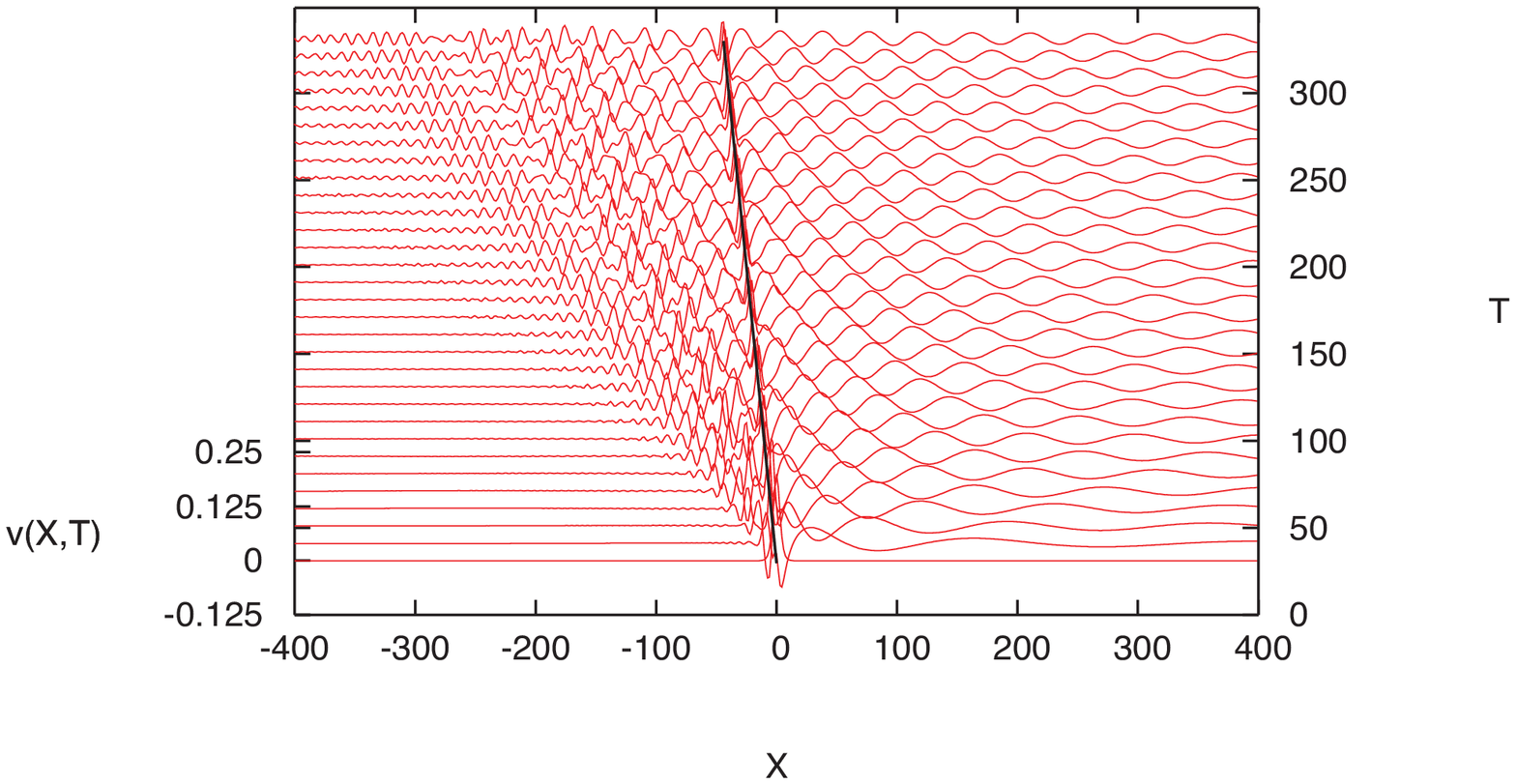}
\caption{Numerical simulations for Case D using a KdV initial condition of weak coupling (\ref{weak}) with $a=0.6$ and $b=0.2$. The solid line in both plots refers to point $A$.}
\label{fig:D1}
\end{center}
\end{figure}

\begin{figure}[htbp]
\begin{center}
\includegraphics[height=3.5cm,width=8cm]{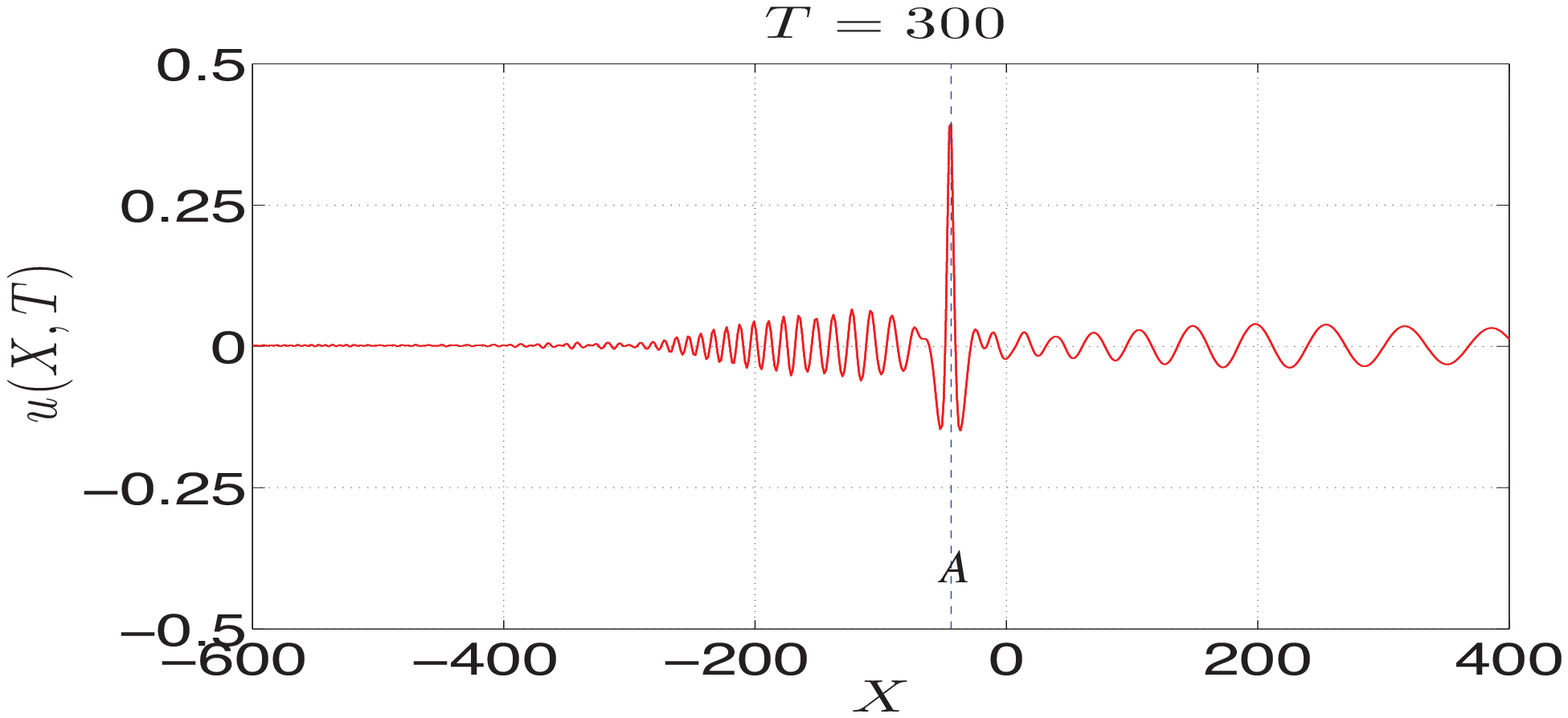}
\includegraphics[height=3.5cm,width=8cm]{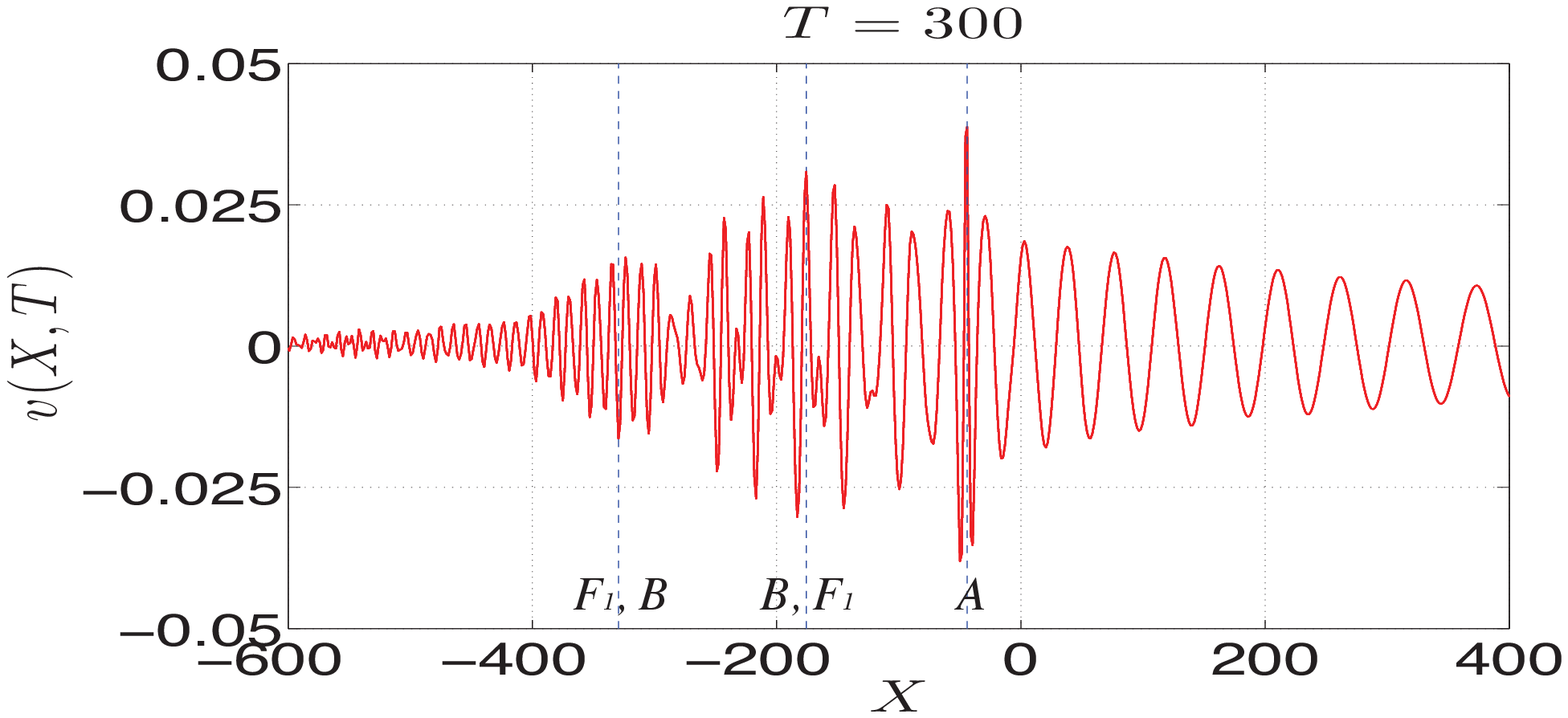}
 \caption{Same as Figure \ref{fig:D1}, but a cross-section at  $T=300$ for both modes.}
\label{fig:D2}
\end{center}
\end{figure}

There are two different wavenumbers to consider when
we use  the wave packet initial condition (\ref{wpic}) corresponding  to the points $A$ and $B$ in Figure \ref{fig:4}.  
First, we choose  $k=k_{s1} =0.3221$ and $R= 10.9729$ corresponding 
to  the point $A$ in Figure \ref{fig:4}, see Table \ref{Table4}. 
The numerical results are shown in Figures \ref{fig:D2a}, \ref{fig:D2b} and we see that the 
solution is dominated by a steady mode $1$ wave packet,
with speed $-0.189$ and ratio $7.934$ in good agreement with the theoretical values.
Another  wave packet can be seen corresponding to the points $B, F_2$  in Figure \ref{fig:4},
  with  speed $-0.846$ and ratio $3.056$. Here there is some interaction between these two points.
 Further, there is a very small wave packet associated with 
the points  $F_2 $ in Figure \ref{fig:4}, with  speed $-1.706$ and  ratio  $0.368$, 
in good agreement to theoretical values, although there may be some contamination here due to the point $D$,
which has a similar speed.  }

\begin{figure}[htbp]
\begin{center}
\includegraphics[height=5cm,width=8cm]{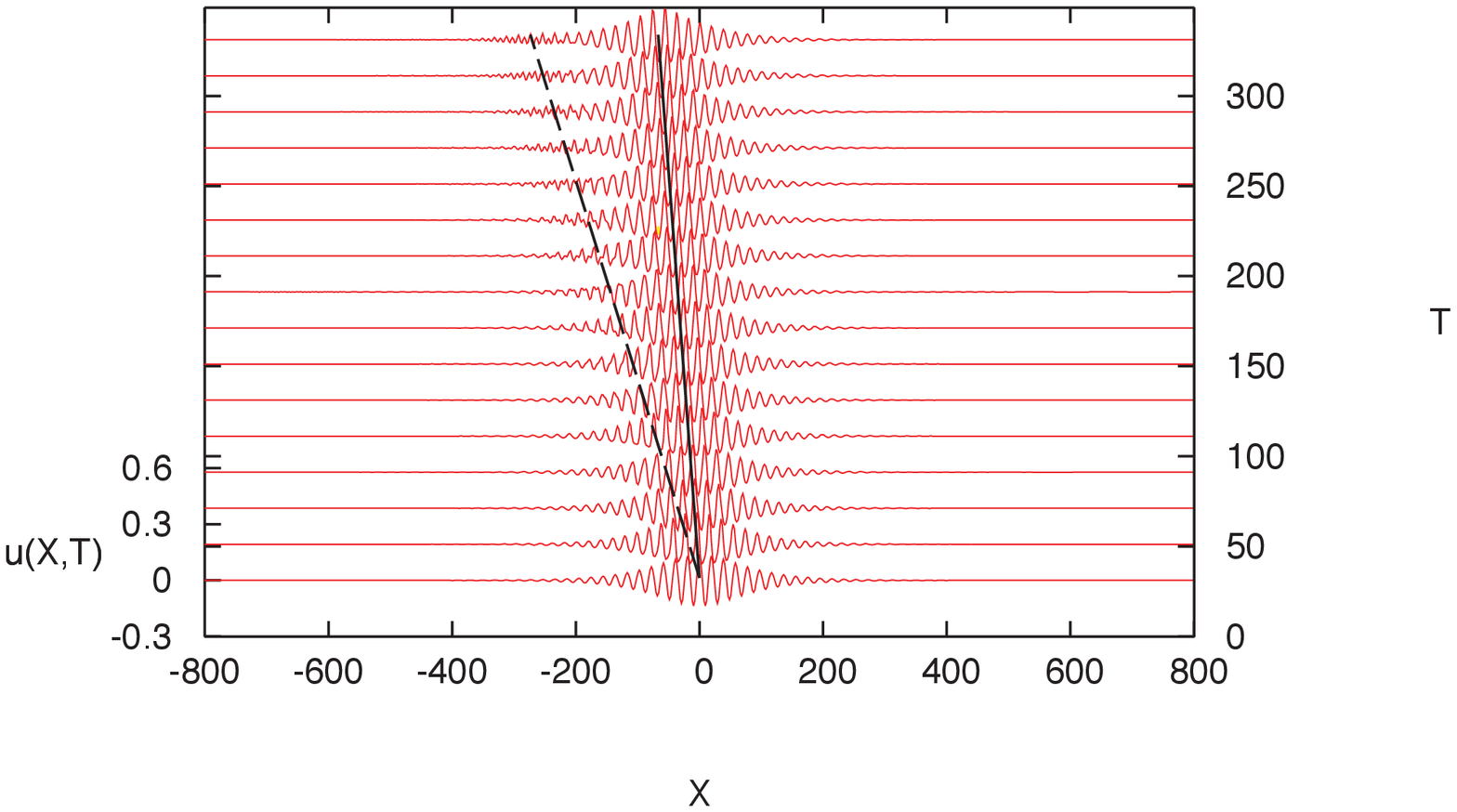}
\includegraphics[height=5cm,width=8cm]{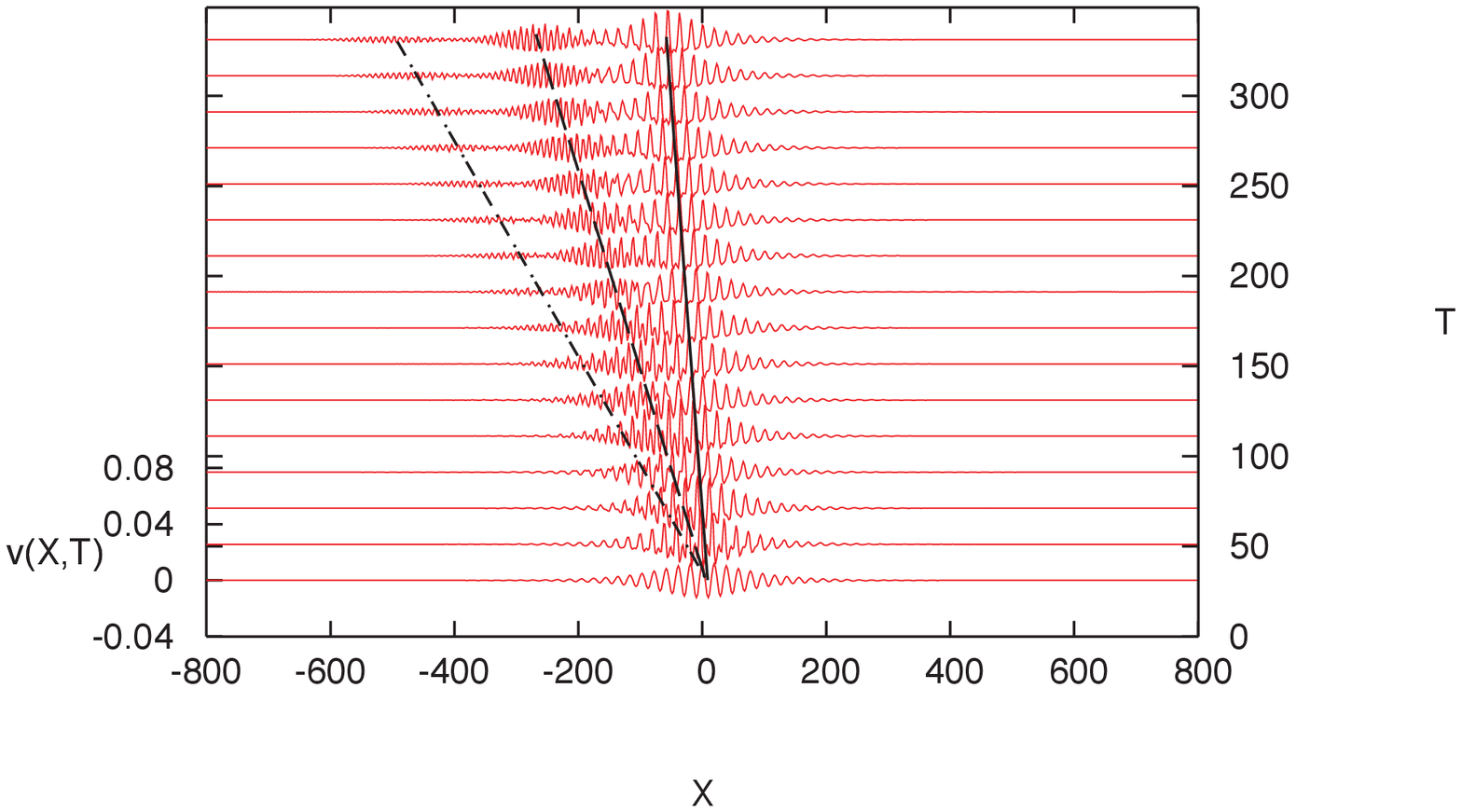}
 \caption{Numerical simulations for Case D using the wave packet initial condition (\ref{wpic}) with $k=k_{m1} = 0.322$ 
 corresponding to the point $A$ in Figure \ref{fig:4}, with $A_0 =0.05, K_0=0.05 \, k, V_0=0.25$. The solid, dashed and dash-dot lines respectively refer to points $A$, ($B, F_1$) and ($F_2,D$).}
\label{fig:D2a}
\end{center}
\end{figure}

\begin{figure}[htbp]
\begin{center}
\includegraphics[height=3cm,width=8cm]{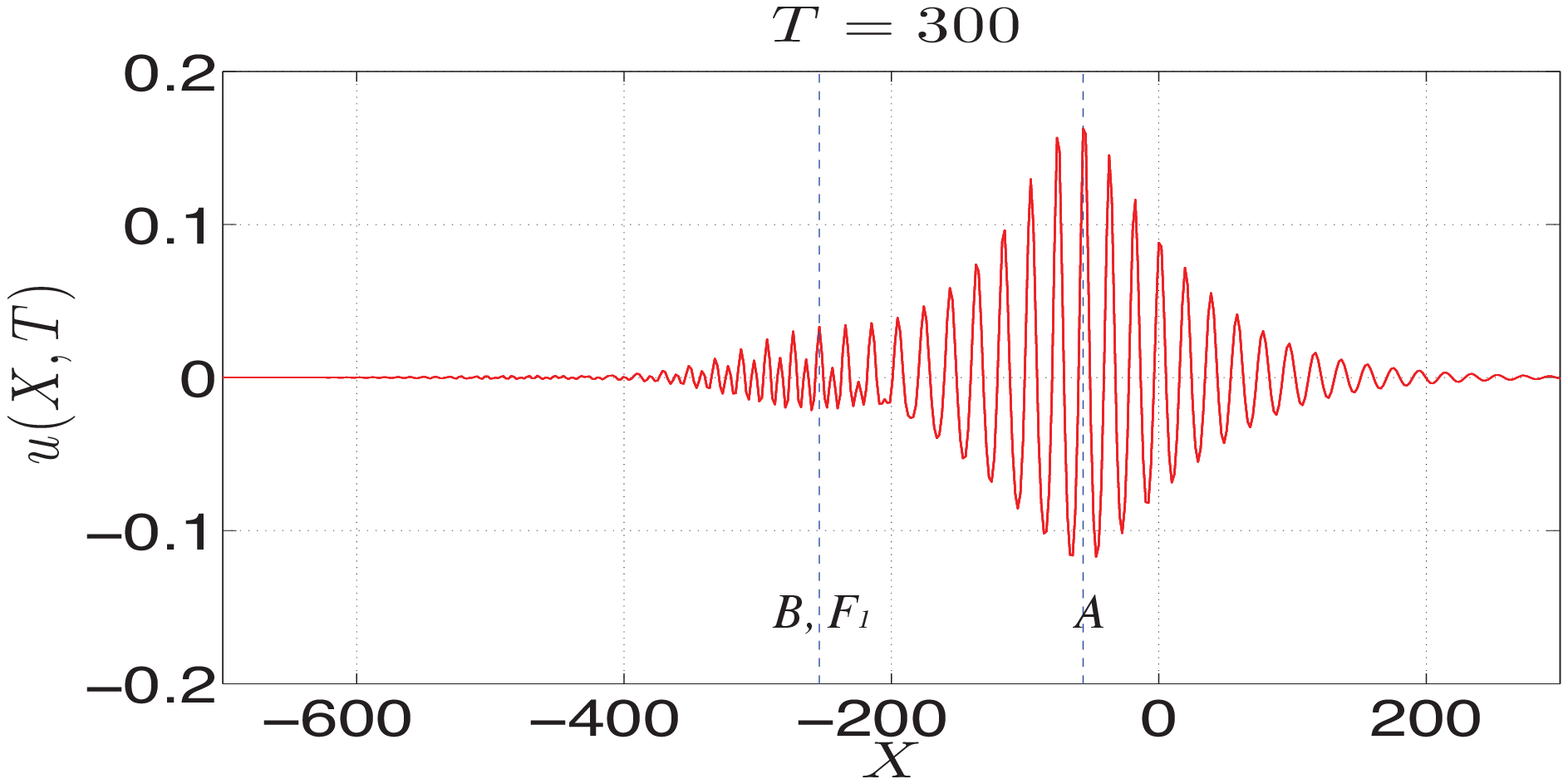}
\includegraphics[height=3cm,width=8cm]{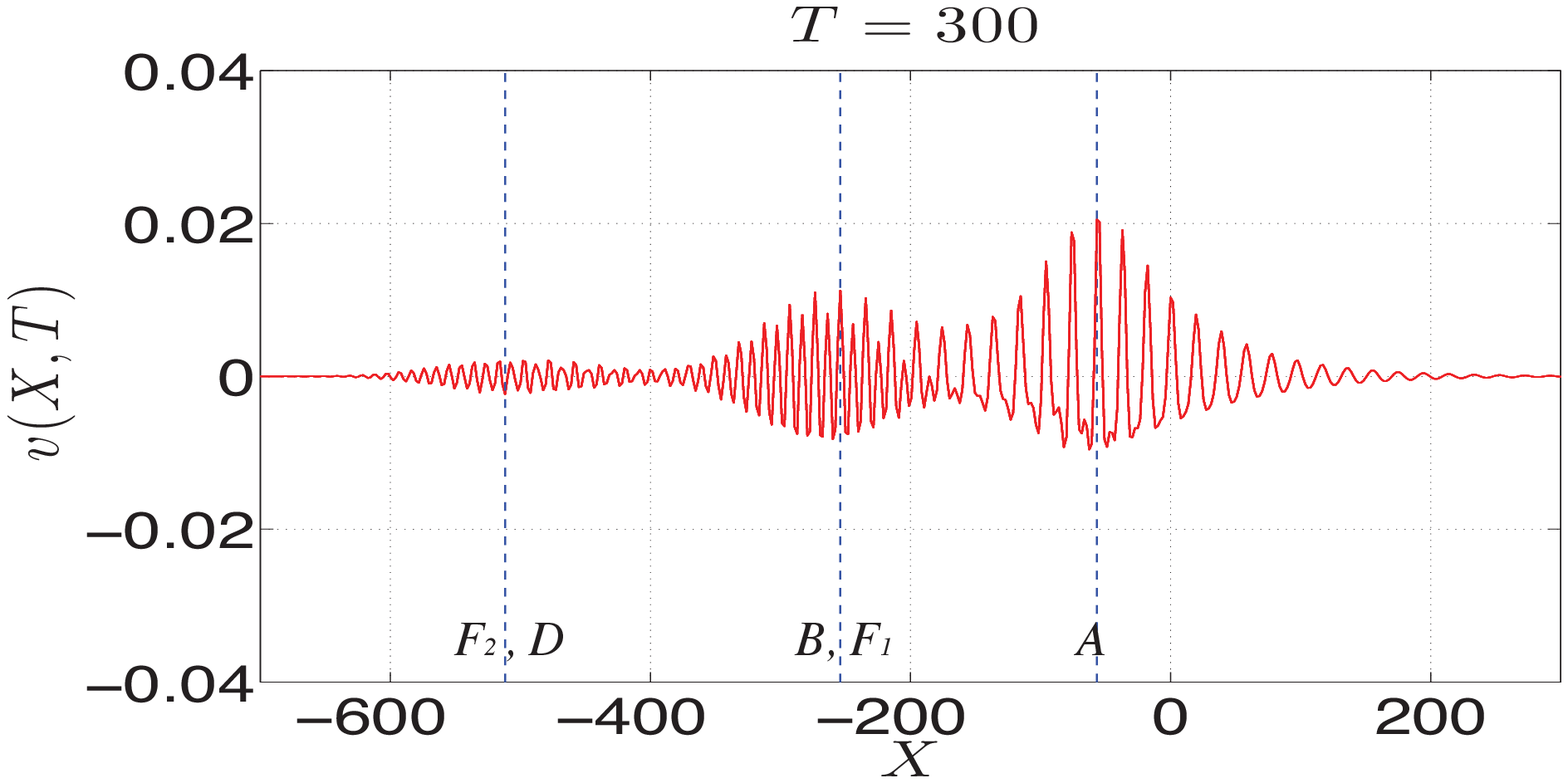}
 \caption{Same as Figure \ref{fig:D2a}, but a cross-section at  $T=300$ for both modes.}
\label{fig:D2b}
\end{center}
\end{figure}

\begin{figure}[htbp]
\begin{center}
\includegraphics[height=5cm,width=8cm]{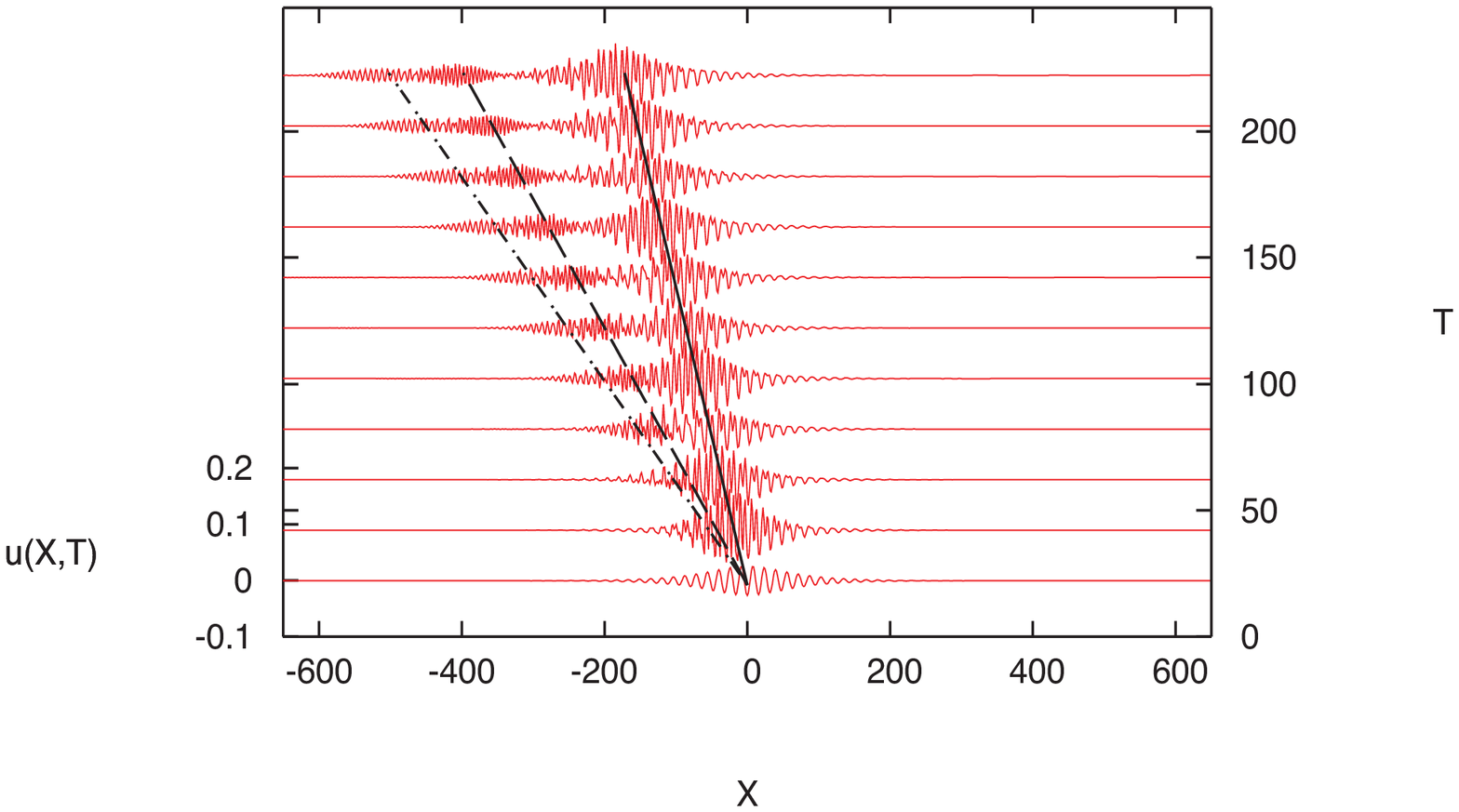}
\includegraphics[height=5cm,width=8cm]{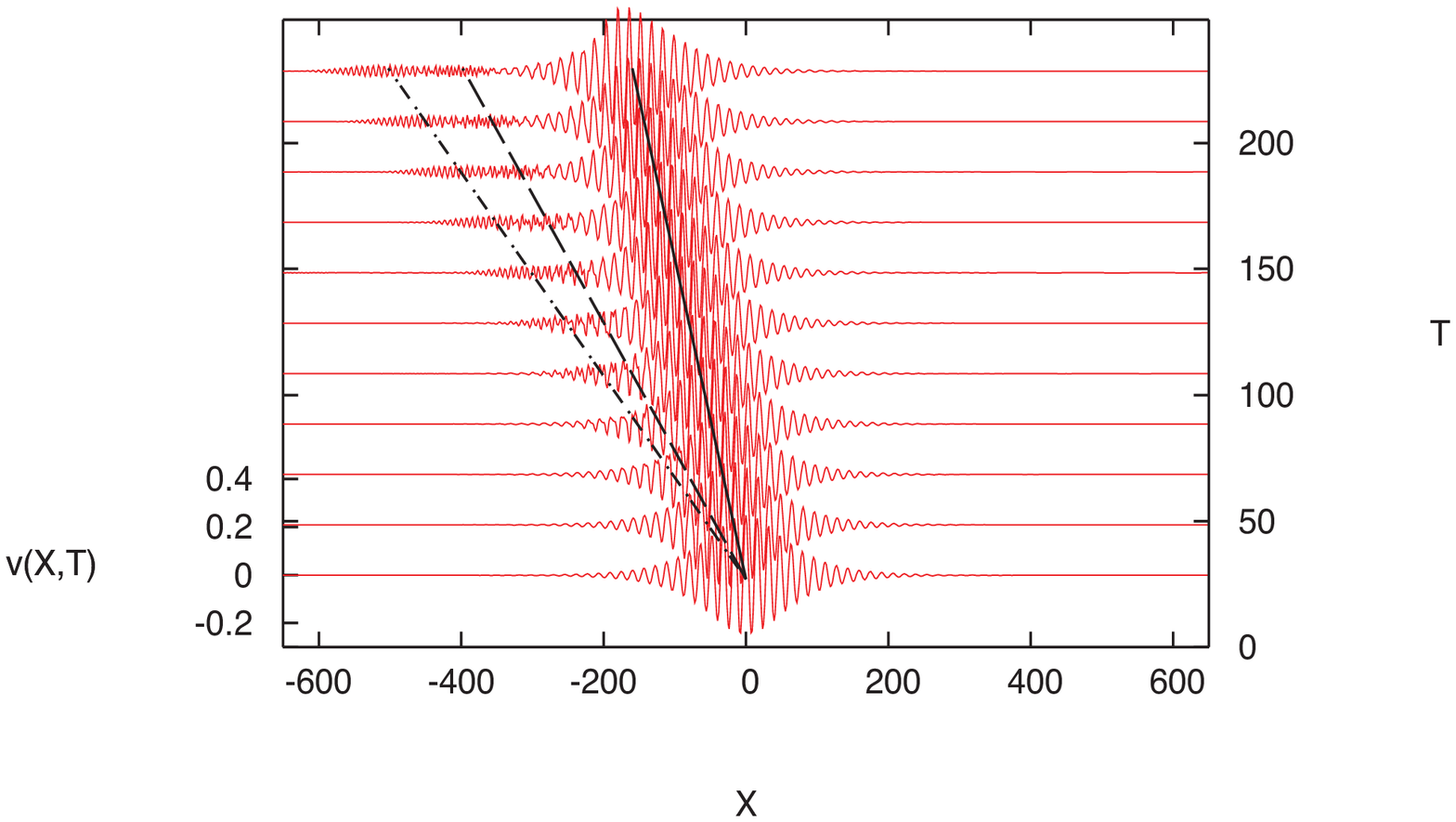}
 \caption{Numerical simulations for Case D using the wave packet initial condition (\ref{wpic}) with $k=k_{s2} = 0.395$,
  corresponding to the point $B$ in Figure \ref{fig:4} with $A_0 =0.25, K_0=0.05 \,k, V_0=1$. The solid, dashed and dash-dot lines respectively refer to points ($B,F_1$), ($F_2, D$) and $G_1$.}
\label{fig:D3a}
\end{center}
\end{figure}

\begin{figure}[htbp]
\begin{center}
\includegraphics[height=3cm,width=8cm]{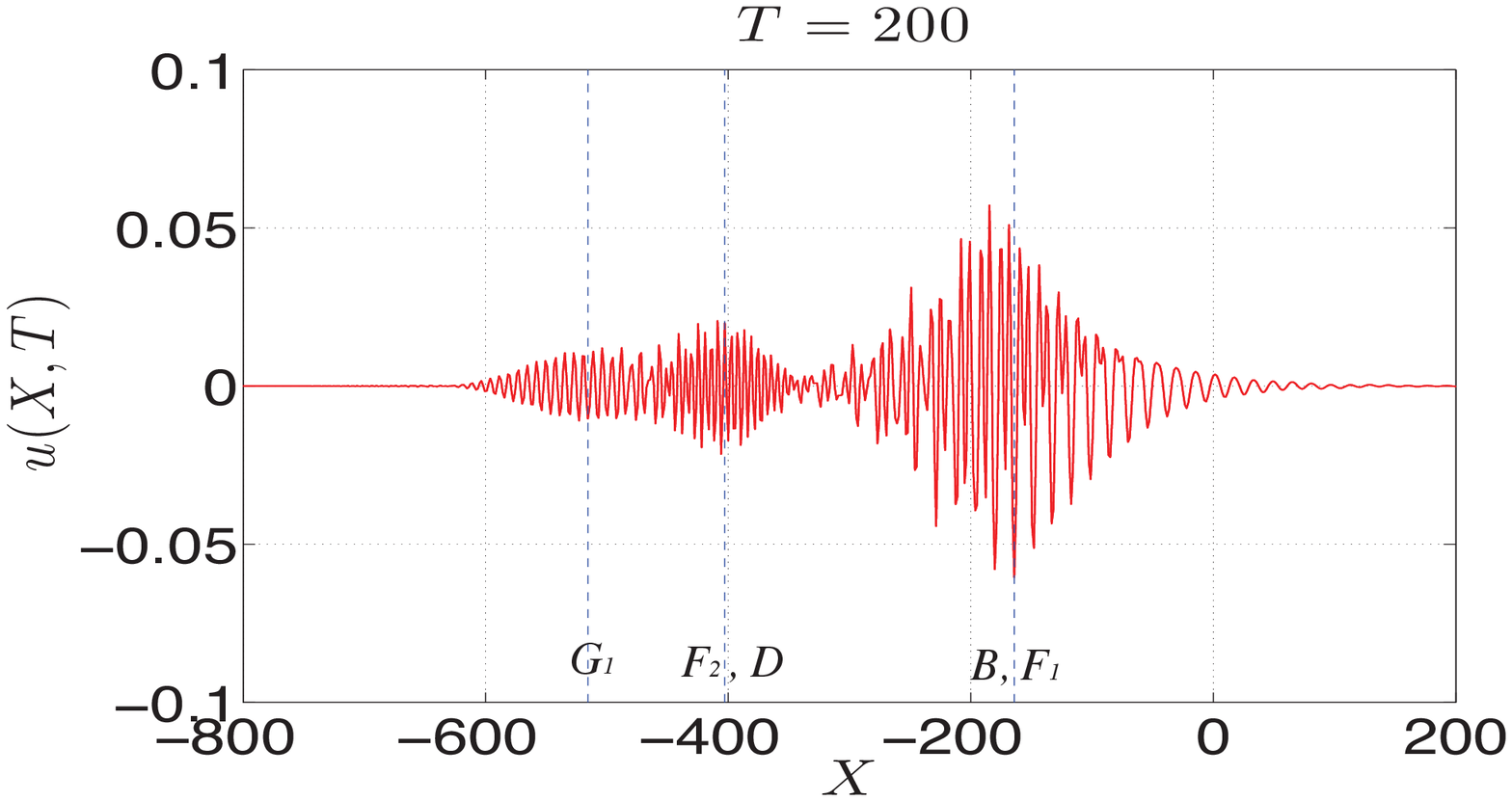}
\includegraphics[height=3cm,width=8cm]{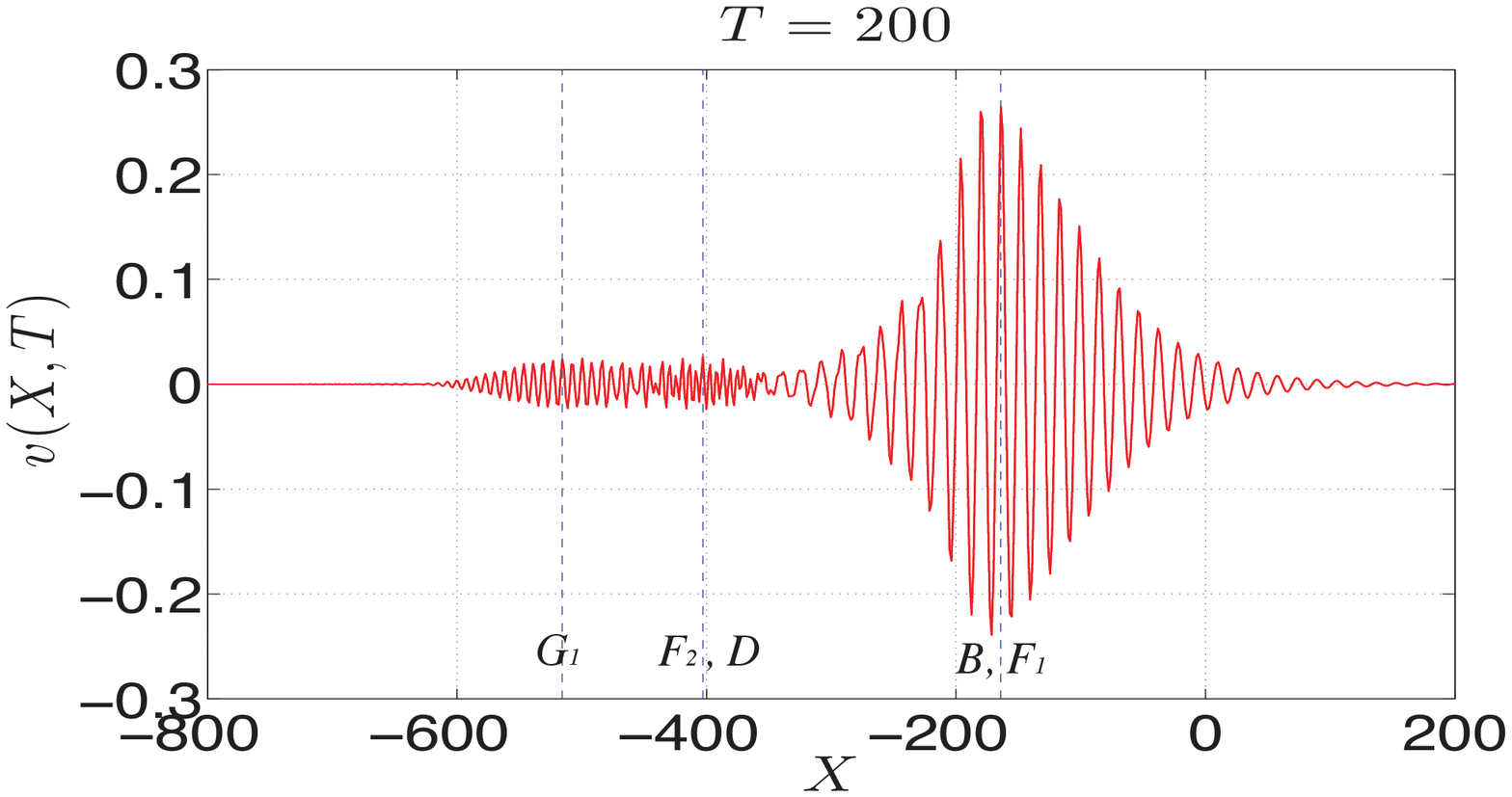}
 \caption{Same as Figure \ref{fig:D3a}, but a cross-section at  $T=200$ for both modes.}
\label{fig:D3b}
\end{center}
\end{figure}

Second, we use the wave packet initial condition (\ref{wpic}) with $k=k_{s2} =0.395$ and ratio, $R= -0.105$ 
corresponding to the point $B$ in Figure \ref{fig:4}, see Table \ref{Table4}.  
The numerical results are shown in Figures  \ref{fig:D3a}, \ref{fig:D3b} and the solution is now dominated by a 
steady mode $2$ wave packet, as expected, with speed $-0.820$ and ratio $-0.229$,
 in good agreement with the theoretical values. There is also some interaction with the 
 point $F_1 $ here, seen in the $u$-plot where two wavenumbers can be seen. 
However, the dispersion curves in Figure \ref{fig:4}
show that here there are potential resonances with mode $1$  at $k=0.1168$ and $k=1.0657$, 
associated with the points $C$ and $D$, see Table \ref{Table4}. 
There is no discernible evidence here of radiation into the wavenumber $k=0.1168$ due to the large ratio of $O(200)$
needed, but a wave packet  is seen with wavenumber $k=1.0657$, indicated by blue vertical line $D$ in Figure \ref{fig:D3b}, 
with the speed $-2.014$ and ratio $0.814$,  in reasonable agreement with the theoretical prediction, although there
could also be some interaction with the point $F_2 $ here, which has quite similar values. 
Another small wave packet can be seen, possibly corresponding to point $G_1 $ in Figure \ref{fig:D3b} with the speed
 $-2.578$ and ratio $0.503$.

\section{Summary and discussion}
\label{discussion}

In this paper, we have  briefly reviewed the derivation of coupled Ostrovsky equations for resonantly 
interacting  weakly nonlinear long oceanic internal waves, presented in detail in our previous work \citet{Alias2013}. 
The resulting system (\ref{O1}, \ref{O2}) describes the evolution of the amplitudes of two linear long wave modes
whose linear long wave phase speeds are nearly coincident.
In an extension of our previous work, here we focus on the effect of a background shear flow, 
 using a three-layer model as a guide to the possible values that the normalised coefficients may take.
 The  significant difference that emerges is that the coefficients $\beta , \mu $ of the
rotation terms in the coupled Ostrovsky equations (\ref{CO1}, \ref{CO2}), are not necessarily equal, or indeed
positive, as is the case in the absence of a background shear flow. Instead,  there are
four essentially different cases corresponding to different  sign combinations of $\beta$ and $\mu$.

Then the system was examined numerically, using two different initial conditions. First,  the initial condition was a
 solitary wave type, based on an approximation to the coupled KdV systems obtained when the
 rotation terms are removed, and for which there is no {\it a priori} wavenumber selection. 
Second, the initial condition was a wave packet  based on certain predicted wavenumbers, obtained 
 from the linear dispersion relation where either the phase velocity, or the group velocity, has a turning point.
 The former can be associated with the possible emergence of a nonlinear steady wave packet, and the 
 latter with the possible emergence of an unsteady nonlinear wave packet. These two contrasting 
 scenarios were examined numerically for each of the four cases. In each case we can identify
 these predicted wave packets as the dominant feature of the numerical solution. However,
 in many cases there was also evidence of nonlinear interactions generating other wave packets associated
 with some of the possible resonant points identified on each linear dispersion curve. Thus,
 in comparison with the simulations of the single Ostrovsky equation reported by \citet{Grimshaw08}
 where only a single unsteady nonlinear wave packet typically emerges, the coupled system
 (\ref{O1}, \ref{O2}) can support a wide variety of nonlinear wave packets. 
  Importantly, we have shown that the dominant features of the observed dynamical behaviours 
  can be classified and interpreted in terms of the main features of the relevant dispersion curves. 
  This is a first step towards predicting the long-time asymptotic behaviour  of solutions of the initial-value problems 
  for this coupled system of equations. 
  
  Although we have used a particular three-layer model to illustrate the range of possible scenarios, 
  based in particular on the signs of the rotational coefficients $\beta ,  \mu $, we suggest that similar combinations of
  stratification and current shear will lead to the same range of possible sign combinations, and hence to the same range of
  complex dynamical behaviour. Thus we expect that these kinds of nonlinear wave packets may be found under 
  certain oceanic conditions, and could be possibly observed in laboratory experiments, similar to that of
  \citet{Grimshaw13b} for the generation of the unsteady wave packets described by the single Ostrovsky equation. 
 Of course, in reality in the ocean the wave packets found here may be affected by dissipation and the competing 
 effects of topography as the waves shoal shoreward, see \citet{Grimshaw14}. Nevertheless, they 
 can provide a useful framework for the interpretation of the observed wave phenomena.

\section{Acknowledgements}
One of the authors, A.Alias, is supported by Universiti Malaysia Terengganu and the Ministry of Higher Education of Malaysia. 


\end{document}